\documentclass[a4paper,11pt]{article}
\pdfoutput = 1
\usepackage{jheppub}
\usepackage{amsmath,amssymb,amscd,braket,amsfonts}
\usepackage{color}
\usepackage{graphicx}
\usepackage{slashed}
\usepackage{mathtools}
\usepackage{amsthm}
\usepackage{amsmath}
\usepackage{tikz}
\usetikzlibrary{calc}
\usetikzlibrary{positioning}
\usepackage{booktabs}
\usepackage{makecell}
\usepackage{subcaption}
\usepackage{comment}
\usetikzlibrary{decorations.markings}
\usetikzlibrary{patterns.meta}
\usetikzlibrary { decorations.pathmorphing, decorations.pathreplacing, decorations.shapes,calc }

\newcommand{\bea}{\begin{eqnarray}}
\newcommand{\eea}{\end{eqnarray}}
\newcommand{\be}{\begin{equation}}
\newcommand{\ee}{\end{equation}}
\newcommand{\ba}{\begin{align}}
\newcommand{\ea}{\end{align}}

\newcommand{\rmi}{\mathrm{i}}
\newcommand{\rme}{\mathrm{e}}

\title{Analytic approaches to perturbations of strongly coupled Yang-Mills plasma}

\date{}

 \author[a]{In\^es Aniceto,}
 \author[a]{Paolo Arnaudo,}
 \author[a]{Alex Ratcliffe,} 
 \author[b,c]{Micha\l\ Spali\'nski} 
 \affiliation[a]{School Mathematical Sciences and STAG Research Centre, University of Southampton, Highfield, Southampton SO17 1BJ, UK}
\affiliation[b]{National Centre for Nuclear Research, 02-093 Warsaw, Poland}
\affiliation[c]{Physics Department, University of Białystok,
15-245 Bia\l ystok, Poland}

 \emailAdd{i.aniceto@soton.ac.uk}\emailAdd{p.arnaudo@soton.ac.uk}
 \emailAdd{a.ratcliffe@soton.ac.uk}\emailAdd{michal.spalinski@ncbj.gov.pl}

\abstract{
We study perturbations of Yang-Mills plasma, represented by scalar quasinormal modes of AdS black branes, as functions of the wave number $q$ in the entire range from zero to infinity. At finite $q$, these modes can be computed by classical spectral methods based on truncating the boundary value problem. We show that this truncation admits a natural analytic interpretation in terms of quantum Seiberg--Witten periods in the Nekrasov--Shatashvili limit, with the spectral condition organised as an instanton expansion around small values of the counting parameter. The physical black-brane problem corresponds to evaluating this series at a finite value of the counting parameter, and the Seiberg--Witten formulation provides a systematic way to analyse when the truncation is under control. In particular, it reveals that, as $q$ or the mode number $N$ increases, the physical point approaches the boundary of the domain of convergence of the instanton expansion, limiting the validity of the truncation approach. 
We overcome this limitation through an exact WKB analysis in which $q^{-1}$ acts as the expansion parameter. 
The resulting exact quantisation conditions, expressed in terms of period integrals and the associated Stokes geometry, incorporate both perturbative and non-perturbative corrections. The resummed quasinormal modes remain accurate far beyond the strict large-$q$ regime and can be analytically continued all the way to $q=0$, notably by using the Seiberg--Witten approach, providing a consistent description of the QNM spectrum.
}

\begin{document}

\maketitle

\section{Introduction}

Quasinormal modes (QNMs) appear in a number of physical contexts in which dissipative systems are perturbed, including black holes, whose horizons act as dissipative boundaries, as well as viscous fluids~\cite{Berti:2009kk}. These two classes of problems are in fact united by the AdS/CFT correspondence, which implies that thermal phenomena in certain gauge theories in Minkowski space can be described in terms of black objects in asymptotically AdS spacetimes~\cite{Son:2002sd, Policastro:2002se, Starinets:2002br, Nunez:2003eq, Kovtun:2005ev}. This correspondence has provided a powerful tool to study properties of nonequilibrium systems in four dimensions using calculations in classical gravity in five dimensions. In particular, the quasinormal modes of  black branes are related to the dispersion relations of strongly coupled supersymmetric Yang-Mills theory.

The interest in QNMs follows from their role in the generation of gravitational waves, whose observational study is currently an area of intense activity~\cite{LIGOScientific:2016aoc, Berti:2025hly, LIGOScientific:2025rid}. In the AdS/CFT context, QNMs have been the focus of great interest in studies of thermalisation in strongly coupled gauge theories. Finally, the analytic properties of QNMs are of intrinsic importance because of the role they play in maintaining causality in relativistic systems~\cite{Heller:2022ejw,Heller:2023jtd}.

The task of determining quasinormal mode frequencies reduces to an eigenvalue problem for a second-order linear ODE obeying a specific set of boundary conditions. In the case of the black branes of interest in this paper, these conditions reflect the presence of a horizon, which acts as a dissipative boundary at one end, and the AdS conformal boundary at the other. While eigenvalue problems are very familiar from quantum mechanics, determining QNMs is mathematically more delicate, since the relevant differential operators are typically not self-adjoint, and the eigenvalues are complex resonances rather than bound states. Nevertheless, there is a common strategy which involves determining local solutions near the boundaries, each satisfying one of the boundary conditions, and then analytically continuing and matching them (see \cite{10.1093/oso/9780198505730.001.0001} for a general picture in the context of ODEs relevant for our discussion).

At finite values of the wave number $q$, the boundary value problem can be rewritten as a complex curve $F(\omega,q)=0$, where $F$ is given by an infinite series whose terms are rational functions of both $\omega$ and $q$ \cite{Withers:2018srf, Grozdanov:2019kge, Grozdanov:2019uhi, Heller:2020uuy}. This complex curve implicitly defines the QNM frequency $\omega$ as a function of $q$. In practice, however, solving it requires truncating the number of terms entering $F$. This immediately raises a number of questions. Do higher truncations necessarily lead to better results for the QNM frequencies? Since higher truncations typically produce a larger number of candidate roots, how does one identify the physical solutions? More generally, can one determine the QNM frequencies as analytic functions of $q$, for example through their Taylor series expansions?

These questions have been partially answered in the literature. The seminal work of Starinets \cite{Starinets:2002br} focused on the scalar mode equation at $q=0$, where the QNM frequencies are expressed in terms of continued fractions. In that setting, truncating the complex curve corresponds to truncating the associated continued fraction, and it was shown that this procedure converges and reproduces the correct scalar QNMs at $q=0$. The same procedure can be generalised to determine the coefficients of the Taylor expansion of $\omega(q)$ around $q=0$. In principle, one can also apply this approach at finite $q$, but in that case the convergence of the corresponding continued fraction must be reassessed. Other works have studied non-hydrodynamic QNMs as functions of $q$, obtaining analytic continuations of their small-$q$ expansions by means of Padé approximants, and successfully predicting crossings between modes in the complex plane \cite{Withers:2018srf, Grozdanov:2019kge, Heller:2020uuy, Heller:2020hnq}. Even so, a systematic understanding of when a given truncation of the complex curve can be trusted is still missing: is the truncation order high enough to determine the QNM frequencies and their Taylor coefficients correctly? Does the sequence of truncations  converge at all? In this paper we will show that mapping this problem to a rather different physical system leads to new insights and substantial developments.

We begin by observing that the analytic continuation of  local solutions can be expressed in terms of a basis of periods in the geometry underlying the problem. These periods, together with their quantisation, will be of crucial importance for our analysis. Allowing for a slight generalisation, we show that the geometry relevant to our QNM problem is an instance of a Seiberg--Witten (SW) quantum curve. 
This correspondence was previously studied in Refs.~\cite{Aminov:2020yma, Bonelli:2021uvf, Bianchi:2021mft, Fioravanti:2021dce, Dodelson:2022yvn, daCunha:2022ewy, Aminov:2023jve, Lei:2023mqx, Jia:2024zes, Aminov:2024mul, Arnaudo:2024sen, Ren:2024hwf, Arnaudo:2025kof}. In the language of $\mathcal{N}=2$ supersymmetric gauge theory in the Nekrasov--Shatashvili (NS) limit, the QNM quantisation condition can then be expressed in terms of the quantum periods of the underlying SW geometry as a series expansion in the instanton counting parameter $t$, which is usually regarded as small. Interestingly, this quantisation condition can also be rewritten, much as in Starinets original treatment, as a continued fraction, with the small parameter $t$ controlling the convergence of the truncation. The original QNM problem is recovered by setting $t=1/2$. The convergence of the instanton series in $t$, or equivalently of the associated continued fraction, therefore sets a natural limit on the effectiveness of the method and the truncation of the complex curve. In particular, we will show that for large-$q$ or large mode number $N$ one approaches the radius of convergence of this series, so that very high truncation orders are required. Moreover, although the SW formalism provides a controlled framework in which to study the complex curve and its truncations, it does not by itself remove the ambiguity in selecting the physically relevant branch when one solves the truncated problem directly. Indeed, at leading order a truncation of the SW expansion typically produces as many candidate solutions as the truncation order itself, including roots that do not correspond to genuine QNMs. From this viewpoint, the appearance of spurious solutions is not an accident but an inherent feature of a direct truncated analysis.

It is therefore natural to ask whether there is a different approach which identifies the physical branch more directly and avoids this ambiguity from the outset, while also determining the analytic behaviour of the QNM frequencies. In this paper we show that such an approach is provided by studying the large-$q$ regime rather than expansions valid for small $q$. To do so, we switch to an asymptotic analysis based on the Liouville--Green/WKB method. Calculations of this type have already been used in a variety of contexts for quasinormal modes, both in asymptotically flat and in asymptotically AdS spacetimes, and have led to important results. In our work we cast the problem of determining the QNMs as a singularly perturbed problem where the small parameter playing the role of $\hbar$ in a WKB calculation is $q^{-1}$. Such a framing has appeared earlier in the work of Festuccia and Liu \cite{Festuccia:2008zx} and in later developments by Fuini, Uhlmann, and Yaffe \cite{Fuini:2016qsc}.

Our treatment goes further by incorporating non-perturbative corrections that are essential for extending the validity of the asymptotic description far beyond the strict large-$q$ regime. More precisely, the exact WKB framework \cite{AIHPA_1983__39_3_211_0,kawai2005algebraic} allows us to write exact quantisation conditions for the QNM frequencies, formulated in terms of periods and of the Stokes geometry underlying the problem, rather than only formal asymptotic expansions in restricted limits. These quantisation conditions involve the QNM mode number $N$ and can be solved recursively, analogously to Bohr--Sommerfeld quantisation conditions valid at large $N$. However, here they generate expansions in both $N$ and $q$ that include perturbative and non-perturbative contributions. 
A key advantage of the WKB description is that it does not suffer from the ambiguities associated with solving a truncated complex curve directly. Rather than producing multiple competing leading-order roots, the WKB analysis selects the relevant branch through the geometry of Stokes lines and the associated quantisation conditions.

The resulting asymptotic series require resummation, which we explain in detail in section \ref{sec:resurgenceofQNMs}. Once resummed, they can be evaluated at finite values of $q$ and $N$ and used to extract the Taylor expansion of QNM frequencies at finite $q$. Although the resummation procedure does not itself allow one to set $q=0$ directly, the combination with the SW approach provides precisely the missing bridge: it furnishes a way to continue analytically from the finite-$q$ regime, where WKB gives an efficient and unambiguous description, to the small-$q$ region and ultimately to $q=0$. In this sense, the WKB method provides a natural seed for the computation around small values of $q$: once the correct leading coefficient is identified, the subleading coefficients are generated recursively and unambiguously by the SW method. This makes the link between the WKB and SW approaches particularly powerful. On the one hand, the WKB method naturally identifies the physical solution and organises its large-$q$ expansion. On the other hand, the SW construction provides a complementary framework in which one can analytically continue that solution from a finite value of $q$, where the WKB description is accurate, all the way to $q=0$.

After reviewing the problem of determining the scalar QNMs of the AdS black brane in section \ref{sec:setup}, we describe the SW approach in section \ref{sec:gauge} and show how it can be used to calculate QNM frequencies and to analyse the convergence of the corresponding truncation scheme. We then identify the natural limitations of this method at large values of $q$, which motivate the Liouville--Green/WKB approach developed in section \ref{sec:WKB method}. In section \ref{sec:resurgenceofQNMs} we discuss the resummation procedure, including the non-perturbative contributions. Finally, in section \ref{sec:results}, we continue our results all the way to $q=0$ by combining the two approaches.

\section{Black brane quasinormal modes in the scalar channel}\label{sec:setup}

We study the dynamics of helicity $\pm 2$ linearised perturbations on the background geometry of an AdS5 black brane, which is dual to $\mathcal{N}=4$ SYM theory. We follow the notations used in \cite{Fuini:2016qsc}.

The background metric in infalling coordinates $(t,\vec{x},u)=(t,x_1,x_2,x_3,u)$ reads
\begin{equation}
\mathrm{d}s^2=u^{-2}\left[-2\mathrm{d}t\mathrm{d}u-\left(1-\left(\pi T\right)^4u^4\right)\mathrm{d}t^2+\mathrm{d}\vec{x}^2\right],
\end{equation}
where $T$ is the temperature at the horizon, and we normalised the AdS radius $R=1$.
The horizon is located at $u=\left(\pi T\right)^{-1}$, and the conformal boundary at $u=0$. For simplicity, we will work in units of $\pi T$.

We denote with $q$ the magnitude (in units of $\pi T$) of the wavevector $\vec{q}$ which is assumed to lie along the direction $x_3$. We consider a metric perturbation of the form
\begin{equation}
\delta g=2u^2\,h(u)\,\rme^{\rmi(q\,x_3-\omega\,t)}\mathrm{d}x_1\mathrm{d}x_2,
\end{equation}
for some undetermined function $h(u)$, and where $\omega$ is the frequency of the perturbation. In terms of the variable $z=u^2$, and redefining the wave function $h$ as discussed in \cite{Fuini:2016qsc}, the only non-trivial part of the linearized Einstein’s equations is encoded in the differential equation
\begin{equation}\label{HeunforWKB}
\psi''(z)-\left[q^2\,V_0(z)+V_2(z)\right]\psi(z)=0,
\end{equation}
where
\begin{equation} \label{scalarPotential}
V_0(z)=\frac{1-s^2-z^2}{4 z \left(1-z^2\right)^2},\qquad
V_2(z)=\frac{3-z^4-6 z^2}{4 z^2 \left(1-z^2\right)^2},
\end{equation}
and where $s\equiv \omega/q$.

The AdS boundary is at $z=0$ and the BH horizon is at $z=1$. For the quantisation of the quasinormal modes, we impose the ingoing boundary condition at the horizon and the normalisability at the boundary. In terms of $\psi(z)$, these boundary conditions require that
\begin{equation}\label{boundcond2}
\psi(z)= \begin{cases}
    (1-z)^{\frac{1}{2}-\frac{\rmi\,q\,s}{4}}\left[1+\mathcal{O}(1-z)\right],&\quad z\to 1,\\
z^{\frac{3}{2}}\left[1+\mathcal{O}(z)\right],&\quad z\to 0.
\end{cases}
\end{equation}

The differential equation \eqref{HeunforWKB} is a Heun differential equation
\cite{heun1888theorie, ronveaux1995heun}
with singularities at $z=0,\pm 1,\infty$.
For the analysis we perform in the following section, it will be convenient to analyse it in a new variable
\begin{equation}
y=\frac{1-z}{2},
\end{equation}
so that the ODE is of the general form
\begin{equation}\label{heunnormal}
\begin{aligned}
0=\,&\psi''(y)+ \\
&+\left[\frac{a_0^2+a_1^2+a_t^2-a_{\infty}^2-\frac{1}{2}}{y (y-1)}+\frac{\frac{1}{4}-a_0^2}{y^2}+\frac{\frac{1}{4}-a_1^2}{(y-1)^2}+\frac{\frac{1}{4}-a_t^2}{(y-t)^2}+\frac{(t-1) c}{y (y-1) (y-t)}\right]\psi(y)\,,
\end{aligned}
\end{equation}
with\footnote{A priori 8 different dictionaries can be obtained, because only the square of the $a$-parameters are determined. Moreover, we denote the accessory parameter with $c$ instead of the usual notation $u$, to avoid clash of notations.}
\begin{equation}\label{dictioHeun}
\begin{aligned}
t&=\frac{1}{2},\qquad c= \frac{q^2}{4} \left(1-s^2\right),\\
a_{\infty}&= 0,\qquad a_{t}= 1,\qquad a_0= \frac{\rmi\,q\,s}{4},\qquad
a_1= \frac{q\,s}{4}.
\end{aligned}
\end{equation}
In this variable, the horizon is at $y=0$ and the boundary is at $y=1/2$.

The local solutions of the Heun equation around a singular point $\hat{y}$ are defined by convergent Frobenius expansions
\begin{equation}\label{frobeniussol}
\psi_{\pm}^{(\hat{y})}=\left(y-\hat{y}\right)^{\frac{1}{2}\pm a_{\hat{y}}}\left[1+\mathcal{O}\left(y-\hat{y}\right)\right],
\end{equation}
whose coefficients satisfy a three-term recurrence relation. 
In turn, any solution of a three-term recurrence relation naturally gives rise to a continued fraction representation for the ratio of consecutive terms. 
The latter representation was extensively used to study boundary value problems. For example, in \cite{Leaver:1985ax}, the quasinormal mode wavefunctions around a Kerr black hole -- which can be described in terms of confluent Heun functions -- were represented in this way to study the values of the quasinormal mode frequencies. A further example, highly related to our analysis, can be found in \cite{Starinets:2002br}, where the scalar mode equation at $q=0$ was considered, and the QNM
frequencies were expressed in terms of continued fractions.
The possibility of rewriting the solution of a three-term recurrence relation in terms of a continued fraction representation is subject to convergence assumptions. The convergence properties are usually not straightforward to analyse, and a possibility is to truncate the representation at different orders and study the convergence of the result, as done in \cite{Starinets:2002br}. We will see in the next section that realising that the three-term relation arises from a Heun differential equation permits to have control on the convergence properties of the continued fraction.

\section{The Seiberg-Witten perspective}\label{sec:gauge}

In this section, we study the gravitational problem from the point of view of the Heun differential equation and its connection to Seiberg-Witten theory, writing the quantisation condition in terms of connection formulae for the Heun solutions, and analysing the corresponding convergence properties.

\subsection{The Heun equation and its connection formula}

The Heun differential equation arises in the contexts of Liouville conformal field theory (CFT) and four-dimensional $\mathcal{N}=2$ supersymmetric gauge theory, which are dual to each other under the AGT correspondence \cite{Alday:2009aq}. 

More precisely, in Liouville CFT a correlation function with four primary insertions and a degenerate field insertion satisfies a PDE equation, known as the BPZ equation, that becomes a Heun equation in the semiclassical limit. The gauge theory dual to Liouville CFT on a four-punctured sphere is $\mathcal{N}=2$ $SU(2)$ gauge theory with $N_f=4$ fundamental hypermultiplets. The connection between the Heun equation and its confluent forms with supersymmetric gauge theory and Seiberg-Witten theory was already studied in \cite{Aminov:2020yma}. We recall the relations between the Heun equation and the $N_f=4$ quantum Seiberg-Witten curve in appendix \ref{app:SW}.
In \cite{Bonelli:2022ten}, the Liouville CFT technology was used to compute the connection formulas relating the local solutions of the Heun equation around different singular points.

In terms of the ODE \eqref{heunnormal} and the basis of solutions \eqref{frobeniussol}
the boundary conditions in \eqref{boundcond2} select the local solutions 
\begin{equation}\label{boundcond}
\begin{aligned}
\psi_-^{(0)}(y)&=y^{\frac{1}{2}-a_0}\left[1+\mathcal{O}(y)\right],\quad y\to 0,\\
\psi_+^{(t)}(y)&=(y-t)^{\frac{1}{2}+a_t}\left[1+\mathcal{O}(y-t)\right],\quad y\to t.
\end{aligned}
\end{equation}
Following the conventions in \cite{Bonelli:2022ten}, the relevant connection formula for our problem reads
\begin{equation}\label{connformula}
\psi_{\theta}^{(t)}(y)= \rme^{\rmi\pi\left(\frac{1}{2} + \theta a_{t}\right)} 
\sum_{\theta'=\pm} \frac{\Gamma\left(1+2\theta\,a_{t}\right)\Gamma\left(-2\theta'\,a_0\right)}{\prod_{\sigma=\pm}\Gamma\left(\frac{1}{2}+\theta\,a_{t}-\theta'\,a_0+\sigma\,a\right)}
t^{\theta a_{t}-\theta' a_0} \rme^{\frac{\theta}{2} \partial_{a_{t}} F\left(t\right)-\frac{\theta'}{2} \partial_{a_0} F\left(t\right)} \psi_{\theta'}^{(0)}(y),
\end{equation}
where $F(t)$ denotes the instanton part of the Nekrasov-Shatashvili free energy of $\mathcal{N}=2$ $SU(2)$ supersymmetric gauge theory with $N_f=4$, and $a$ is the v.e.v. of the adjoint scalar in the vector multiplet that parametrizes the composite monodromy around the singularities at $y=0$ and $y=t$. We refer to appendix \ref{app:NS} for the conventions we use.

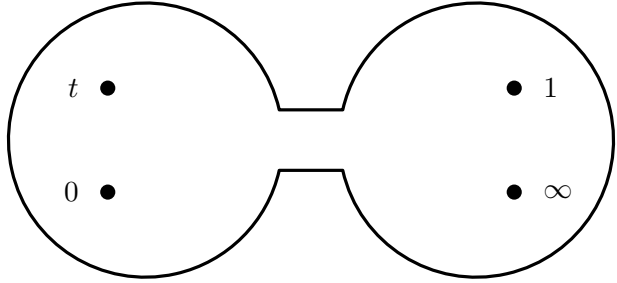
\begin{figure}[h]
    \centering
\begin{tikzpicture}[scale=1.25, line join=round, line cap=round]

\draw[line width=1.2pt]
  (-0.336,0.320)
  arc[start angle=12.75, end angle=347.25, radius=1.45]
  -- (0.336,-0.320)
  arc[start angle=192.75, end angle=527.25, radius=1.45]
  -- cycle;

\fill (-2.15,0.55) circle (0.08);   
\fill (-2.15,-0.55) circle (0.08);  
\fill (2.15,0.55) circle (0.08);    
\fill (2.15,-0.55) circle (0.08);   

\node[left] at (-2.35,0.55) {$t$};
\node[left] at (-2.35,-0.55) {$0$};
\node[right] at (2.35,0.55) {$1$};
\node[right] at (2.35,-0.55) {$\infty$};

\end{tikzpicture}
    \caption{The singularity structure decomposition of the 4-punctured sphere describing the $y$ space.}
    \label{fig:4point}
\end{figure}

An important remark is needed. To derive the connection formula \eqref{connformula}, we have to assume a specific pants decomposition of the four-punctured sphere describing the $y$ space of the Heun differential equation. In particular, we work in the regime of small $t$, even though in the gravitational problem $t=1/2$, see figure \ref{fig:4point}. The dependence of the connection coefficients on the parameter $t$, which corresponds to the instanton counting parameter in the gauge theory language, permits us to consistently organise and truncate the continued fraction representation needed for the quantisation condition of QNMs, as we discuss in the next subsection.

\subsection{QNMs from the quantisation condition }

By selecting the local solutions satisfying the boundary conditions in \eqref{boundcond}, the quantization condition for the QNMs is obtained by imposing that the connection coefficient in front of the discarded local solution in \eqref{connformula} to vanish. This reads\footnote{This quantisation condition was first recognised as the quantisation of the A period of the SW quantum curve in \cite{Aminov:2020yma} for the Kerr angular problem.}
\begin{equation}\label{quantcondHeun2}
\pm a=\frac{1}{2}-a_{t}+a_0-N,\quad N\in\mathbb{Z}_{> 0}.
\end{equation}

To make use of this quantisation condition, we substitute \eqref{quantcondHeun2} in the Matone relation \cite{Matone:1995rx}, which relates the parameters $a$ and $c$:\footnote{We remark that only $a^2$ appears in the right hand side of \eqref{Matonerel}, so that the choice for the sign in front of $a$ in \eqref{quantcondHeun2} does not make any difference.}
\begin{equation}\label{Matonerel}
c =-\frac{1}{4} - a^2 + a_{t}^2 + a_0^2 + t\, \partial_{t} F(t).
\end{equation}
This relation provides an expansion of the parameter $c$, which in the gauge theory language is related to the energy of the Seiberg-Witten curve, in powers of the instanton counting parameter $t$. 
To the best of our knowledge, no rigorous results exist for the convergence properties of this series in $t$ in the Nekrasov-Shatashvili limit. For a generic choice of the parameters in the $\Omega$-background, the radius of convergence has been studied, for instance, in \cite{guillarmou2024conformal, Arnaudo:2022ivo}.
Finally, since we know from the dictionary \eqref{dictioHeun} the actual value of $c$ for the gravitational problem, we truncate the instanton series in $t$ in \eqref{Matonerel} at some finite order, and extract the solution for $\omega$ for fixed values of $q$ and $N$.

We describe in appendix \ref{app:3term} an efficient way of obtaining the instanton series expansion of $c$, which is based on the three-term recurrence relation satisfied by the coefficients of the Heun Floquet solution around the singular points $y=0$ and $y=t=1/2$. The existence of this three term relation permits to build a convergent continued fraction, weighted by powers of $t$. This gives an effective way to extract the coefficients of $c$ in the instanton expansion. Moreover, the convergence properties of the series expansion in $t$ make it possible to safely truncate the continued fraction representation, rigorously generalising approaches such as the one presented in Ref.~\cite{Starinets:2002br}.

To give a taste of the results, we begin by considering small values for the overtone number $N$. 
By defining
\begin{equation}
s=1+\frac{\omega_{\text{table}}}{q},
\end{equation}
and setting $N=2$, we find the numerical results in table \ref{table_omega_inst}, which are compared with table IV in \cite{Fuini:2016qsc}.
\begin{table}[ht!]
    \centering
    \begin{tabular}{|c|c|c|}\hline
        value of $q$ & $\omega_{\text{table}}$ from instantons & $\omega_{\text{table}}$ from table IV in \cite{Fuini:2016qsc}  \\  \hline
         10&  $2.2037\, -3.49811\,\rmi$ &  $2.20415\, - 3.49828\,\rmi$ \\ \hline
         20 &$1.71622\, -2.86244\,\rmi$ &  $1.71646\, -2.86252\,\rmi$  \\ \hline
         40 & $1.35471\, -2.30307\,\rmi$ &  $1.34867\, -2.29987\,\rmi$\\ \hline
         80 &$1.03273\, -1.86081\,\rmi$ &  $1.06566\, - 1.83424\,\rmi$ \\ \hline 
         160 &$0.917763\, -1.43041\,\rmi$ &  $0.84425\, -1.45862\,\rmi$ \\ \hline     
    \end{tabular}
    \caption{Solutions for $\omega_{\text{table}}$ with 12 instantons and $N=2$ in \eqref{quantcondHeun2}}
    \label{table_omega_inst}
\end{table}

As can be seen, when increasing the value of $q$, we lose accuracy on the result. 
We expect this to be related to a shrinking of the radius of convergence of the instanton series \eqref{Matonerel} as the parameter $q$ is increased.

\subsection{The limitations of the AC truncation at large wave number/overtone}

\begin{figure}[h!]
\centering

\begin{minipage}{0.48\textwidth}
    \centering
    \includegraphics[width=\linewidth]{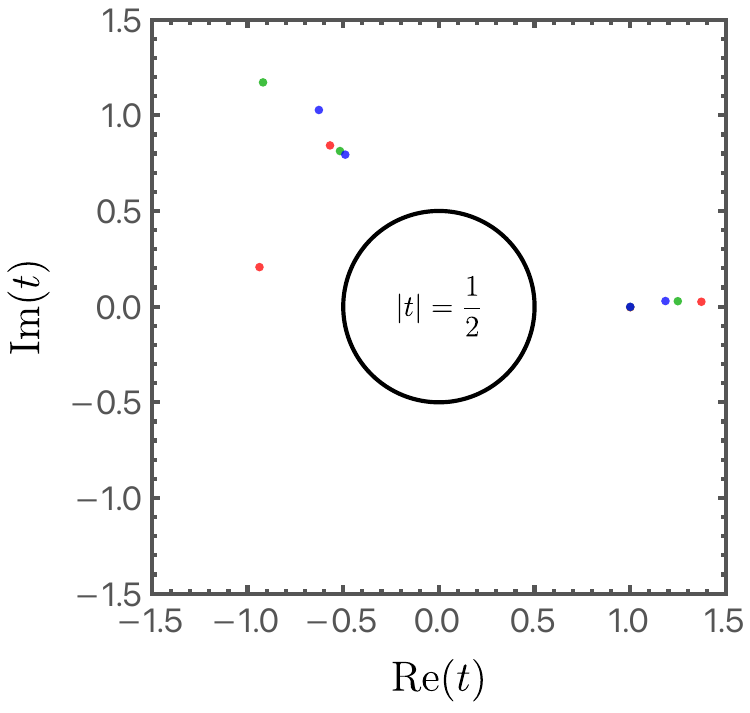}
\end{minipage}
\hfill
\begin{minipage}{0.48\textwidth}
    \centering
    \includegraphics[width=\linewidth]{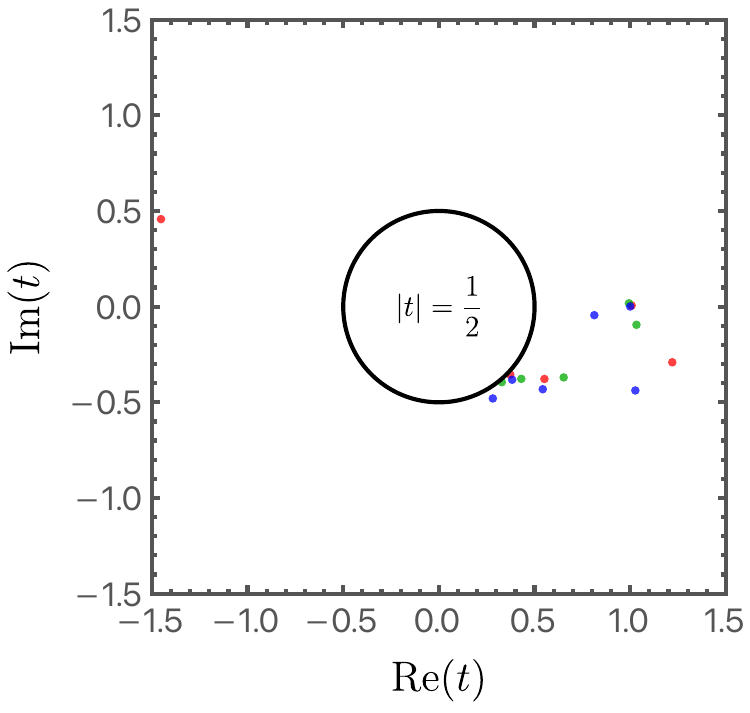}
\end{minipage}

\vspace{0.5cm}

\begin{minipage}{0.48\textwidth}
    \centering
    \includegraphics[width=\linewidth]{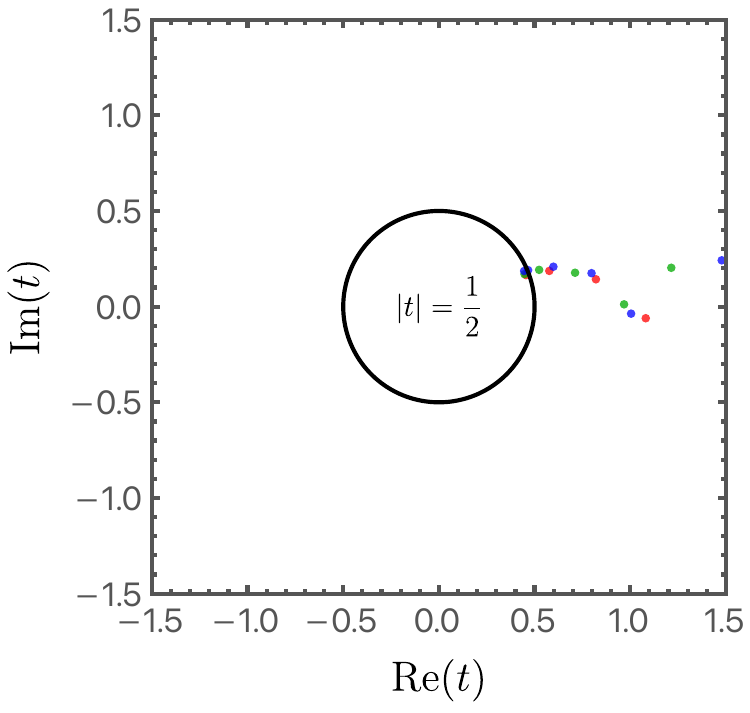}
\end{minipage}
\hfill
\begin{minipage}{0.48\textwidth}
    \centering
    \includegraphics[width=\linewidth]{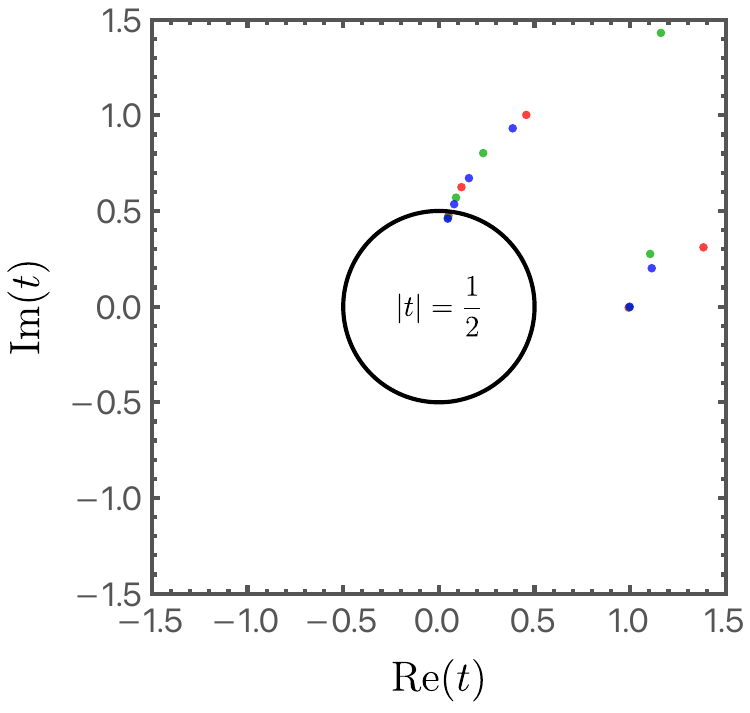}
\end{minipage}

\caption{Plot of the singularities of the diagonal Padé approximant of the instanton series for $c$. We used 16 instanton orders in the $t$-series for $c$; the red dots are the singularities at order 12, the green dots are the singularities at order 14, and the blue dots are the singularities at order 16. Upper left panel: $q=10$ and $N=2$. Upper right panel: $q=10$ and $N=15$. Lower left panel: $q=160$ and $N=2$. Lower right panel: $q=160$ and $N=15$.}
\label{fig:pade_combined}
\end{figure}

To investigate this issue, we consider these series at different values of $q,N,\omega$ provided in table IV in \cite{Fuini:2016qsc} and leaving $t$ undetermined, and we study the singularities in the $t$ plane of a diagonal Padé approximant of the resulting series. 
We find that 
as $q$ increases, the singularities of the diagonal Padé approximant approach the circumference at $|t|=1/2$. Moreover, fixing $q=10$ and increasing $N$ the same phenomenon appears.
Results for $q=10,160$ and $N=2,15$ are shown in figure \ref{fig:pade_combined}.

Furthermore, we effectively see a competition when both the parameters $q$ and $N$ increase. More precisely, the numerical values we can find with the instanton series (up to 13 instanton contributions) for $\omega_{\text{table}}$ for $q=10,20$ and $N=15$ are not accurate compared to the results in table IV in \cite{Fuini:2016qsc}.
However, with 12 instanton contributions, for $q=80$ and $N=15$ we find 
$\omega_{\text{table}}=12.6957 - 20.601\,\rmi$ and for $q=160$ and $N=15$ we find $\omega_{\text{table}}=9.9567 - 16.744\,\rmi$, which approximate the results in \cite{Fuini:2016qsc} rather well. 

A closely related, alternative approach to deal with the Heun connection problem at hand has been introduced in \cite{Lisovyy:2022flm} and is reviewed in appendix \ref{app:alternativeHeun}. This method reorganises the connection coefficient in terms of a recurrence relation and permits to consider many orders in the instanton expansion, so that for any not too large values of $q$ and $N$ accurate numerical results can be obtained.
For the latter ranges of parameters, an alternative description in terms of WKB methods, as well as a new ansatz for the wave functions, is needed. We will see that the competition between the regimes of large-$q$ and large $N$ becomes apparent in such a description
\footnote{We remark, in particular, that from the dictionaries \eqref{dictioHeun} and \eqref{fundmasses}, all the fundamental masses of the $N_f=4$ gauge theory scale with $q$. When $q$ becomes large, the effective regime becomes equivalent to reintroduce a small $\epsilon_1$ parameter, which was set to 1 in the notations introduced in appendix \ref{app:NS}. When both $\epsilon_1,\epsilon_2\to 0$, the regime we recover from the gauge theory viewpoint is the one of the classical Seiberg-Witten curve.}.

To summarise, the method presented in this section provides a continued fraction expansion in the variable $t$, described in appendix \ref{app:3term}, from which the QNM condition can be extracted by truncating the analytic representation at finite order. This is conceptually close to the approach of \cite{Starinets:2002br}, with the additional advantage that here the convergence properties of the expansion are explicitly under control. For every finite value of $q$ the continued fraction can, in principle, be truncated at sufficiently high order to approximate the corresponding QNMs with arbitrary accuracy. However, the analysis of the Padé singularities indicates that, as either $q$ or $N$ increases, the relevant value of $t$ is driven towards the boundary of the domain of convergence, namely $|t|=1/2$. In this regime, the convergence becomes progressively slower, so that obtaining even an approximate result requires a truncation of increasingly high order. The loss of efficiency of the method at large-$q$ and/or large $N$ should therefore not be interpreted as a breakdown of the construction itself, but rather as the signal that the small-$t$ representation is no longer the most convenient one for that subregion of the parameter space relevant to the QNMs. 

From the point of view of the eigenfunctions of the original differential equation, this phenomenon reflects the fact that the Frobenius expansions around the regular singular points cease to provide a convenient asymptotic description in the large-$q$ and large-$N$ regimes. In that region, a WKB  
is more natural
and this will be the subject of the next section. As we shall see, the WKB analysis is not merely a technical device to access the large-$q$ regime: owing to the asymptotic nature of the problem, it also furnishes an analytic framework from which the results at finite $q$ and $N$ can be recovered by continuation, potentially all the way to the  vicinity of $q=0$. In this sense, the WKB description should be regarded as a complementary representation of the same spectral problem, one that becomes essential when the instanton expansion approaches the edge of its domain of effectiveness.

\section{QNMs from exact WKB}
\label{sec:WKB method}
In this section we discuss the exact WKB method \cite{AIHPA_1983__39_3_211_0,kawai2005algebraic,iwaki2014exactwkbanalysiscluster}, an approach which, unlike the Seiberg-Witten approach, is valid for large-$q$. 
When $q$ is large, the differential equation \eqref{HeunforWKB} is singularly perturbed and we have to consider a description of our solutions $\psi(z)$ that differs from the Frobenius expansion. 
More precisely, we consider a Liouville-Green ansatz for $\psi(z)$:
\begin{equation}\label{WKBansatz}
\psi(z)= \exp\left( \int^z S(z') \,dz'\right),
\end{equation}
where $S(z)$ can be expanded perturbatively in $1/q$ as 
\begin{equation}\label{wkbSexpansion}
S(z) = \sum_{n=-1}^\infty q^{-n} S_n(z).
\end{equation}
Note that $\psi(z)$ is an eigenfunction of the original ODE with two different representations (Frobenius or Liouville-Green) in different regions of $q$. Any solution can be written as a linear combination of eigenfunctions in either representation, and the equivalence can be checked via resummation  \cite{sauzin2014introduction1summabilityresurgence, Aniceto_2019,Dorigoni_2019}. This approach is well-known in quantum mechanics as the WKB method/approximation where $1/q$ plays the role of $\hbar$. The WKB approximation also has a long history of use in  QNM calculations (for example, see \cite{Berti:2009kk} and references therein). In these cases, typically only the first few orders are considered. The WKB approximation does not capture all orders, which limits its use. 
To go beyond leading orders, the exact WKB method \cite{AIHPA_1983__39_3_211_0} is necessary (see \cite{kawai2005algebraic, iwaki2014exactwkbanalysiscluster} for more detailed reviews of this method than we will present in this section). The exact WKB method has had extensive use in quantum mechanics and gauge theories e.g \cite{Delabaere1997ExactSE,Kashani-Poor:2015pca,Ito:2018eon,Sueishi:2021xti,Grassi:2021wpw,Bucciotti:2023trp}, as well as some recent applications in QNM calculations \cite{Imaizumi_2022, Imaizumi_2023, Miyachi:2025ptm, Miyachi:2025dyk,Hatsuda:2026ghx}. This method enables one to derive exact quantisation conditions (EQCs) which capture \emph{all} perturbative and non-perturbative contributions to the energy or, in our case, the QNM frequency. The following subsections discuss how, using the exact WKB method, we can derive an EQC capturing all perturbative and non-perturbative contributions to the QNM frequency for real and positive values of $q$.

\subsection{The Liouville-Green expansion: lightening view} \label{the exact WKB method}

By inserting the ansatz \eqref{WKBansatz} in the ODE \eqref{HeunforWKB},
one find that $S(z)$ obeys the Riccati equation\footnote{Here, we assume that no terms in $V_0$ compete with $q^2$, and we will discuss the validity of this assumption later in this section.}
\begin{equation}\label{riccati}
S(z)^2+ S'(z) -\left[q^2\,V_0(z)+V_2(z)\right]=0,
\end{equation}
which can be solved perturbatively as
\begin{equation}\label{WKB equations}
\begin{aligned}
S_{-1}(z)^2 &=V_0(z)\\
S_0(z)&=-\frac{S_{-1}'(z)}{2S_{-1}(z)},\\
S_{1}(z) &= -\frac{1}{2S_{-1}(z)}\left( S_0'(z)+S_{0}(z)^2 -V_{2}(z)\right),\\
S_{j+1}(z) &= -\frac{1}{2S_{-1}(z)} \left( S_j'(z) + \sum_{k=0}^j S_{j-k}(z) S_k(z)\right), \quad j\ge 1.
\end{aligned}
\end{equation}
The first equation in \eqref{WKB equations} gives two possible solutions, according to the selected branch of the square root:
\begin{equation}
S^{\pm}_{-1}(z)=\pm\sqrt{V_0(z)}.
\end{equation}
We denote by $S^\pm$ the solutions for $S$ obtained by starting with $S^\pm_{-1}$ and solving the subsequent recurrence relations \eqref{WKB equations}.

Moreover, it is useful to write 
\begin{equation}
\begin{aligned}
S_\text{odd} &=\frac{1}{2}( S^+ -S^-)\\
S_\text{even} &=\frac{1}{2}( S^+ +S^-).
\end{aligned}
\end{equation}
and we note that this choice of $S_\text{odd}$ selects the $+$ branch of $\pm\sqrt{V_0(z)}$.
Using \eqref{riccati}, it is possible to show that 
\begin{equation}
S_\text{even} = -\frac{1}{2}\frac{\mathrm{d}}{\mathrm{d} z}\left(\log S_\text{odd}\right),
\end{equation}
and therefore the solutions \eqref{WKBansatz} can be written in terms of only $S_\text{odd}$ as 
\begin{equation} 
\label{WKB ansatz sodd}
\psi^{\pm}(z) = \frac{1}{\sqrt{S_\text{odd}}}\exp\left(\pm\int^z S_\text{odd} \, dz'\right).
\end{equation}
To compute $\psi^{\pm}(z)$ one swaps the $z$-integral and the summation of powers in $1/q$ defining $S_{\text{odd}}$. After this operation, the solutions $\psi^\pm(z)$ should be understood as formal series in $1/q$ which typically diverge and need to be resummed. However, we can obtain the EQC in terms of formal series and leave the details of resummation to section \ref{sec:resurgenceofQNMs}. Note that we will often abuse notation and use $\psi^\pm(z)$ to represent the formal series in powers of $1/q$ obtained by exchanging the order of integral and summation in \eqref{WKB ansatz sodd}; furthermore, we will refer to these formal solutions as WKB solutions.

\subsection{Exact WKB and connection formulas}\label{quantisation condition section}

In close analogy with the Seiberg--Witten approach discussed in the previous section, our goal is to relate the basis of WKB solutions \eqref{WKB ansatz sodd} defined near the AdS boundary at $z=0$ to its analytic continuation near the horizon at $z=1$. Once this connection problem is understood, the quasinormal-mode condition follows by imposing the boundary conditions \eqref{boundcond2} at the two endpoints. However, there is an important new feature in the WKB formulation. The formal series \eqref{WKB ansatz sodd} does not provide a single uniform representation of the solutions throughout the $z$-plane. As we shall discuss later in this section, when the solutions are analytically continued from one region to another, the formal series representing the solutions may change. Thus, the connection problem requires keeping track of how the formal WKB expansions change as we attempt to analytically continue from $z=0$ to $z=1$.

More precisely, the formal series representing $\psi(z)$ and obtained by exchanging the order of integral and summation in the solutions will jump discontinuously by exponentially suppressed terms, a well-known feature of asymptotic expansions known as \emph{Stokes jumps} \cite{1864TCaPS..10..105S}. The lines delimiting the different sectors where a given expansion is valid and across which Stokes jumps occur are called \emph{Stokes lines}. The collection of all Stokes lines form a graph in the complex $z$-plane which is called the \emph{Stokes graph}. The special points $z_*$ in the complex $z$-plane from which Stokes lines emanate are referred to as \emph{turning points}. These are the singularities of $S_\text{odd}$, the points at which our formal series representing $\psi^\pm(z)$ becomes ill-defined.
By looking at \eqref{WKB equations}, one can see that the turning points are either poles or zeros of the potential $V(z)$. The zeros of $V_0(z)$ become poles of $S_i$ for $i\ge 0$, due to the factor of $S_{-1}$ in the denominators.

At a Stokes line, one WKB solution is maximally exponentially suppressed compared to the other.
This occurs when the leading-order exponential factor of \eqref{WKBansatz} is purely real. Thus, given a turning point $z_*$, its associated Stokes lines are the set of points $z$ obeying the following integral equation
\begin{equation} \label{Stokes lines equation}
    \mathrm{Im}\, q\int_{z_*}^z \sqrt{V_0(z')}\,dz' =0.
\end{equation}
The condition \eqref{Stokes lines equation} also implies that along a Stokes line the quantity
\begin{equation} \label{real part of Stokes line}
    \mathrm{Re}\, q\int_{z_*}^z \sqrt{V_0(z')}\,dz' 
\end{equation}
has a definite sign. We call a Stokes line a \emph{positive Stokes line / negative Stokes line} if the above quantity \eqref{real part of Stokes line} is positive/negative along the Stokes line. The sign of this quantity dictates which WKB solution is maximally enhanced and which is maximally suppressed. For example, near a positive Stokes line $\psi^-$ is maximally suppressed but $\psi^+$ is maximally enhanced. 

These Stokes lines separate the sectors in which a given formal WKB solution provides an asymptotic representation of the actual solution. When one crosses a Stokes line, the formal expression associated with a given solution generally changes: a series that is sufficient on one side of the line may need to be supplemented by the second independent formal solution (exponentially suppressed along that line) on the other side. For instance, if a solution is represented by $\psi^{+}(z)$ on one side of a positive Stokes line, then after crossing that line the same solution is represented by $\psi^{+}(z)+c\,\psi^{-}(z)$, where $c$ denotes the coefficient associated with the crossing. Depending on the nature of the turning point, it may be either a constant or a $z$-independent formal series in powers of $q^{-1}$.

\begin{figure}[!h]
    \centering
   \begin{tikzpicture}[scale =3]
   \draw[black,thick,domain=0:180,smooth,variable=\t,
        postaction={decorate},
        decoration={markings, mark=at position 0.8 with {\arrow{<}}}]
    plot ({1.2*cos(\t)+0.5},{0.2*sin(\t)});
      \draw[black,thick,dashed, domain=180:360,smooth,variable=\t,
        postaction={decorate},
        decoration={markings, mark=at position 0.8 with {\arrow{<}}}]
    plot ({1.2*cos(\t)+0.5},{0.2*sin(\t)});
    
    \node[below] at (0.5,0.4) {$\gamma_{z,z_*}$};
    
    \filldraw[black] (1.7, 0) circle (0.8pt) node[black,above right] {$z$};
    
    \draw[decorate,blue,decoration={zigzag,segment length = 2mm,amplitude=0.5mm}] (1, 0)--(-1, 0);
       
       \filldraw[red] (1, 0) circle (1pt) node[black,above right] {$z_*$};
   \end{tikzpicture}
    \caption{The path $\gamma_{z,z_*}$ defining our normalisation around a turning point $z_*$. We draw with a solid line paths on the first sheet and with a dashed line paths on the second sheet. The zig-zagged blued line represents the branch cut.}
    \label{fig:WKB normalised at turning point}
\end{figure}
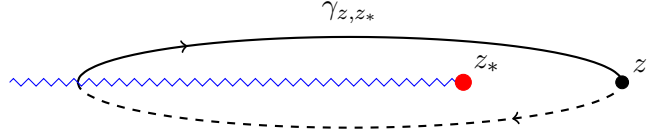

The precise way the Stokes jump occurs when crossing a Stokes line depends on the type of turning point and the normalisations of $\psi^{\pm}(z)$. Notice that the definition \eqref{WKB ansatz sodd} does not specify the path along which the integral is computed, therefore does not fix the normalisation of our solutions. In principle, we could choose any path to conduct our analysis, but there are certain canonical choices which make the analysis more systematic. In our case\footnote{We recall that the curve characterising the geometry of our problem is a torus, which is a double cover of the 4-punctured sphere, with branch points at the turning points.}, all turning points have a branch cut emanating from them and together these yield two possible sheets\footnote{Note that in defining $S_{\text{odd}}$, we must pick a sheet and that we picked the sheet corresponding to $+\sqrt{V_0(z)}$.}. Using these branch cuts, we can define the following normalisation: given a turning point $z_*$, we denote the WKB solutions \emph{normalised at a turning point $z_*$} by $\psi^{\pm}_{z_*}$ and define them as 
\begin{equation} \label{normalisationturningpoint}
\psi^{\pm}_{z_*}(z) \equiv \frac{1}{\sqrt{S_\text{odd}}}\exp\left({\pm\frac{1}{2}\int_{\gamma_{z, z_*}} S_\text{odd} \, dz'}\right),
\end{equation}
where $\gamma_{z, z_*}$ is the contour starting from $z$ on the second sheet, encircling $z_*$ once passing through to the first sheet, and ending at $z$ on the first sheet, see figure \ref{fig:WKB normalised at turning point}. 

With this choice of normalisation, we can make the nature of the Stokes jumps more precise. Let $z_*$ be a turning point, then let $I$ and $II$ be two regions delimited by a Stokes line emanating from $z_*$ and denote the basis of WKB solutions normalised at the turning point $z_*$ defined in the region $I$/$II$ by $\psi^\pm_{z_*, I}(z)$/$\psi^\pm_{z_*, II}(z)$ -- see figure \ref{fig:Stokes jump of simple turning point}. Then the analytic continuation of the WKB basis $\psi^\pm_{z_*, I}(z)$ from region $I$ to region $II$ across the Stokes line in the anticlockwise direction is related to the WKB basis $\psi^\pm_{z_*, II}(z)$ via the following connection matrices 
\begin{equation} \label{general stokes jump}
    \begin{cases}
        \left( \begin{matrix}
           \psi^+_{z_*,I}\\
           \psi^-_{z_*,I}
       \end{matrix}\right)=\left( \begin{matrix}
           1 & A\\
           0 & 1
       \end{matrix}\right)\left( \begin{matrix}
           \psi^+_{z_*,II}\\
           \psi^-_{z_*,II}
       \end{matrix}\right),  \quad \mathrm{Re} \, q\int_{z_*}^z\sqrt{V_0(z')}\,dz'>0\\
   
        \left( \begin{matrix}
           \psi_{z_*,I}^+\\
           \psi^-_{z_*,I}
       \end{matrix}\right)=\left( \begin{matrix}
           1 & 0\\
           B & 1
       \end{matrix}\right)\left( \begin{matrix}
           \psi^+_{z_*,II}\\
           \psi^-_{z_*,II}
       \end{matrix}\right),  \quad \mathrm{Re} \, q\int_{z_*}^z\sqrt{V_0(z')}\,dz'<0,
    \end{cases}
\end{equation}
where the integrals on the right hand side are taken along the Stokes line crossed by the solutions, and $A/B$ are $z$-independent constants, which depend on the details of the turning point.
The most relevant turning point for our problems is the \emph{simple turning point}. This is a turning point $a$ that  obeys $V_0(a)=0$ and $V_0'(a)\neq 0$. The other type of turning point is located at $z=0$, the simple pole of $V_0(z)$. Although this line is not crossed in the derivation of the EQC, it does play a role in the relation between the two quantisation conditions we will obtain. For more details on potentials with simple poles, see \cite{Iwaki_2015,Koike_2000}.

The Stokes jumps of WKB solutions normalised at simple turning points can be given explicitly. Let $\psi^{\pm}_{a,I}$ be the WKB solutions normalised at a simple turning point $a$, defined in the region $I$, then the connection matrices when crossing a Stokes line in the anticlockwise direction that emanates from $a$ and delimitates two regions $I$ and $II$ are of the  form \cite{AIHPA_1983__39_3_211_0, iwaki2014exactwkbanalysiscluster}
\begin{equation}
 \label{Stokes jump of simple turning point}
    \begin{cases}
        \left( \begin{matrix}
           \psi^+_{a,I}\\
           \psi^-_{a,I}
       \end{matrix}\right)=\left( \begin{matrix}
           1 & \mathrm{i}\\
           0 & 1
       \end{matrix}\right)\left( \begin{matrix}
           \psi^+_{a,II}\\
           \psi^-_{a,II}
       \end{matrix}\right),  \quad \mathrm{Re} \, q\int_{a}^z\sqrt{V_0(z')}\,dz'>0\\
   
        \left( \begin{matrix}
           \psi_{a,I}^+\\
           \psi^-_{a,I}
       \end{matrix}\right)=\left( \begin{matrix}
           1 & 0\\
          \mathrm{i} & 1
       \end{matrix}\right)\left( \begin{matrix}
           \psi^+_{a,II}\\
           \psi^-_{a,II}
       \end{matrix}\right),  \quad \mathrm{Re} \, q\int_{a}^z\sqrt{V_0(z')}\,dz' <0.
    \end{cases}
\end{equation}
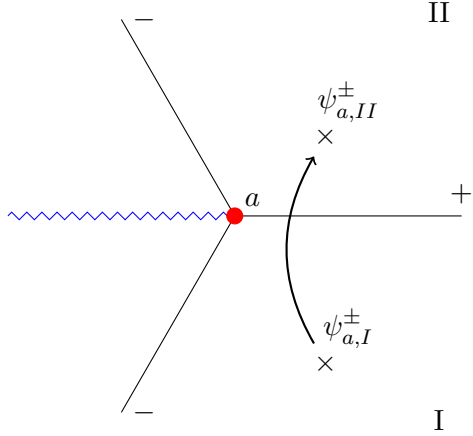
\begin{figure}[!h]
    \centering
   \begin{tikzpicture}[scale=3]

\draw[] (-0.5, 0.866)--(0, 0);
    \draw[] (-0.5, -0.866)--(0, 0);
       \draw[] (1, 0)--(0, 0);
    \draw[decorate,blue,decoration={zigzag, segment length = 2mm,amplitude=0.5mm}] (-1, 0)--(0, 0);
       \node[] at (-0.4, 0.866) {$-$};
        \node[] at (-0.4,-0.866) {$-$};
         \node[] at (1,0.1) {$+$};
             \node[] at (0.9,0.9) {II};
              \node[] at (0.9,-0.9) {I};
               \node at (0.5,-0.5) {$\psi^\pm_{a,I}$};
                \node at (0.5,+0.5) {$\psi^\pm_{a,II}$};
                 \node (A) at (0.5-0.1,0.5-0.15) {$\times$};
              \node (B) at (0.5-0.1,-0.5-0.15) {$\times$};
  \draw[->, thick, bend left=30] (B) to (A);

       \filldraw[red] (0, 0) circle (1pt) node[black,above right] {$a$};
   \end{tikzpicture}
    \caption{Diagram depicting the analytic  continuation and Stokes jump near a simple turning point $a$. $\psi^\pm_{a,I}$/$\psi^\pm_{a,II}$ is the basis of WKB solutions defined in the regions $I$/$II$ and normalised at the turning point $a$. The solid black lines represent Stokes lines and the blue zig-zagged line represents the branch cut. The labels $+/-$ denote the sign of of the Stokes lines.}
    \label{fig:Stokes jump of simple turning point}
\end{figure}

\subsection{Exact quantisation conditions and connection to Seiberg-Witten}

\begin{figure}[h!]
    \centering
    
   \begin{tikzpicture}[scale=3]
 
   \node (A) at (-1,0) {} ;
  \node (B) at (1,0) {} ;

  \draw[black,
  thick,postaction={
    decorate,
    decoration={
      markings,
      mark=at position 0.8 with {\arrow{<}}
    }}] plot coordinates {(0,-0.5) (-0.8, 0)};
    
    \draw[black, smooth, dashed, thick, postaction={
    decorate,
    decoration={
      markings,
      mark=at position 0.5 with {\arrow{<}}
    }}] plot coordinates {(-0.8,0) (-1, 0.1) (-1.1,0) (-1, -0.12) (-0.7, -0.24) (0,-0.5)};
    
 \draw[black, dashed, thick,postaction={
    decorate,
    decoration={
      markings,
      mark=at position 0.8 with {\arrow{<}}
    }}] plot coordinates {(0,-0.5)  (0.8, 0)};

    \draw[black, smooth,tension =0.7, thick,postaction={
    decorate,
    decoration={
      markings,
      mark=at position 0.5 with {\arrow{<}}
    }}] plot coordinates {(0.8,0) (1, 0.1) (1.1,0)  (1, -0.12) (0.7, -0.24) (0,-0.5)} ;

          \draw[decorate,blue,decoration={zigzag, segment length = 2mm,amplitude=0.5mm}] (A)--(B);
       
       \filldraw[red] (B) circle (1pt);
       \node at (1.1,0.15) {$b$};
        \filldraw[red] (A) circle (1pt);
        \node at (-1.1,0.15) {$a$};
        \filldraw[black] (0,-0.5) circle (0.8pt);
           \node  at (0,-0.6) {$z$};
             \node  at (0.7,-0.4) {$-\gamma_{z,b}$};
            \node  at (-0.7,-0.4) {$\gamma_{z,a}$};
   \end{tikzpicture}
     \begin{tikzpicture}[scale=3]
 
   \node (A) at (-1,0) {} ;
  \node (B) at (1,0) {} ;

  \draw[black, dashed,thick,postaction={
    decorate,
    decoration={
      markings,
      mark=at position 0.8 with {\arrow{>}}
    }}] plot coordinates {(0,-0.5)  (-0.8, 0)};
    \draw[black, smooth, thick,tension =0.7, postaction={
    decorate,
    decoration={
      markings,
      mark=at position 0.5 with {\arrow{>}}
    }}] plot coordinates {(-0.8,0) (-1, 0.1) (-1.1,0) (-1, -0.12) (-0.7, -0.24) (0,-0.5)};
    
 \draw[black, dashed, thick,postaction={
    decorate,
    decoration={
      markings,
      mark=at position 0.8 with {\arrow{<}}
    }}] plot coordinates {(0,-0.5)  (0.8, 0)};

    \draw[black, smooth,tension =0.7, thick,postaction={
    decorate,
    decoration={
      markings,
      mark=at position 0.5 with {\arrow{<}}
    }}] plot coordinates {(0.8,0) (1, 0.1) (1.1,0)  (1, -0.12) (0.7, -0.24) (0,-0.5)} ;

          \draw[decorate,blue,decoration={zigzag, segment length = 2mm,amplitude=0.5mm}] (A)--(B);
       
       \filldraw[red] (B) circle (1pt);
       \node at (1.1,0.15) {$b$};
        \filldraw[red] (A) circle (1pt);
        \node at (-1.1,0.15) {$a$};
        \filldraw[black] (0,-0.5) circle (0.8pt);
           \node  at (0,-0.6) {$z$};
            
   \end{tikzpicture}
    \begin{tikzpicture}[scale=3]
 
   \node (A) at (-1,0) {} ;
  \node (B) at (1,0) {} ;

          \draw[decorate,blue,decoration={zigzag, segment length = 2mm,amplitude=0.5mm}] (A)--(B);
       \filldraw[red] (B) circle (1pt);
       \node at (1.15,0.15) {$b$};
        \filldraw[red] (A) circle (1pt);
        \node at (-1.15,0.15) {$a$};
          
            \draw[
  black,
  thick,
  tension=0.2,
  postaction={
    decorate,
    decoration={
      markings,
      mark=at position 0.1 with {\arrow{<}},
      mark=at position 0.6 with {\arrow{<}}
    }
  }
]
plot[smooth cycle] coordinates
{(1.05,0.1) (1.05,-0.1) (-1.05,-0.1) (-1.05,0.1)};
   \end{tikzpicture}
    \caption{Figure depicting how the contour defining the change in normalisation from a turning point at $b$ to the turning point at $a$ can be written as a period. In going from the upper left image to the upper right we note that the integral of $S_\text{odd}$ along $\gamma_{z,a}$ is invariant if we swap the sheets and the direction of the integral. The blue zig-zagged line represents a branch cut, dashed black lines represent paths on the second sheet and solid black lines represent paths on the first sheet.}
    \label{fig:change of normalisation}
\end{figure}
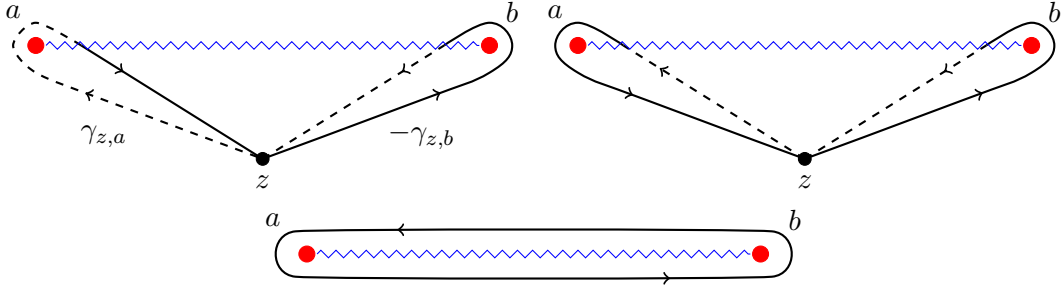

Our goal now is to analytically continue the WKB solutions $\psi^{\pm}(z)$ from $z=0$ to $z=1$. In the analytic continuation procedure, the WKB solutions will cross Stokes lines and the formal series will jump discontinuously\footnote{We emphasise that the formal series representing the actual function jumps discontinuously, not the function itself.}. To calculate the Stokes jumps correctly, our WKB solutions need to be normalised at the turning point from which the Stokes line emanates. However, our WKB solutions may not be normalised at the correct turning point. More precisely, this means that we will need to find the Stokes jumps associated to a turning point $b$ for WKB solutions $\psi_a^\pm$ normalised at some different turning point $a$. The formulas \eqref{Stokes jump of simple turning point} apply only to WKB solutions that are normalised at the turning point from which the Stokes line emanates; therefore, we need to express $\psi_a^\pm$ in terms of $\psi^\pm_b$. The factors which change one normalisation to another can be expressed in terms of a \emph{period} integral: 
\begin{equation} \label{change of norm matrix}
   \left( \begin{matrix}
        \psi_a^+\\
        \psi_a^-
    \end{matrix}\right) = \left(\begin{matrix}
       \mathrm{e}^{ \frac{1}{2}\oint_a^b S_\text{odd}} & 0 \\
       0 &\mathrm{e}^{ - \frac{1}{2}\oint_a^b S_\text{odd}}
    \end{matrix} \right) \left( \begin{matrix}
        \psi_b^+\\
        \psi_b^-
    \end{matrix}\right),
\end{equation}
as represented in figure \ref{fig:change of normalisation}.

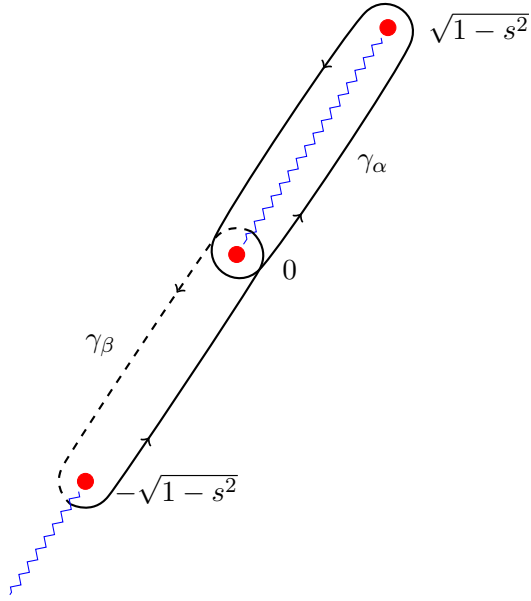
\begin{figure}[!h]
    \centering

 \begin{tikzpicture}[scale=1]
 
   \node (A) at (2,3) {} ;
  \node (B) at (0,0) {} ;
   \node (C) at (-2,-3) {} ;

    \draw[decorate,blue,decoration={zigzag, segment length = 2mm,amplitude=0.5mm}] (A)--(B);
       
       \filldraw[red] (B) circle (3pt);
        \filldraw[red] (A) circle (3pt);
         \node at (3.2,3) {$\sqrt{1-s^2}$};
        \filldraw[red] (C) circle (3pt) ;
        \node at (-0.8,-3.1) {$-\sqrt{1-s^2}$};
        \node at (0.7,-0.2) {$0$};
        \node at (1.8,1.2) {$\gamma_\alpha$};
         \node at (-1.8,-1.2) {$\gamma_\beta$};
\draw [black,thick,tension=0.4,postaction={
    decorate,
    decoration={
      markings,
      mark=at position 0.1 with {\arrow{>}},
       mark=at position 0.6 with {\arrow{>}}
    }
  }] plot [smooth cycle] coordinates {(-0.3,0.2) (0.3,-0.2) (2.3,2.8) (1.7,3.2)};

  \draw[black, smooth, thick,dashed, tension=0.3, postaction={
    decorate,
    decoration={
      markings,
      mark=at position 0.3 with {\arrow{>}}
    }
  }] plot coordinates
    {(0.2,0.3)(-0.3,0.2) (-2.3,-2.8) (-2.2,-3.3)};
  \draw[black, smooth,thick,tension =0.3, postaction={
    decorate,
    decoration={
      markings,
      mark=at position 0.7 with {\arrow{<}}
    }
  }] plot coordinates
    {(0.2,0.3)(0.3,-0.2) (-1.7,-3.2)  (-2.2,-3.3)};
       \draw[decorate,blue,decoration={zigzag, segment length = 2mm,amplitude=0.5mm}] (C)--(-3,-4.5);
   \end{tikzpicture}

    \caption{Figure depicting the cycles $\gamma_\alpha$ and $\gamma_\beta$ defining $\alpha$ and $\beta$ respectively. The blue zig-zagged lines represent the branch cuts, dashed black lines represent paths on the second sheet and solid black lines represent paths on the first sheet.}
    \label{fig:alpha and beta cycle}

\end{figure}

In general, all changes of normalisation can be expressed in terms of a basis of period integrals along cycles of the curve characterising the underlying geometry of the problem, see appendix \ref{app:SW} for its description. In our cases, this curve has always genus one, and two periods are particularly relevant, which are the standard A and B cycles\footnote{In quantum mechanics, the A-cycle is a period integral around the interval of allowed classical motion and the B-cycle is a period integral around the interval associated to tunnelling.}. If we denote with $\gamma_\alpha$ and $\gamma_\beta$ the cycles encircling the turning points $0$ and $\sqrt{1-s^2}$, and $0$ and $-\sqrt{1-s^2}$, respectively, see figure \ref{fig:alpha and beta cycle}, we define the periods\footnote{We note here that periods in exact WKB analysis are typically referred to as Voros coefficients and their exponentials as Voros symbols \cite{delabaere:hal-01886535}, we will avoid this nomenclature and refer to $\alpha$ and $\beta$ as periods.}
\begin{equation}\label{ABperiods}
 \begin{aligned}
     \alpha &= \oint_{\gamma_\alpha} S_\text{odd}dz',\\
     \beta &= \oint_{\gamma_\beta} S_\text{odd}dz'.
 \end{aligned}
 \end{equation}

The process of finding the connection formula is now easy to describe: first we find the Stokes lines, then we start with the basis of WKB solutions $\psi_0^\pm$ normalised at $z=0$, and finally we analytically continue this to $z=1$, tracking the effect of the Stokes jumps using \eqref{Stokes jump of simple turning point} and \eqref{change of norm matrix}.

The leading term of the potential \eqref{scalarPotential} in the large-$q$ expansion, $V_0(z)$, has turning points located at $z=0, \pm \sqrt{1-s^2}$ from which the Stokes lines emanate. Clearly, to plot the Stokes graphs we need to pick a value of $s$. Different restrictions on $s$ may lead to topologically distinct Stokes graphs and, in principle, we should consider the connection problem for all topologically distinct Stokes graphs. Each topologically distinct Stokes graph would lead to a different quantisation condition, some of which should be discarded if the resulting value of $s$ is not compatible with the original restrictions imposed on $s$. However, inspired by the computations in the previous section, and by the leading order results in the large-$q$-expansion provided in \cite{Fuini:2016qsc}, we make some assumptions on $s$. These assumptions reduce the number of Stokes graphs to consider.

We assume 
\begin{equation} \label{s assumptions}
     1-s^2 \ll 1.
\end{equation}
For this to hold, we require that $s = \pm1 + \epsilon s_r + \rmi \epsilon s_i$ where $\epsilon$ is real and obeys $0< \epsilon \ll 1$, $s_r$ is the real part of $s\mp1$ and $s_i$ is the imaginary part of $s$. If $\omega$ is a QNM frequency then $-\omega^*$ is also a QNM frequency; therefore, it is enough to consider the plus sign only. Secondly, based on the results of the previous section, we restrict to $s_i <0$\footnote{The fact that the imaginary part of the QNM frequencies are negative ensures the stability of the system against the perturbation. Here, we are taking this as an assumption, thanks to results found by the previous method.} and $s_r > 0$. 

\begin{figure}[!h]
    \begin{subfigure}{0.45\textwidth}
        \centering
        \includegraphics[width=\linewidth]{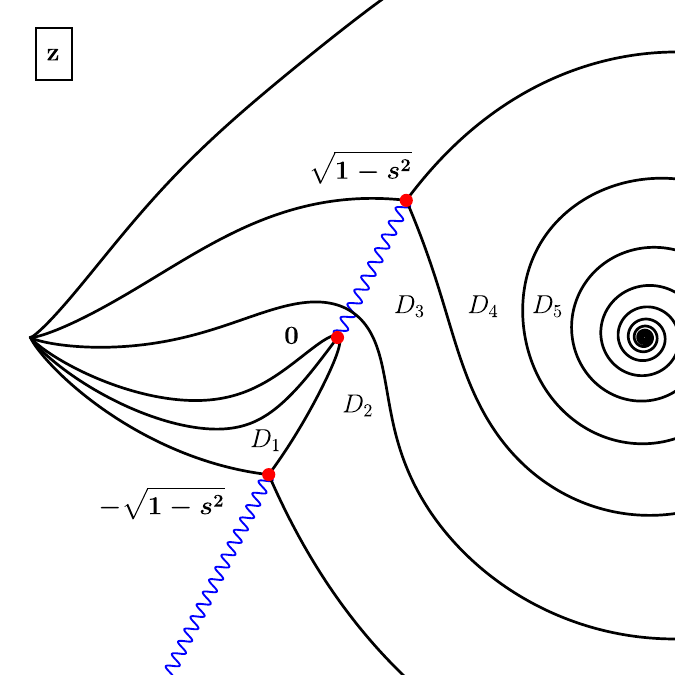}
        
    \end{subfigure}
    \hfill
        \begin{subfigure}{0.45\textwidth}
        \centering
        \includegraphics[width=\linewidth]{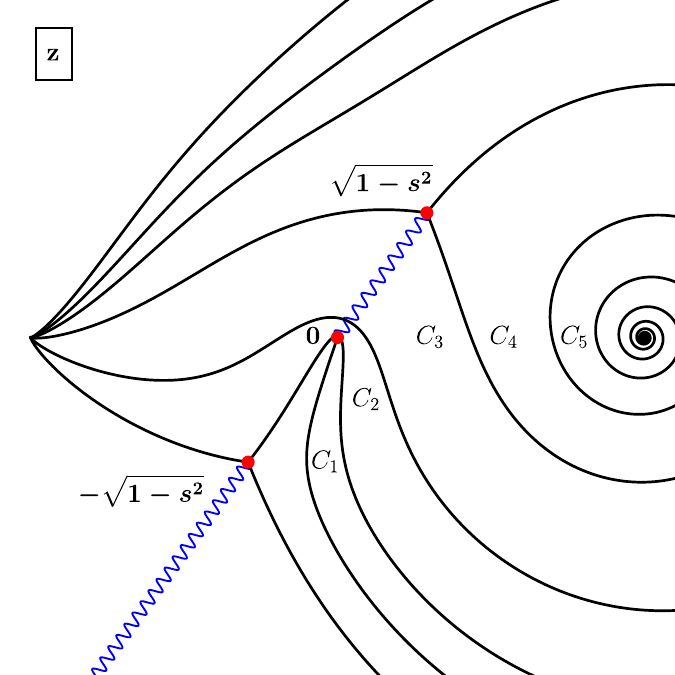}
    \end{subfigure}
   
    \label{plusandminus Stokes graphs}
    \caption{The two topologically distinct Stokes graphs for $\arg(1-s^2) \in(\pi/2 + \delta, \pi)/\{\theta_c\}$. The two points at which the Stokes lines end are $z=-1$ and $z = 1$. \textbf{Left Panel}: The plus Stokes graph. The WKB solutions are analytically continued from the region $D_1$ to the region $D_5$ and then to $z=1$. \textbf{Right Panel}: The minus Stokes graph. The WKB solutions are analytically continued from the region $C_1$ to the region $C_5$ and then to $z=1$.}
    \label{fig:Stokes graphs plus and minus}
\end{figure}

With these assumptions, we can restrict the possible arguments of $1-s^2$ that need to be considered, thus reducing the allowed number of topologically distinct Stokes graphs. The real part and imaginary part of $1-s^2$ can be written as 
\begin{equation}
    \mathrm{Re}(1-s^2) = -2\epsilon s_r +\epsilon^2( s_i^2-s_r^2), \quad\mathrm{Im}(1-s^2) = -2\epsilon s_i-2\epsilon^2 s_r s_i.
\end{equation}
Then, ignoring the subleading terms in $\epsilon$, the restrictions on $s_r$ and $s_i$ impose that $\mathrm{Im}(1-s^2)>0$ and $\mathrm{Re}(1-s^2)< 0$.
Thus, we consider Stokes graphs for $\arg(1-s^2) \in (\pi/2, \pi)$, for which there are only three topologically distinct Stokes graphs. One of these graphs occurs for $\arg(1-s^2)$ close to $\pi/2$. However, the results of table \ref{table_omega_inst} suggest that $s_r$ and $s_i$ are of a similar order, so we rule out that $|s_i|\gg |s_r|$ and consider $\arg(1-s^2)\in (\pi/2 +\delta, \pi)$ for some positive $\delta$. This leaves two topologically distinct graphs as shown in figure \ref{fig:Stokes graphs plus and minus}.

\begin{figure}
    \centering
    \includegraphics[width=0.6\linewidth]{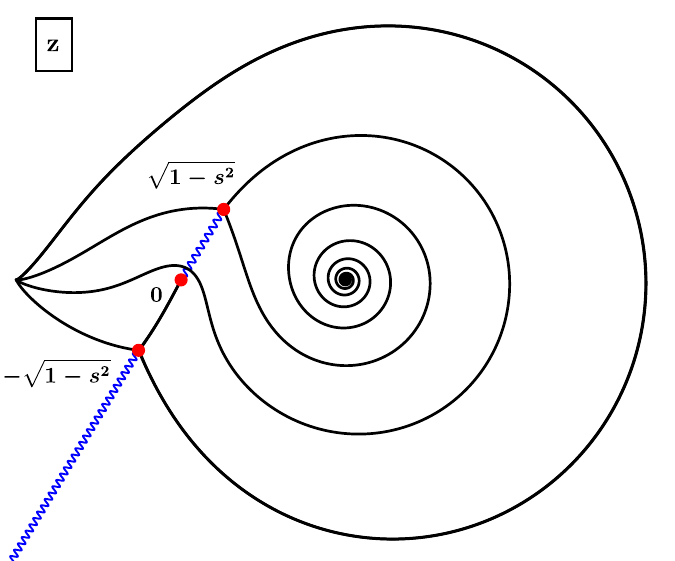}
    \caption{The critical Stokes graph occurring for $\arg(1-s^2) =\theta_c$. The Stokes lines are in drawn in black. These emanate from the red turning points and end at the points $z=-1$ and $z=1$. Note the bounded Stokes line between $-\sqrt{1-s^2}$ and $0$. The blue zig-zagged lines are branch cuts.}
    \label{fig:Stokesgraphbounded}
\end{figure}

As we increase $\arg(1-s^2)$ from $\pi/2+\delta$ to $\pi$, there will be a critical Stokes graph, occurring at some critical argument $\arg (1-s^2) =\theta_c$, which interpolates between the two topologically distinct Stokes graphs - see figure \ref{fig:Stokesgraphbounded}. We call this graph the \emph{critical Stokes graph}. This is characterised by a \emph{bounded Stokes line}, which is a Stokes line starting and ending at turning points, in this case the simple pole at $z=0$ and the simple turning point $z = -\sqrt{1-s^2}$.

Critical Stokes graphs containing bounded Stokes lines play an important role in exact WKB analysis (see \cite{delabaere:hal-01886535}, \cite{iwaki2014exactwkbanalysiscluster}, \cite{Iwaki_2015} for more detailed discussions). These Stokes graphs require more care for the reasons we will now explain. Periods in exact WKB analysis with large parameter $q$ often inherit the asymptotic nature of the WKB solutions and also diverge as series expansions in $q$. This means that periods should be understood via Borel-Laplace resummation. As we will make concrete in the next section, divergent series are resummed in a particular direction or, in other words, for a particular $\arg(q)$. If a Stokes graph contains a bounded Stokes line for a particular $\arg(q)=\phi_c$, then (in general) any period whose path of integration intersects this bounded Stokes line will not be directly resummable in the direction $\phi_c$. Instead, we have to resum for $\arg(q) = \phi_c \pm \delta$, where $0<\delta\ll1$. However, the resummations are not completely distinct and are related by a Stokes jump. In fact, exact WKB analysis dictates that this Stokes jump can be expressed in terms of a period integral around the bounded Stokes line. Although this fact is intimately related to our problem, we note that our Stokes graphs have $\arg(q)$ fixed and that the presence of the bounded Stokes line depends on $\arg(1-s^2)$. Despite this, we will see that a period appearing in the EQC will intersect the bounded Stokes line and we are ultimately forced to consider the analysis for $\arg(1-s^2) \neq \theta_c$. 
We call the Stokes graph for which $\arg (1-s^2) \in (\theta_c, \pi)$ the \emph{plus Stokes graph} and the graph for which $\arg(1-s^2) \in ( \pi/2+\delta, \theta_c)$ the \emph{minus Stokes graph}.


We will therefore consider two connection procedures, which we describe in full detail in appendix \ref{app:WKBquantcond}. By imposing the boundary conditions \eqref{boundcond2}, as explained in appendix \ref{app:boundary conditions}, we can obtain the following two EQCs obeyed by $s$: 
\begin{equation}\label{quantcondWKBscalar}
\alpha = 2\pi \rmi \left(N+\frac{1}{2}\right) \mp \log(1+\rme^{\beta}), \qquad N \in \mathbb{N},
\end{equation}
where we take the minus sign for the plus Stokes graph and the plus sign for the minus Stokes graph.

The periods $\alpha$ and $\beta$ are defined as in \eqref{ABperiods}. Due to the symmetry of the problem, the two periods have a simple relation 
\begin{equation}\label{PNP}
    \beta(q)= \alpha(\rmi q)
\end{equation}
which we exploit throughout section \ref{sec:resurgenceofQNMs}.

We remark that the two quantisation conditions are exactly related to each other via a Stokes jump of the period $\alpha$. The Stokes jumps of periods that intersect bounded Stokes lines connecting simple poles to simple turning points have been well studied \cite{Koike_2000,Iwaki_2015}. In our case, if one exponentiates the EQCs \eqref{quantcondWKBscalar},  the relation between the two conditions is an example of the Stokes jump described in equation (2.32) of \cite{Iwaki_2015} with the parameter $t$ of that equation set to $1$.

We end this subsection by discussing the relations between the quantisation conditions obtained from the Seiberg--Witten approach, in \eqref{quantcondHeun2}, and the above one from the WKB analysis \eqref{quantcondWKBscalar}.
Even though they might look different, the two quantisation conditions are equivalent, as will become apparent from the final results for the frequencies in both approaches. However, they are expressed by different analytic expansions. More precisely, the quantisation condition in \eqref{quantcondHeun2} is perturbative in the instanton counting parameter $t$, which is fixed to its value $t=1/2$ only in the end of the computation, and it is exact in $q$. The WKB condition \eqref{quantcondWKBscalar}, instead, is exact in $t$, but is expressed as a series in $q$.

This latter approach makes it important to understand the structure of the non-perturbative corrections to the quantisation condition, and the relations to the gauge theory approach. 
In the first regard, we remark that the relation \eqref{PNP} can be interpreted as a PNP relation. Indeed, as expressed by the quantisation condition \eqref{quantcondWKBscalar} and as will be discussed in the following section, the B-period $\beta$ encodes the non-perturbative corrections to the A-period $\alpha$. But since the $q$-expansion of $\beta$ is simply related to the $q$-expansion of $\alpha$ as in \eqref{PNP}, we conclude that the information about the non-perturbative corrections -- which, as we will describe more in detail, in the present case are non-perturbative corrections in the overtone number $N$ -- is already encoded in the perturbative ones. Furthermore, this is not a standard resurgence relation, where the large-order growth of perturbative coefficients around one non-perturbative sector controls the low-order data of perturbative expansions around other sectors. Instead, it relates low-order perturbative coefficients between the two expansions.

The existence and properties of PNP relations have been discussed at length in the literature~\cite{Alvarez2000, 10.1063/1.1767988, Dunne:2013ada, Basar:2015xna, Codesido:2017dns, Basar:2017hpr, Codesido:2017jwp, Cavusoglu:2024usn, Meynig:2025lnk}. In our context, the relation is simply a consequence of the symmetries of the problem. Indeed, the boundary value problem is symmetric under the transformation
\begin{equation}\label{symmetryframe}
z\mapsto -z\quad\text{and}\quad q\mapsto \rmi\,q.
\end{equation}
The relation \eqref{PNP} between the A-period, encircling $z=0$ and $z=\sqrt{1-s^2}$, and the B-period, encircling $z=0$ and $z=-\sqrt{1-s^2}$, can be understood in light of the symmetry \eqref{symmetryframe} by looking at the relation between the indices of the singular points $z=1$ and $z=-1$, which are $\rmi\,q\,s/4$ and $q\,s/4$ (these are the indices appearing in the dictionary \eqref{dictioHeun}). This is related to the generalisation of the standard PNP relation discussed in \cite{Cavusoglu:2024usn}, which explicitly depends on the residues at singular points. 

The expected existence of such a relation is even better understood from the point of view of the gauge theory quantisation condition analysed in the previous section, which was expressed as a quantisation condition for the A-period of the underlying quantum SW curve. Indeed, in section \ref{sec:gauge}, we used a decomposition of the 4-punctured sphere characterising the singularity structure of the spectral problem that sets the two singular points where the boundary conditions are imposed on the same side of the pants decomposition of the 4-punctured sphere (this is in fact related with the definition of the variable $z$ used in section \ref{sec:gauge}). This choice of frame makes the connection formula particularly simple, ending up as a quantisation for the composite monodromy parameter $a$ as written in \eqref{quantcondHeun2}. Using the Matone relation \eqref{Matonerel}, this condition is particularly efficient at finding the QNM frequencies.\footnote{The Matone relation itself is recognised as a PNP relation in the context of simpler cases such as the (modified) Mathieu equation \cite{Codesido:2017jwp}.}

One could in principle consider an alternative decomposition of the singularity structure and put the singular points where the boundary conditions are imposed on different patches of the pants decomposition of the 4-punctured sphere. This would involve a more complicated connection formula of the form of equation (4.1.22) in \cite{Bonelli:2022ten}.
We checked that this reproduces the correct frequencies in a concrete choice of variable $z$. However, the need to compute the instanton expansion of the composite monodromy parameter $a$ makes including several instanton corrections more involved, and the results less accurate as a consequence. 
The symmetry of the problem is precisely the key aspect that allows us to choose a simple frame in which the quantisation condition is efficient, and the existence of a PNP relation in the WKB approach follows from the same perspective, since the change of frame that swaps the A and B periods is realised by the map \eqref{symmetryframe}.

\subsection{Leading order results}
\label{leading order result for the WKB method}

In this subsection, we examine the first subleading behaviour of $s$ implied by the EQC  \eqref{quantcondWKBscalar}, making contact with the previous work of \cite{Fuini:2016qsc}. As explained in appendix \ref{app:genericS}, it is possible to write the first term of the large-$q$ expansion of the period $\alpha$ (defined in equation \eqref{ABperiods}) in terms of a hypergeometric function, and all the subsequent terms of this expansion as linear combinations of Beta functions. The large-$q$ expansion of $\alpha$ can then be organised as 
\begin{equation} \label{largeqexpansionofalpha}
\alpha =(1-s^2)^{3/4}\left( \, q\,\nu\,_2F_1\left(1,\frac{1}{4}, \frac{7}{4},1-s^2\right) + \sum_{m=1}^\infty q^{1-2m} \sum_{k=0}^{2m-1}c_{m,2m-k}\,(1-s^2)^{k-3m/2}\right),
\end{equation}
where 
\begin{equation}
\nu = \frac{\Gamma(1/4) \Gamma(3/2)}{2\Gamma(7/4)},
\end{equation}
and $c_{m, 2m-k}$ are constant coefficients obtained from the Beta functions and can be found systematically as detailed in appendix \ref{app:genericS}.
Ignoring the contribution of the logarithm containing $\beta$ in \eqref{quantcondWKBscalar}\footnote{We will see later that these terms encode non-perturbative contributions, and thus are not relevant for the leading order analysis, but for now we take this fact for granted.}, and taking only the first term of the large-$q$ expansion of $\alpha$, we have
\begin{equation} \label{leading order quantisation condition}
q(1-s^2)^{3/4}\frac{\Gamma(1/4)\Gamma(3/2)}{2\Gamma(7/4)}\, _2F_1(1, 1/4;7/4, 1-s^2)+\text{h.o.t.}= 2\pi \rmi \left(N+\frac{1}{2}\right).
\end{equation}
This is the leading order quantisation condition. In the previous subsection, we derived the EQC \eqref{quantcondWKBscalar} under assumption \eqref{s assumptions} and $\arg(1-s^2)$ subject to restrictions. Solving \eqref{leading order quantisation condition} yields a leading order solution for $s$ which is consistent with the restrictions imposed on $\arg(1-s^2)$:
\begin{equation}\label{q expansion of s}
s = 1 + \omega_1(N,\infty)\, q^{-4/3} + \mathcal{O}(q^{-8/3}).
\end{equation}
The convenience of this notation will become apparent in section \ref{sec:results}. 
To leading order in large $N$ one finds
\begin{align}\label{s1}
\omega_1(N,\infty) &= \frac{\rme^{-\rmi\pi/3}}{2}\left( \frac{\sqrt\pi \Gamma(7/4)}{\Gamma(5/4)}( 2N+1)\right)^{4/3} + \text{h.o.t}.
\end{align}
The behaviour of $\omega_1(N,\infty)$ is in agreement with the large-$N$ behaviour of the coefficient presented in \cite{Fuini:2016qsc}. Astute readers may observe that if $s$ has the large-$q$ asymptotics of \eqref{q expansion of s}, then $1-s^2  = \mathcal{O}(q^{-4/3})$. This would invalidate the original assumptions of equation \eqref{riccati} unless restrictions are placed on $\omega_1(N,\infty)$. Specifically, we require that $\omega_1(N,\infty)\,q^{-4/3}$ is larger than $q^{-1}$ and this can be achieved if $ q^{-1}  \ll (N/q)^{4/3}$. Together with assumption  \eqref{s assumptions}, our analysis is now valid for 
\begin{equation} \label{N and q inequal}
    q^{-1}  \ll (N/q)^{4/3} \ll 1.
\end{equation}
Thus, the validity of this assumption requires a large-$N$, large-$q$ analysis which cannot be disentangled. One could in principle analyse the problem for finite $N$, changing our original assumptions about the $q$-expansion of the potential $V(z)$, obtaining a very different asymptotic problem, where $N$ and large-$q$ would not be entangled. This has been seen in other well known problems, for example in Reissner-Nordstr\"{o}m \cite{Dias:2018ufh}. In our case, the resulting problem would have increased difficulty, thus our choice of an entangled expansion in the current paper. Both should provide consistent results with the boundary value problem.

To compute further terms in the large-$q$ expansion of $s$, one has to include additional terms in the large-$q$ expansion of the period $\alpha$. A priori, one might expect that the leading term of $\alpha$ determines $\omega_1(N,\infty)$, the next term in $\alpha$ should determine the next coefficient in $s$, and so on. This expectation is not correct. Instead, \emph{every} term in the large-$q$ expansion of $\alpha$ contributes to \emph{every} coefficient in the large-$q$ expansion of $s$\footnote{This situation contrasts to EQCs appearing in quantum mechanics \cite{zinn-justinMultiInstantonsExactResults2004,dunneUniformWKBMultiinstantons2014}. In these scenarios, the energy $E$ is found as a small-$\hbar$ expansion and the first term of the small-$\hbar$ expansion of the EQC is sufficient to calculate \emph{exactly} the first term of the energy. In general, including  $M$ terms in the EQC is sufficient to calculate $M$ terms of $E$.}. 
As we will make explicit in the next subsection, this follows from the scaling implied by the leading-order quantisation condition \eqref{leading order quantisation condition}. That condition requires $(1-s^2)^{3/4}\sim q^{-1}$, so that the leading contribution to $\alpha$ is of order $\mathcal{O}(q^0)$. The next term in the formal large-$q$ expansion of $\alpha$ carries an explicit factor of $q^{-1}$, but it also contains an overall factor of $(1-s^2)^{-3/4}\sim q$. As a result, it is again of order $\mathcal{O}(q^0)$ once evaluated on the scaling dictated by the quantisation condition. The same mechanism persists at all orders: each term in the large-$q$ expansion of $\alpha$ contains the corresponding power of $(1-s^2)$ needed to compensate its explicit power of $q^{-1}$, so that it contributes already at order $\mathcal{O}(q^0)$ to the leading coefficient $\omega_1(N,\infty)$. The same structure then propagates to all higher coefficients in the large-$q$ expansion of $s$.

One might hope to determine the large-$q$ coefficients of $s$ numerically by truncating the large-$q$ expansion of $\alpha$ at some finite order, substituting an appropriate ansatz for $s$, and solving the resulting polynomial equations for the coefficients of the expansion of $s$. One would then hope that, as the truncation order is increased, the corresponding numerical solutions converge in the sense that the extracted coefficients of $s$ stabilise to a certain number of digits. This is analogous to the truncation procedure used in the SW approach, where truncating the small-$t$ expansion at a given order leads to a polynomial equation of the same order for $\omega$. In the present case, however, if one attempts this procedure, the values obtained for the coefficients of $s$ initially appear to stabilise, but only up to a certain truncation order; beyond that point they cease to improve, and eventually drift away. This behaviour is naturally interpreted in terms of optimal truncation for an asymptotic expansion \cite{dingle1973asymptotic}. As the truncation order increases, the approximation improves only up to an optimal order, after which the neglected asymptotic nature of the series manifests itself and the accuracy deteriorates. This can be seen explicitly already for $\omega_1(N, \infty)$, see table \ref{tab:poly solve s1}, for which an independent numerical value is available from \cite{Fuini:2016qsc} in table IV. This suggests that the coefficients of $s$ should not be viewed as the limits of roots of polynomial equations at increasing truncation order, but rather as asymptotic quantities to be determined order by order. This motivates solving the coefficients of $s$ as asymptotic series, rather than expecting the polynomial truncation procedure itself to converge.
\begin{table}[h!]
    \centering
    \begin{tabular}{|c|c|}
    \hline
         Number of terms in expansion of $\alpha$& $\omega_1(N,\infty)\, \rme^{\rmi \pi/3}$  \\
\hline
1  & 4.727122889 \\
\hline
5  & 4.463978747 \\
\hline
10 & 4.465120159 \\
\hline
15 & 4.945596942 \\
\hline
20 & 7.083015495 \\
\hline
25 & 9.700829321 \\
\hline
\end{tabular}
    
    \caption{Values of $\omega_1(N,\infty)\,\rme^{\rmi \pi /3}$ calculated from truncating the large-$q$ expansion of $\alpha$ and solving the resulting polynomial obeyed by $\omega_1(N,\infty)$. The numerical value of $\omega_1(N,\infty)\rme^{\rmi \pi /3}$ taken from $\cite{Fuini:2016qsc}$ is $4.46404110...$ . Indeed, the values found initially converge to the numerical value before drifting away.}
    \label{tab:poly solve s1}
\end{table}

\subsection{Beyond leading order} \label{beyond leading order}
To make the discussion of the previous subsection concrete, it will prove useful to introduce new variables. In the large-$q$ expansion of $\alpha$ (see equation \eqref{largeqexpansionofalpha}) powers of $(1-s^2)^{3/4}$ appear frequently, motivating the definition
\begin{equation} \label{E defn}
E\mathcal{E}\equiv q(1-s^2)^{3/4},
\end{equation}
where $\mathcal{E}(N)$ is a $q$ independent quantity. In terms of the original variable $s$, $\mathcal{E}$ is a simple function of the subleading behaviour $\omega_1=\omega_1(N,\infty)$:
\begin{equation} \label{curlyE}
    \mathcal{E} = (-2\omega_1)^{3/4}.
\end{equation}
We discuss how to compute $\mathcal{E}$ from the EQC \eqref{quantcondWKBscalar} in section \ref{sec:resurgenceofQNMs}. We also introduce a rescaling of $q$,
\begin{equation}\label{hbar}
    \hbar\equiv( q/\mathcal{E})^{-4/3}\,.
\end{equation}
We then organise the expansions of the period $\alpha$ and the quantity $E$ with respect to the following inequality
\begin{equation}
    \hbar\ll\mathcal{E}^{-1}\ll1,
\end{equation}
taking a double expansion, first for small $\hbar$, and then for large $\mathcal{E}$.

Together with the inequality \eqref{N and q inequal} we now have the following range of validity for our analysis
\begin{equation} \label{full N and q inequality}
    q^{-1} \ll (N/q)^{4/3} \ll N^{-1}.
\end{equation}
However, as we shall see in section \ref{sec:results}, the resulting expansions yield accurate results beyond what one might expect from \eqref{full N and q inequality}. Thus, 
inequality \eqref{full N and q inequality} should be thought of as an organising principle rather than a strict range of validity.

With these new variables, the large-$q$ expansion \eqref{largeqexpansionofalpha} of the period $\alpha$ can be organised as a small-$\hbar$ expansion
\begin{equation} \label{alpha period with T}
    \alpha = \sum_{n=0}^\infty \hbar^n E^{4n/3} \left(\nu \frac{(1/4)_n}{(7/4)_n} E\mathcal{\mathcal{E}} + T_n(E  \mathcal{\mathcal{E}})\right),
\end{equation}
where
\begin{equation}\label{Tnx}
T_{n}(x) \equiv\sum_{m=\lceil (n+1)/2\rceil }^\infty x^{1-2m}c_{m,2m-n}.
\end{equation}
To solve for the coefficients of $E$, we substitute the small-$\hbar$ expansion of $E$
\begin{equation} \label{hbar expansion of E}
    E=E_0+\sum_{n=1}^\infty E_n\hbar^n, \qquad E_0 =1,
\end{equation}
into the small $\hbar$ expansion of $\alpha$ \eqref{alpha period with T}. Then, by considering only the perturbative part of the EQC \eqref{quantcondWKBscalar} and using the formula for the $n^{th}$ term of a composition of series, see appendix \ref{app:CompOfseries}, we obtain the following relation obeyed by the coefficients $E_n$:
\begin{equation}\label{hbar expansion of alpha}
    \sum_{n=0}^\infty\alpha_{n}(\mathcal{E},E_1, \dots, E_n)\hbar^n = 2\pi  \rmi \left(N+\frac{1}{2}\right),
\end{equation}
where the coefficients $\alpha_n$ are given by\footnote{We note that $E_0$ also appears in $\alpha_n$ but its value is $1$.}
\begin{equation} \label{alphanexpanded}
    \alpha_{n}(\mathcal{E}, E_1, \dots, E_n)\equiv \sum_{k=0}^n\sum_{l=0}^{n-k}
D_l\left(E^{4k/3}\right)\, D_{n-k-l}\left( \nu \frac{(1/4)_k}{(7/4)_k}E\mathcal{E}+T_k(E\mathcal{E})\right), 
\end{equation}
and $D_l$ is the following differential operator
\begin{equation}
    D_l \equiv \frac{1}{l!}\left.\frac{d^l}{d\hbar^l}\right \vert_{\hbar=0}.
\end{equation}

As discussed in the previous subsection, each term $\alpha_n$ receives infinitely many contributions. However, these contributions are ordered by a large $\mathcal{E}$ series captured by $T_{k}(E\mathcal{E})$ and its derivatives evaluated at $\hbar=0$ or $E=1$. To compute $\alpha_n$, we must compute $D_l\left(E^{4k /3}\right)$ and $D_l(T_k(E\mathcal{E}))$. We discuss $D_l\left(E^{4k /3}\right)$ in appendix \ref{app:CompOfseries} but note that
\begin{equation}
    \frac{d^{n}T_{k}}{d\hbar^n}(E \mathcal{E}) =  \sum_{i=1}^n\mathcal{E}^i \left.\frac{\partial^iT_k(x)}{\partial x^i}\right|_{x=E \mathcal{E}} B_{n,i}\left( \frac{dE}{d \hbar}, \dots, \frac{d^{n-i+1} E}{d \hbar^{n-i+1}}\right),
\end{equation}
where $B_{n,i}$ are partial exponential Bell polynomials \cite{bell}. These derivatives can be further simplified once evaluated at $\hbar =0$ yielding the following expression for $D_n\left(T_k(E \mathcal{E})\right)$:
\begin{equation}
n! D_n(T_k(E\mathcal{E})) =  \sum_{i=1}^n \mathcal{E}^i \left. \frac{\partial^iT_k(x)}{\partial x^i} \right \vert_{x=\mathcal{E}} B_{n,i}(E),
\end{equation}
where
\begin{equation} \label{BellpolyE}
  B_{n,i}(E) \equiv B_{n,i}\left(  E_1, \dots,E_{n-i+1}(n-i+1)!\right).
\end{equation}
Then \eqref{alphanexpanded} implies that for $n\geq 1$ we have $\alpha_n(\mathcal{E}, E_1, \dots, E_n)=0$ and by rearranging we obtain a recurrence relation for $E_n$:
 \begin{equation}\label{E recur}
 \begin{aligned}
     E_n = \frac{-1}{\mathcal{E}\left[\nu + D_1(T_0(E\mathcal{E}))\right]}
\Bigg(
& \sum_{k=1}^n \sum_{l=0}^{n-k}
D_l(E^{4k/3})\, D_{n-k-l}\!\left( \nu \frac{(1/4)_k}{(7/4)_k}E\mathcal{E} + T_k(E\mathcal{E}) \right) \\
& + \frac{1}{n!} \sum_{i=2}^n \mathcal{E}^i
\left.\frac{\partial^i T_0(x)}{\partial x^i}\right|_{x=\mathcal{E}}
B_{n,i}(E)
\Bigg), \qquad n\geq 1,
\end{aligned}
\end{equation}
where the sums over infinitely many terms appearing in the coefficients $\alpha_n$ are contained solely in the partial derivatives $\partial^iT_n(x)/\partial x^i$ evaluated at $x =\mathcal{E}$. The series $T_n(x)$ and their derivatives are not convergent power series in $x$. Rather, they are asymptotic expansions with factorially growing coefficients and hence vanishing radius of convergence. As a result, the recursive determination of the coefficients $E_n$ through \eqref{E recur} is only formal until a precise summation prescription is specified for $T_n(x)$ and its derivatives evaluated at $x=\mathcal{E}$. This is where resummation becomes essential.

The next section is devoted to giving a precise meaning to these divergent series by means of resummation. In particular, we will explain how to resum $T_n(x)$ and their derivatives, and how the period $\beta$ in \eqref{ABperiods} enters this construction and the recursive determination of the coefficients $E_n$.

\section{The resurgence of quasinormal modes} \label{sec:resurgenceofQNMs}

To go beyond the formal manipulations of the previous section, we need a summation procedure for the asymptotic series $T_n(\mathcal{E})$. These series have vanishing radius of convergence, so in order to extract the coefficients $E_n$ from \eqref{E recur} one must first assign a precise meaning to the divergent quantities $T_n(\mathcal{E})$ and to their derivatives. This is achieved by the machinery of Borel--Laplace resummation and resurgence.

As we will describe, this summation procedure is not unique. The Borel transform of $T_n(x)$ develops singularities that obstruct naive Laplace resummation along certain directions, leading to non-perturbative ambiguities in the resummed answer. For the series relevant here, these ambiguities are precisely compensated once one restores the exponentially small contributions $\mp \log(1+e^{\beta})$ that were omitted at the end of the previous section. The role of resurgence is therefore twofold: it provides a consistent summation prescription for the divergent large-$q$ expansions, and it identifies the non-perturbative sectors required to complete the quantisation condition unambiguously.

With this structure in place, the large-$N$ expansion of the quasinormal frequency should be understood as a transseries in which the exponentially small corrections are controlled by the period $\beta$. We now explain this resummation procedure in detail.

\subsection{Resummation and resurgence }
In this subsection, we briefly review how one can associate analytic functions to formal asymptotic expansions through Borel--Laplace resummation. The discussion will be brief and mainly aims to establish notation, for more details see \cite{Aniceto_2019, sauzin2014introduction1summabilityresurgence, Dorigoni_2019}. Given a function $F$ admitting a divergent large-$x$ asymptotic expansion $F(x)$, we now outline the corresponding Borel--Laplace procedure.

Let the asymptotic expansion $F(x)$ be given by 
\begin{equation} \label{app: large z expansion}
    F(x) = x^\mu \sum_{n=0}^\infty a_n x^{-n},
\end{equation}
where $\mu$ and $a_n$ are complex constants. We can define its Borel transform via the term-by-term transformation $x^{-\alpha}\to \xi^{\alpha-1}/\Gamma(\alpha)$:
\begin{equation} \label{boreltransform}
    \mathcal{B}[F](\xi) \equiv \sum^{\infty}_{n=0}a_n \frac{\xi^{n-\mu-1}}{\Gamma(n-\mu)}.
\end{equation}
The $\xi$-plane is called the \emph{Borel plane}. If the coefficients $a_n$ exhibit factorial growth, the division by $\Gamma(n-\mu)$ regularises this behaviour, and the Borel transformed series will have a finite radius of convergence around $\xi=0$. 
To reconstruct the function in the direction $\arg(x)=\theta$, one first analytically continues $\mathcal{B}[F](\xi)$ away from its disk of convergence and then performs the Laplace transform along the ray of angle $-\theta$ in the Borel plane. The resulting \emph{Borel--Laplace resummation} of $F(x)$ in the direction $\theta$ is 

\begin{equation}
    \mathcal{S_\theta}[F](x) \equiv \int_0^{\infty \rme^{-\rmi\theta}} \rme^{-x\xi}\mathcal{B}[F](\xi)\,d\xi. 
\end{equation}
Whenever this integral is well-defined, it provides an analytic function of $x$ in the corresponding sector, whose asymptotic expansion for large $x$ in direction $\theta$ reproduces the original formal series $F(x)$. In practice, one only has a finite number of terms in \eqref{boreltransform} and one has to approximate the analytic continuation, typically via Pad\'e approximants (see \cite{Aniceto_2019} for the use of Padé approximants in the resurgence context). Moreover, the Borel--Laplace resummation is compatible with the algebraic operations that will be relevant in our analysis. This allows us to work formally at the level of asymptotic expansions and resum only at the end.

Depending on the behaviour of $\mathcal{B}[F](\xi)$ one may not be able to perform the above procedure on $F$. The above construction can fail if the analytic continuation of $\mathcal{B}[F](\xi)$ develops singularities along the integration ray.
In that case, we need to choose a path that avoids these singularities. The most obvious choice of paths is obtained by deforming the contour slightly above or slightly below the original ray. This motivates defining the \emph{lateral Borel-Laplace summations}
\begin{equation} \label{lateral summations}
\mathcal{S}_{\theta^\pm}[F](x) \equiv \int_0^{\infty \rme^{-\rmi(\theta\pm \epsilon)}} \rme^{-x\xi}\mathcal{B}[F](\xi)\,d\xi, 
\end{equation}
for some small $\epsilon>0$. We will often abuse notation and write $\mathcal{S}_{\theta^\pm}[F](x)$ =  $\mathcal{S}_{\pm}[F](x)$ when $\theta$ is clear from context.

If no singularity lies on the integration ray, the two lateral resummations coincide. If instead the ray intersects singular points of the Borel transform, the two resummations generally differ, and this difference is the manifestation of the Stokes phenomenon.
Specifically, if we assume that near a singularity $\xi_*$, $\mathcal{B}[F](\xi)$ can be written as 
\begin{equation}
     \mathcal{B}[F](\xi) = \frac{1}{2\pi i} \mathcal{B}[G](\xi-\xi_*)\log(\xi-\xi_*) + \text{reg}(\xi),
\end{equation}
where $\text{reg}(\xi)$ is a function regular at $\xi_*$, and  $G(x)$ is an asymptotic series of the same form as \eqref{app: large z expansion} (potentially with a different $\mu$), then the lateral resummations will differ by terms schematically given by
\begin{equation}
    \rme^{- x \xi_*} \mathcal{S}_{\theta^{\pm}}[G](x).
\end{equation}
These are non-perturbative in $x$ and are not included in the expansion \eqref{app: large z expansion}. To account for the appearance of these non-perturbative terms, we should instead consider the \emph{transseries} expansion of \eqref{app: large z expansion}
\begin{equation}\label{transseries}
\begin{aligned}
    \mathcal{F}(x) = \sum_{m=0}^\infty  \rme^{-mA x} F^{(m)}(x),\\
    F^{(m)}(x) 
    =x^{\mu_m}\sum_{n=0}^\infty a_n^{(m)}x^{-n},
\end{aligned}
\end{equation}
where we are assuming that $A$ is the closest singularity to the origin in the direction $-\theta$ and that the other singularities in this direction are multiples of $A$\footnote{If there are singularities appearing in different directions we can extend the definition of our transseries to include these as well, see \cite{Aniceto_2019, Dorigoni_2019} for more details.}.

Typically, the definition of a transseries includes a \emph{transseries parameter} $\sigma$ and is given by $\mathcal{F}(x) = \sum_{m=0}^\infty \sigma^m\rme^{-mA x} F^{(m)}(x) $. This parameter is then usually set to a specific value by the initial/boundary conditions in the context of differential equations, while the exact quantisation condition (EQC) is an implicit equation that automatically sets the value of $\sigma$ in the region of interest for our problem. Thus, in what follows, the value of $\sigma$ is already implicitly included in the results. See \cite{spaendonckExactInstantonTransseries2024} for further details.

We call $F^{(m)}(x)$ the \emph{$m^{th}$ transseries sector}. Borel-Laplace resummation can then be applied sector-wise, yielding
\begin{equation}
    \mathcal{S}_\theta[\mathcal{F}](x) = \sum_{m=0}^\infty e^{-mAx}\mathcal{S}_\theta [\mathcal{F}^{(m)}](x).
\end{equation}
The difference between the two summations can be written as follows 
\begin{equation}
      \mathcal{S}_{\theta^{+}}[\mathcal{F}](x) = \mathcal{S_{\theta^-}} [\mathcal{F}](x)-S_{\theta^-}[\text{Disc}_\theta[\mathcal{F}]](x)
\end{equation}
where $\text{Disc}_\theta$ is an operator that acts on transseries and captures the discontinuity between the two summations.
\subsection{The large \texorpdfstring{$N$}{} expansion of \texorpdfstring{$\mathcal{E}$}{}} \label{large N of mathcal E}

With the machinery to handle divergent series, we discuss one method to calculate the quantity $\mathcal{E}$, given in equation \eqref{curlyE},  as a transseries for large $N$. This serves two purposes. Firstly, it is necessary to calculate $\mathcal{E}$ in order to compute the coefficients $E_n$ of \eqref{hbar expansion of E}, which are equivalent to the coefficients of the large-$q$ expansion of the QNMs. Secondly, it serves to illustrate how the non-perturbative corrections $\mp\log(1+\rme^{\beta})$ are necessary to find the correct value of $\mathcal{E}$. The discussion in this subsection is analogous for the series $T_n(x)$ in \eqref{Tnx}. 

\begin{figure}[h!]
    \centering
    \includegraphics[width=0.7\linewidth]{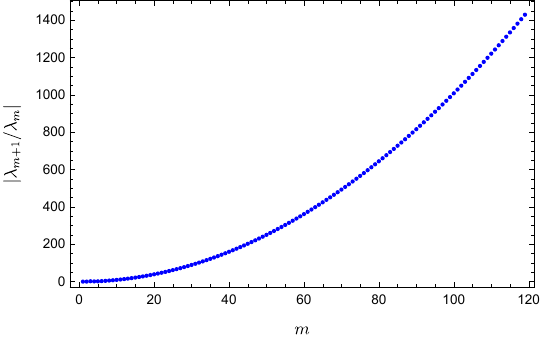}
    \caption{Plot of the absolute value of $\lambda_{m+1}/ \lambda_m$, the ratio clearly does not converge suggesting the coefficients $\lambda_m$ grow factorially.}
    \label{fig:factorialgrowthofE_0}
\end{figure}

We start by finding the perturbative part of $\mathcal{E}$. This series can be calculated by inverting the relation obeyed by $\alpha_0(\mathcal{E})$ of equation \eqref{alphanexpanded}:
\begin{equation}
\nu \mathcal{E} + \sum_{m=1}^\infty c_{m,2m} \mathcal{E}^{1-2m}= 2\pi \rmi \left(N+\frac{1}{2}\right),
\end{equation}
which yields a large $N$ series for $\mathcal{E}$
\begin{equation}\label{eq:Large_N_expansion_of_mathcal_E}
    \mathcal{E} = \sum_{m=0}^{\infty} \lambda_m\left(N+\frac{1}{2}\right)^{1-2m}.
\end{equation} 
Plotting $|\lambda_{m+1}/\lambda_m|$ we see that the ratio of coefficients grows quadratically -- see figure \ref{fig:factorialgrowthofE_0}. This confirms that the coefficients $\lambda_m$ grow factorially in $2m$ for large values of $m$.
The factorial growth of the coefficients $\lambda_m$ means that $\mathcal{E}$ is a divergent asymptotic series and that Borel-Laplace resummation must be used to obtain a value for $\mathcal{E}$.

\begin{figure}[h!]
    \centering
    \includegraphics[width=0.6\linewidth]{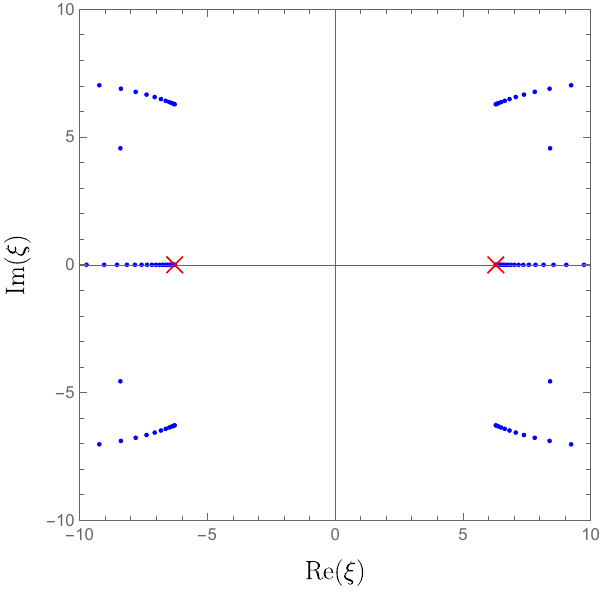}
    \caption{The poles of the Pad\'e approximant of $\mathcal{B}[\mathcal{E}](\xi)$ represented as blue dots. The clustering of poles indicates the location of a singularity for $\mathcal{B}[\mathcal{E}](\xi)$. The red crosses are located at $\pm 2\pi$.}
    \label{fig:BorelPlaneOfL0expansion}

\end{figure}

Since $N+1/2$ is real and positive, we must take the Laplace transform of $\mathcal{B}[\mathcal{E}]$ along the positive real axis to obtain $\mathcal{S}[\mathcal{E}](N+1/2)$. However, as shown in figure \ref{fig:BorelPlaneOfL0expansion}, there are singularities in the Borel plane of $\mathcal{E}$ along the real axis that obstructs the integral in this direction. For this reason, we must instead use lateral Borel-Laplace resummation, specified in equation \eqref{lateral summations}, and integrate along a contour slightly above or slightly below the real axis. Alternatively, we can consider the resummation of \eqref{eq:Large_N_expansion_of_mathcal_E} along the real axis but with $N+1/2$ replaced by $(N+1/2)e^{\pm i\epsilon}$, for some small $\epsilon>0$. This leads to an ambiguous result as $\mathcal{S}_{0^+}[\mathcal{E}]\neq \mathcal{S}_{0^-}[\mathcal{E}]$. However, the choice of $(N+1/2)e^{\pm i\epsilon}$ selects the sign present in the EQC \eqref{quantcondWKBscalar}. Recall that the EQCs were derived subject to assumptions on $\arg(1-s^2)$. This led to two topologically distinct graphs, the plus Stokes graph and the minus Stokes graph shown in figure \ref{fig:Stokes graphs plus and minus}. Each Stokes graph yielded a different EQC. If $N+1/2$ corresponds to the critical Stokes graph (shown in figure \ref{fig:Stokesgraphbounded}) and to $\arg(1-s^2) = \theta_c$, then the lateral summations $\mathcal{S}_{0^\pm}$ correspond to resumming for $(N+1/2)e^{\pm i\epsilon}$ which then leads to a choice of sign in the EQC. Therefore, although the choice of lateral summation will lead to an ambiguity, the ambiguity will be cancelled by including the terms $\mp\log(1+e^\beta)$ with the sign dictated by the choice of resummation direction.

To show that taking $N+1/2$ real and positive corresponds to the critical Stokes graph, we recall that this Stokes graph is characterised by a bounded Stokes line starting at $z = 0$ and ending at $z=-\sqrt{1-s^2}$. Therefore, this Stokes line is present if
\begin{equation}
 \mathrm{Im} \,\left(  q \int_{-\sqrt{1-s^2}}^0 \sqrt{V_0(z')}\,dz' \right) =0.
\end{equation}
The above integral is the leading term of the $\beta$ period; and, by using the PNP relation \eqref{PNP} as well as the leading term of the $\alpha$ period \eqref{leading order quantisation condition}, is given by
\begin{equation} \label{Leading beta term}
     \rmi q(1-s^2)^{3/4}\frac{\Gamma(1/4)\Gamma(3/2)}{2\Gamma(7/4)}\, _2F_1(1, 1/4;7/4, 1-s^2).
\end{equation}
Then, if $s$ obeys the leading order quantisation condition \eqref{leading order quantisation condition}, the quantity \eqref{Leading beta term} is real when $N+1/2$ is real and choosing  $(N+1/2)\rme^{+ \rmi \epsilon}/(N+1/2)\rme^{- \rmi \epsilon}$ selects the plus/minus Stokes graph.

This ambiguity also exists if one attempts to resum the series $T_n(x)$. The large $x$ series of $T_n(x)$ are also divergent, and their Borel transforms also contain singularities in the direction we wish to resum, in this case $\arg(\mathcal{E})$. Again, the resulting ambiguities will be exactly cancelled by the previously omitted term $\mp\log(1+\rme^{\beta})$.

We end this subsection by commenting that this is an example of \emph{median summation}. Median summation appears naturally in the context of differential equations obeyed by a real function, see \cite{delabaere:hal-01886535, Aniceto:2013fka} for more details. If the real function has an asymptotic series with factorially growing coefficients, one might find singularities along the real axis in the Borel plane, which leads to an imaginary ambiguity when attempting to resum above or below the real axis. This ambiguity can be cancelled by including a non-perturbative correction multiplied by a suitable constant. However, this non-perturbative correction will also contain an ambiguity, which then must be cancelled by the second non-perturbative correction, again multiplied by a suitable constant. In fact, there is a prescribed choice of transseries parameter that ensures exact cancellation among resummed transseries sectors. This prescription is called median summation. In our case, there are no transseries parameters. However, the EQC automatically encodes this prescription; and, as we shall see in the next subsection, the ambiguities we find are precisely cancelled.

\subsection{Calculating non-perturbative corrections}
As discussed in the last section, the resummed $\mathcal{E}$ and $T_n(x)$ will be ambiguous. To cancel these ambiguities we need to include the contributions from the $\beta$ period. As anticipated, these corrections are non-perturbative in the overtone number $N$. To see this, we note that the leading term of the large-$q$ expansion of $\beta$ is given by \eqref{Leading beta term}. When $s$ solves $\alpha = 2\pi \rmi (N+1/2)$, the leading term \eqref{Leading beta term} equals $-2 \pi (N+1/2)$. Then, ignoring subleading corrections in large $N$ and $q$, we have 
\begin{equation}
\rme^{\beta} \sim \rme^{-2\pi \left(N+\frac{1}{2}\right)},
\end{equation} 
which is non-perturbative for large real $N$.

To find the non-perturbative corrections, we note that the series $T_n$ appears in the small-$\hbar$ expansion of $\alpha$ \eqref{alpha period with T} before explicitly expanding $E$. Thus, the non-perturbative corrections to $T_n$ must be captured by the $k^{th}$ term of the small-$\hbar$ expansion of $\mp \log(1+\rme^{\beta})$ before $E$ is explicitly expanded. The small-$\hbar$ expansion of $\beta$ is straightforward to calculate by utilising the P/NP relation \eqref{PNP}:
\begin{align}
     \beta &= \sum_{n=0}^\infty \hbar^{n}E^{4n/3}\beta_n, 
\end{align}
where 
\begin{equation}
    \beta_n \equiv\left( \rmi \nu \frac{(1/4)_n}{(7/4)_n} E\mathcal{ E} +T_n(\rmi E\mathcal{E})\right).
\end{equation}
Then the small-$\hbar$ expansion of $\rme^\beta$ is then given by 
\begin{equation}\label{exp beta expansion}
    \rme^{\beta} = \rme^{\beta_0} \sum_{n=0}^\infty \frac{\hbar^n E^{4n/3}}{n!} B_n(\beta_1, \dots, n!\beta_n),
\end{equation}
where $B_n$ are complete exponential Bell polynomials \cite{bell}, see appendix \ref{app:large q and N expansion} for more details.

For convenience, we will drop $B_n$'s explicit dependence on $\beta_i$. Using the expansion of $\exp(\beta)$, \eqref{exp beta expansion}, the non-perturbative corrections to the period $\alpha$ can be written as
\begin{align}
\log\left(1+\rme^\beta\right)&= \log\left(1+\rme^{\beta_0}\right) + \log\left(1+ \eta \sum_{n=0}^\infty \frac{\hbar^n E^{4n/3}}{n!} B_n\right)\\
&= \log\left(1+\rme^{\beta_0}\right) + \sum_{n=1}^\infty\hbar^nE^{4n/3} \sum_{k=1}^n \frac{(-1)^{k+1}(k-1)!}{n!} \eta^kB_{n,k}(B_1, \dots, B_{n-k+1}),
\end{align}
where $\eta$ is non-perturbative for large real $N$ and is given by 
\begin{equation} \label{eq:eta}
\eta(\mathcal{E}) \equiv \frac{\rme^{\beta_0(\mathcal{E})}}{1+\rme^{\beta_0(\mathcal{E})}}.
\end{equation}
Including the small-$\hbar$ expansion of the logarithm, our EQCs \eqref{quantcondWKBscalar} expanded for small-$\hbar$ (again, not expanding $E$) can be written as 
\begin{equation} \label{transseries EQC}
\sum_{n=0}^\infty\hbar^{n}E^{4n/3}\left(\nu\frac{(1/4)_n}{(7/4)_n}E\mathcal{E} + \mathcal{T}^\pm_n(E\mathcal{E})\right) =2 \pi \rmi \left(N+\frac{1}{2}\right),
\end{equation}
where $\mathcal{T}^\pm_n(x)$ are transseries given by
\begin{equation} \label{T transseries}
   \mathcal{T}^\pm_n(x) \equiv \begin{cases}
     T_0( x )  \pm \log\left( 1+\rme^{\beta_0(x)}\right), &n=0\\  T_n( x )
       \pm \sum_{k=1}^n \eta(x)^k T_n^{(k)}(x),& n\geq1,
    \end{cases}
\end{equation}
where 
\begin{equation} \label{eq:Tnpcorrection}
    T_n^{(k)}(x) \equiv \frac{(-1)^{k+1}(k-1)!}{n!} B_{n,k}(B_1(x), \dots, B_n(x))
\end{equation}
and $+/-$ corresponds to the contributions derived from the plus/minus Stokes graph.
The choice of $\eta(x)$ as the non-perturbative parameter means that there are finitely many transseries sectors for a given $\mathcal{T}^\pm_n(x)$. If one expanded $\eta(\mathcal{E})$, each transseries would have infinitely many transseries sectors graded by powers of $\rme^{\beta_0(\mathcal{E})}$. To find the non-perturbative corrections to the derivatives of $T_n(x)$ we simply take derivatives of the above expressions with respect to $x$.

The corrections to the large $N+1/2$ expansion of $\mathcal{E}$, \eqref{eq:Large_N_expansion_of_mathcal_E}, require more care as $\beta_0$ depends on $\mathcal{E}$. We discuss the details of this calculation in appendix \ref{app:large q and N expansion}. The result is a transseries expansion for $\mathcal{E}$ of the form 
\begin{equation} \label{mathcal E large N transseries}
    \mathcal{E^\pm}= \sum_{n=0}^\infty\delta^n \mathcal{E}_\pm^{(n)},
\end{equation}
with $\delta$ capturing non-perturbative contributions in large real $N$ and transseries sectors $\mathcal{E}_\pm^{(n)}$. The precise expressions for $\delta$ and $\mathcal{E}_\pm^{(n)}$ can be found in appendix \ref{app:large q and N expansion}. To obtain the correct value for $\mathcal{E}$, one picks a transseries $\mathcal{E}^{\pm}$ and the corresponding lateral summation $\mathcal{S}_{0^\pm}$ evaluated at $N+1/2$:
\begin{equation}
\mathcal{E} = \mathcal{S}_{0^\pm} [\mathcal{E^\pm}]\left(N+\frac{1}{2}\right).
\end{equation}
The result of including the first and second non-perturbative contribution are shown in table \ref{resummed mathcal E table} \footnote{Note that these results were obtained by using uniformising conformal maps in the Borel plane. This process allows one to approximate analytic continuations of Taylor series with more accuracy than standard Pad\'e approximants.    Specifically, before taking the Pad\'e approximant we first map the Borel plane variable $\xi$ to $\zeta = \varphi(\xi)$ where $\varphi(\xi)$ is the inverse elliptic nome function. We then take the small $\zeta$ expansion, truncating at the same order as the original Borel transform. Finally, we take the Pad\'e approximant in the variable $\zeta$ and then reintroduce $\xi$ by replacing $\zeta$ with $\varphi^{-1}(\xi)$. For further details see \cite{Costin:2020pcj,Costin:2021bay}.}. They clearly show that including contributions from the $\beta$ period cancels the ambiguities present.
\begin{table}[h!] 
\centering
\makebox[\textwidth][c]{%

\begin{tabular}{|c| c| c| c|}
\hline
\rule[-1.7ex]{0pt}{5.8ex}\smash{\raisebox{0.45ex}{$N$}}
&
\smash{\raisebox{0.45ex}{$-\rmi \mathcal{S}_{0^-}\left[\mathcal{E}_-^{(0)}\right]$}}
&
\smash{\raisebox{0.45ex}{$-\rmi \mathcal{S}_{0^-}\left[\mathcal{E}_-^{(0)}+\delta \mathcal {E}_-^{(1)}\right]$}}
&
\smash{\raisebox{0.45ex}{$-\rmi \mathcal{S}_{0^-}\left[\mathcal{E}_-^{(0)}+\delta \mathcal{E}_-^{(1)}+\delta^2 \mathcal{E}_-^{(2)}\right]$}}
\\
\hline
1 & $5.16498 + 1.08357\times10^{-4}\,\rmi$ & $5.16498 - 9.99632\times10^{-9}\,\rmi$ & $5.16498 - 1.77882\times10^{-10}\,\rmi$ \\\hline
2 & $8.85159 + 1.40550\times10^{-7}\,\rmi$ & $8.85159 - 1.68361\times10^{-14}\,\rmi$ & $8.85159 + 1.63992\times10^{-16}\,\rmi$ \\\hline
3 & $12.4848 + 2.26987\times10^{-10}\,\rmi$ & $12.4848 - 4.47754\times10^{-20}\,\rmi$ & $12.4848 - 1.05625\times10^{-22}\,\rmi$ \\\hline
4 & $16.1006 + 3.91900\times10^{-13}\,\rmi$ & $16.1006 - 1.33610\times10^{-25}\,\rmi$ & $16.1006 - 4.86500\times10^{-29}\,\rmi$ \\\hline
5 & $19.7085 + 6.96711\times10^{-16}\,\rmi$ & $19.7085 - 4.21680\times10^{-31}\,\rmi$ & $19.7085 + 1.08374\times10^{-33}\,\rmi$ \\
\hline
\end{tabular}
}
\caption{Results of resumming the transseries $\mathcal{E}^-$ with $0$, $1$ and $2$ non-perturbative terms included and truncating the perturbative expansion of each transseries sector after $119$ terms. The error of each value is given approximately by the imaginary part of these values.}
\label{resummed mathcal E table}
\end{table}

\subsection{Resumming the transseries}
Given the transseries $\mathcal{T}^\pm_n(x)$ we can now calculate the coefficients $E_n$ in terms of analytic functions evaluated at specific values rather than formal series. To do this, we use the fact that resummation commutes with all operators used appearing in \eqref{transseries EQC}, and we obtain 
\begin{equation} 
\sum_{n=0}^\infty\hbar^{n}E^{4n/3}\left(\nu\frac{(1/4)_n}{(7/4)_n}E\mathcal{E} + \mathcal{S}_{\pm}[\mathcal{T}^\pm_n](E\mathcal{E})\right) =2 \pi \rmi \left(N+\frac{1}{2}\right).
\end{equation}
Then, we can repeat the derivation of $E_n$ from subsection \ref{beyond leading order} to obtain 
\begin{equation}\label{resummed E recursion}
\begin{aligned} 
E_n = \frac{-1}{\mathcal{E}\left[\nu + D_1(\mathcal{S}_{\pm}[\mathcal{T}^\pm_0](E\mathcal{E}))\right]}
\Bigg(
& \sum_{k=1}^n \sum_{l=0}^{n-k}
D_l(E^{4k/3})\, D_{n-k-l}\!\left( \nu \frac{(1/4)_k}{(7/4)_k}E\mathcal{E} + \mathcal{S}_{\pm}[\mathcal{T}^\pm_k](E\mathcal{E}) \right) \\
& + \frac{1}{n!} \sum_{i=2}^n \mathcal{E}^i
\left.\frac{\partial^i\mathcal{S}_{\pm}[\mathcal{T}^\pm_0](x)}{\partial x^i}\right|_{x=\mathcal{E}}
B_{n,i}(E)
\Bigg),
\end{aligned}
\end{equation}
which is a recurrence relation, with finite sums, depending only on the resummed values of $\mathcal{E}$, $\mathcal{T}^\pm_n(x)$ and its derivatives.

Having obtained the resummed recurrence relation for $E_n$ \eqref{resummed E recursion}, we can now describe the process of calculating $E_n$ given the data $\nu $ and $c_{k,m}$ of the period $\alpha$ defined in \eqref{largeqexpansionofalpha}. First, we calculate $\mathcal{E}$. This can be done by resumming the large-$N$ transseries \eqref{mathcal E large N transseries} of $\mathcal{E}$ or numerically using the methods discussed in appendix B of \cite{Fuini:2016qsc}. Once a suitably accurate value of $\mathcal{E}$ has been obtained, we calculate the resummed transseries $\mathcal{S}_{\pm}[\mathcal{T}_n^\pm](\mathcal{E})$ and their derivatives using \eqref{T transseries}. This calculation only requires $\mathcal{E}$, $\nu$ and $c_{m,k}$, and no coefficients of $E_n$ are necessary. This means that the precision of our calculation depends solely on the resummed values $\mathcal{S}_{\pm}[\mathcal{T}_n^\pm](\mathcal{E})$ and its derivatives. Once $\mathcal{S}_{\pm}[\mathcal{T}_n^\pm](\mathcal{E})$ and its derivatives have been obtained, we can calculate $E_n$ using the recurrence relation \eqref{resummed E recursion}.
Finally, once the coefficients $E_n$ have been obtained,the large-$q$ expansion of $s$ (and thus of $\omega$) is found by replacing $\hbar$ with $(q/\mathcal{E})^{-4/3}$ and rearranging \eqref{E defn}. We discuss these results in the next section.

\section{Naught to all: the spectrum of QNMs for large to small \texorpdfstring{$q$}{}}\label{sec:results}

In this section we show that the resummed large-$q$ expansion is not only an asymptotic description of the QNM spectrum at large wave number, but also a practical tool for accessing much smaller values of $q$. We start by comparing the resummed WKB frequencies with numerical data at large values of $q$, then use the large-$q$ expansion to approach the small-$q$ regime. Finally, we combine it with the Seiberg-Witten approach and explain how the large-$q$ expansion provides an unambiguous seed for the expansion around a finite value of $q$, from which the frequency can be analytically continued all the way to $q=0$. 

\subsection{Large-\texorpdfstring{$q$}{} results from WKB}

In the last section, we obtained the large-$q$ expansion of $\omega(N,q)$, potentially to arbitrary order. An immediate consequence of the last two sections, specifically equations \eqref{E defn} and \eqref{hbar expansion of E}, is that the expansion of $\omega(N,q)$ is of the form
\begin{equation}\label{large q expansion of omega}
    \omega(N,q) =q\left(1 +  \sum_{\ell=1}^\infty \omega_\ell(N,\infty)q^{-\frac{4}{3}\ell}\right).
\end{equation}
We proceed now to evaluate this expansion. To compute the coefficients $\omega_\ell(N, \infty)$ 
we start by first including the first two non-perturbative corrections in the transseries $\mathcal{T}^\pm_n(\mathcal{E})$, given in \eqref{T transseries}, with $119$ terms included in each transseries sector. Using the numerical values of $\mathcal{E}$\footnote{Specifically, for $N=1, \dots, 5$ we obtain $\mathcal{E}$ via the numerical method outlined in appendix B of \cite{Fuini:2016qsc}. For $N > 5$ we find that the transseries calculation discussed in section \ref{sec:resurgenceofQNMs} yields comparable accuracy to this method.}, the resummed values of $\mathcal{T}^\pm_n(\mathcal{E})$ and its derivatives are obtained, we then compute the coefficients $E_n$ using the recurrence relation \eqref{resummed E recursion}. Finally, we compute the coefficients $\omega_l(N,\infty)$ by inverting the expression \eqref{E defn} and restoring $q$ via \eqref{hbar} for fixed $N$

\begin{figure}[h!]
    \centering
    \begin{subfigure}{0.48\linewidth}
        \centering
        \includegraphics[width=\linewidth]{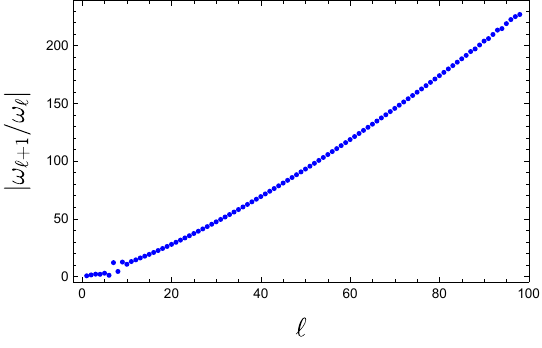}
        
    \end{subfigure}
    \hfill
    \begin{subfigure}{0.48\linewidth}
        \centering
        \includegraphics[width=\linewidth]{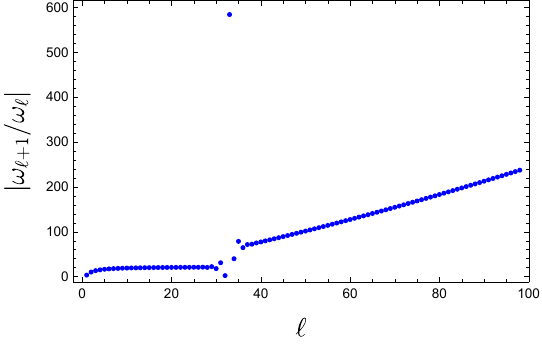}
       
    \end{subfigure}
    
    \caption{The ratio  $\left|\omega_{\ell+1}(N, \infty)/\omega_\ell(N,\infty)\right|$ for the coefficients of the large-$q$ expansion of $\omega$ \eqref{large q expansion of omega}. \textbf{Left panel}: $N=1$. \textbf{Right panel}: $N=5$.}
    \label{fig:E_nRatios}
\end{figure}

A question that naturally arises is whether the series \eqref{large q expansion of omega} is divergent. Plotting the ratio $|\omega_{\ell+1}(N,\infty)/\omega_\ell(N, \infty)|$ we find that large-$q$ coefficients $\omega_\ell(N, \infty)$ diverge factorially for large $\ell$. This is clearly seen in figure \ref{fig:E_nRatios} for $N=1$ and $N=5$. Thus, to evaluate the large-$q$ expansion of $\omega$, we need to resum the divergent series \eqref{large q expansion of omega} using the same asymptotic resummation methods described in section \ref{sec:resurgenceofQNMs}. Following this procedure we compute $\omega(N, q)-q$ for various $N$ and $q$ by resumming the first $100$ terms of the large-$q$ expansion of $\omega$. We compute the numerical values of these QNMs following the method outlined in appendix B of \cite{Fuini:2016qsc} and compare it to our large-$q$ resummation in table \ref{tab:largeqomegaresults}, which shows excellent agreement. For large values of $q$ the error saturates to that of the coefficients $\omega_\ell(N, \infty)$. As expected, increasing $N$ generally increases the accuracy of the results as the coefficients $\omega_\ell(N,q)$ are more accurate for large $N$. However, as can be seen for $q=10$, this is not always true. Recall that our expansion in large-$q$ is effectively an expansion in small $(N/q)^{4/3}$. Consequently, as $N$ increases for fixed $q$, the ratio $N/q$ moves further outside the regime of validity of the expansion. Thus, to maintain the same level of accuracy for larger $N$ we must include more terms in the large-$q$ expansion \eqref{large q expansion of omega}.

\begin{table}[h!]
    \centering

\makebox[\textwidth][c]{%
\begin{tabular}{|c|c|c|c|c|c|}\hline
$N $ & $q= 10$ & $q=20$ & $q=40$ & $q=80$ & $q=160$ \\\hline
1 & $7.59006\times 10^{-10}$ & $1.15515\times 10^{-10}$ & $1.56187\times 10^{-11}$ & $2.01367\times 10^{-12}$ & $2.54803\times 10^{-13}$ \\\hline
2 & $1.89116\times 10^{-14}$ & $1.43251\times 10^{-15}$ & $2.27252\times 10^{-16}$ & $3.12692\times 10^{-17}$ & $4.06017\times 10^{-18}$ \\\hline
3 & $5.84658\times 10^{-16}$ & $2.37433\times 10^{-20}$ & $4.66989\times 10^{-21}$ & $7.10355\times 10^{-22}$ & $9.60908\times 10^{-23}$ \\\hline
4 & $4.11498\times 10^{-15}$ & $5.90751\times 10^{-25}$ & $1.35878\times 10^{-25}$ & $2.34731\times 10^{-26}$ & $3.35639\times 10^{-27}$ \\\hline
5 & $5.26489\times 10^{-14}$ & $1.78055\times 10^{-28}$ & $7.66763\times 10^{-30}$ & $3.18039\times 10^{-30}$ & $1.90949\times 10^{-30}$ \\\hline
\end{tabular}
}
    \caption{Absolute error of the resummation $\mathcal{S}_{0^-}[\omega(N,q)-q]$ of the large-$q$ expansion \eqref{large q expansion of omega} compared with the numerically calculated values, for various values of $N$ and $q$.}
    \label{tab:largeqomegaresults}
\end{table}

We further probe the resurgent structure of the large-$q$ expansion of $\omega$ \eqref{large q expansion of omega} by investigating the type of factorial growth and the singularity structure of the associated Borel plane. By considering the ratio $|\omega_{\ell+3}(N, \infty)/\omega_\ell(N, \infty)|$ we find that $\omega_\ell \sim \Gamma(4\ell/3)$. Then, by taking the appropriate Borel transform for this growth, we find that the Pad\'e approximant of the Borel transform has a line of poles that show a branch cut starting very close to the real axis. With the current accuracy of $\omega_\ell(N, \infty)$, we cannot exclude the possibility that there is a branch point along the real axis and that there are non-perturbative terms for real positive $q$. In any case, in the resummation we avoid the poles around the real axis by integrating slightly below that axis. We remark that whether or not there are additional non-perturbative terms does affect the accuracy of our calculations, as this is limited by the number of terms in the large-$N$ expansion.

However, there is evidence to suggest that this branch cut does not lie along the positive real axis.
Following section \ref{sec:WKB method} we can analyse the Stokes graphs for different values of $\arg q$, and we find a value of $\arg(q)$ close to $0$ for which there is a Stokes graph containing a bounded Stokes line. This bounded Stokes line starts from $\sqrt{1-s^2}$, encircles $z=1$, and ends at $z=0$. Recall that any period whose path of integration intersects a bounded Stokes line for a particular value of $\arg(q)$, will have a singularity in the Borel plane in the direction $-\arg(q)$. Both the paths of integration of $\alpha$ and $\beta$, see \eqref{ABperiods}, intersect this bounded Stokes line; thus the singularity that lies close to the real axis is likely due to this bounded Stokes line which is absent for real positive $q$.

%

In order to evaluate the resummed \eqref{large q expansion of omega}, the accuracy of the coefficients $\omega_\ell(N,\infty)$ plays a major role. Each coefficient $\omega_\ell(N,\infty)$ is determined in the large-$N$ regime, and is effectively described by a (resummed) large$-N$ transseries that we could find explicitly and resum in $N$. However, this large-$N$ expansion is extremely intricate and it is difficult to analyse the limitations to the accuracy of the coefficients. Instead, we calculated the $\omega_\ell(N,\infty)$ through an intermediary object, the transseries $\mathcal{T}^\pm_n(\mathcal{E)}$ (as in \eqref{T transseries}) which is defined naturally from the data of the WKB periods
--- the coefficients $c_{m,k}$ and $\nu$ in equations \eqref{largeqexpansionofalpha}, known to arbitrary accuracy. In the transseries $\mathcal{T}^\pm_n(\mathcal{E)}$, the dependence in $N$ is fully encoded in the numerical variable $\mathcal{E}$. The precision of $\omega_\ell(N,\infty)$ can then be investigated by assessing the accuracy of the resummations $\mathcal{S}_\pm[\mathcal{T}^\pm_n](\mathcal{E})$ and their derivatives. 

\begin{figure}
    \centering
     \begin{subfigure}{0.48\linewidth}
        \centering
    \includegraphics[width=\linewidth]{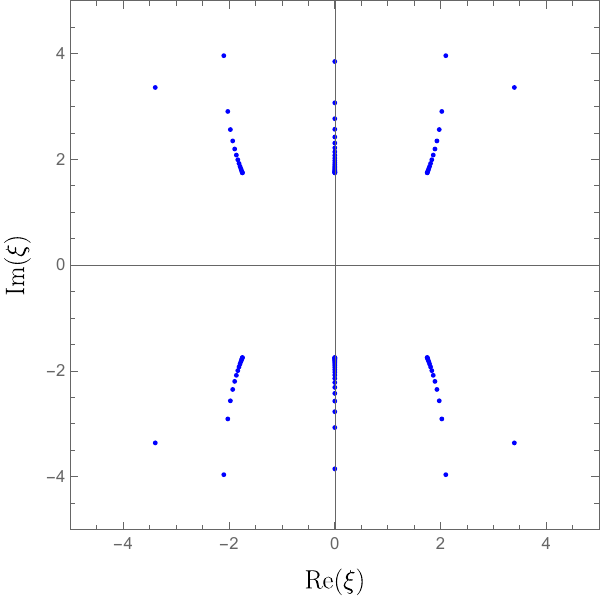}
       
    \end{subfigure}
     \begin{subfigure}{0.48\linewidth}
        \centering
    \includegraphics[width=\linewidth]{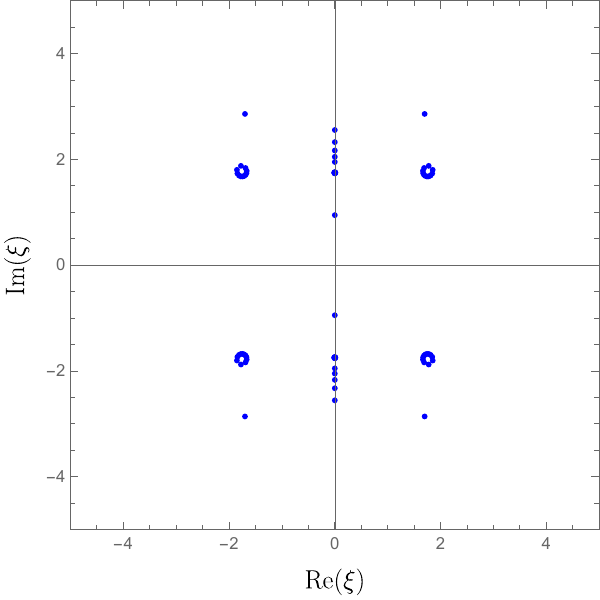}
       
    \end{subfigure}

    \caption{Plot of the poles of the Pad\'e approximant of $\mathcal{B}[T_{n}](\xi)$ for $n=1$ and $n=10$. \textbf{Left panel}: $n=1$. We see a standard picture - singularities in the Borel Plane represented by a line of poles and we expect accurate results when integrating. \textbf{Right panel}: $n=10$. In this case we see irregular behaviour which worsens for large $n$.}
    \label{fig:T_30BorelPlane}
\end{figure}

We find that the resummations of $\mathcal{S}_\pm[\mathcal{T}^\pm_n](\mathcal{E})$ are less accurate as we increase $n$. Starting with the perturbative contributions $T_n(\mathcal{E)}$ of the transseries $\mathcal{T}^\pm_n(\mathcal{E})$ (as defined in \eqref{Tnx}), we can determine their Borel transforms $\mathcal{B}[T_n](\xi)$, and the corresponding singularity structure is shown in figure \ref{fig:T_30BorelPlane}. For $n=1$, the Pad\'e approximant of $\mathcal{B}[T_1](\xi)$ has dense lines of poles suggesting the expected pattern of branch cuts. However, for $n=10$, we already find that the singularities in the Borel plane are less clear, as can be seen by the ring of poles in each quadrant where one would expect branch points. Upon resummation this translates to a decrease in accuracy in the coefficients, and the problem worsens as $n$ increases. We expect that this would be counteracted by simply computing further terms in the large-$\mathcal{E}$ expansions of the transseries $\mathcal{T}^\pm_n(\mathcal{E})$.

More specifically, consider $\mathcal{T}^-_{30}(\mathcal{E})$ for $N=1$. The resummed value of the perturbative part $-\rmi\,\mathcal{S}_{-}\left[T_{30}\right](\mathcal{E})$ should be corrected by the corresponding non-perturbative contribution $\rmi\,\mathcal{S}_{-}\left[\eta\, T^{(1)}_{30}\right](\mathcal{E})$\footnote{Note that we are considering $\mathcal{T}^-$ so the first non-perturbative correction is $-\eta\, T^{(1)}$.} where the first non-perturbative sector $T^{(1)}_{30}$ was defined in \eqref{eq:Tnpcorrection} and $\eta$ is the exponential term \eqref{eq:eta}. However, $-\rmi\,\mathcal{S}_{-}\left[T_{30}\right](\mathcal{E})=4.1\times10^{5} - 8.6 \times 10^{-5} \rmi$ is calculated inaccurately and we find a non-perturbative ambiguity (the imaginary part of this result) which one expects to be cancelled by the first non-perturbative contribution $\rmi\,\mathcal{S}_{-}\left[\eta\, T^{(1)}_{30}\right](\mathcal{E})=-2.2\times10^{-13}- 2.4 \times 10^{-1} \rmi$. This is clearly not the case. 
However, as we increase $N$, $\mathcal{E}$ increases and the resummation becomes more accurate. 
For example, when $N =5$, $-\rmi\,\mathcal{S}_{-}\left[T_{30}\right](\mathcal{E})=5.3\times10^{-15} + 1.6\times10^{-26}\,\rmi$ whose non-perturbative ambiguity (imaginary part) becomes much closer to the imaginary part of the first non-perturbative contribution $\rmi\,\mathcal{S}_{-}\left[\eta\, T^{(1)}_{30}\right](\mathcal{E})= -1.1\times10^{-52}-1.7 \times 10^{-26} \rmi$. 
Thus, to calculate these ambiguities to greater accuracy we would have to compute additional terms in the large-$\mathcal{E}$ expansions of $T_n(\mathcal{E)}$, i.e. additional terms of the WKB periods $\alpha$ and $\beta$, equations \eqref{largeqexpansionofalpha} and \eqref{PNP}. This shows that our computation behaves consistently, with accuracy limited by the number of terms included in the large-$\mathcal{E}$ expansions of  $\mathcal{T}^\pm_n(\mathcal{E})$ and that, for larger $\ell$, $\omega_\ell(N,\infty)$ will be accurate up to the order of the first non-perturbative contributions of $\mathcal{T}^\pm_n(\mathcal{E})$ for larger $n$.

\subsection{Connecting to small \texorpdfstring{$q$}{}}

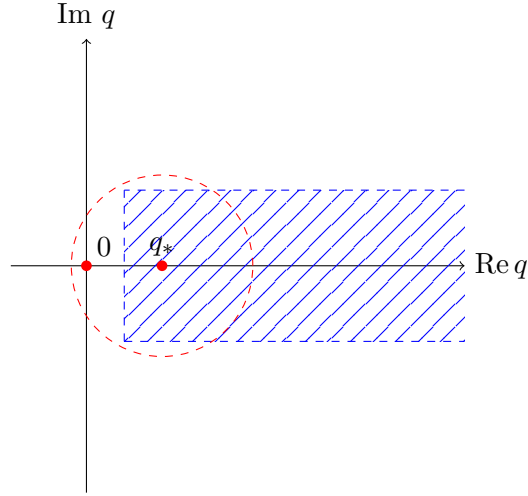
\begin{figure}[h]
\centering
    \begin{tikzpicture}
 
  \draw[->] (-3,0) -- (3,0) node[right] {$\mathrm{Re}\, q$};
\draw[dashed, blue] (-1.5,-1) -- (-1.5,1)  ;
\draw[dashed, blue] (-1.5,-1) -- (3,-1)  ;
\draw[dashed, blue] (3,1) -- (-1.5,1)  ;
 \fill[
    pattern={Lines[angle=45, distance=6pt]},
    pattern color=blue
  ] (-1.5,-1) rectangle (3,1);

  \draw[->] (-2,-3) -- (-2,3) node[above] {Im $q$};
  \fill[red] (-1,0) circle (2pt) node[above, black] {$q_*$};
  \draw[dashed,red] (-1,0) circle (1.2); 
   \fill[red] (-2,0) circle (2pt) node[above right, black] {$0$};
\end{tikzpicture}
    \caption{How to calculate the value of $\omega(0)$. The hatched blue represents the region in which the resummation of the large-$q$ series is valid. Using this, the Taylor series around a point $q_*$ can be calculated and then analytically continued to $q=0$.}
    \label{fig:placeholder}
\end{figure}
\begin{figure}
        \centering
    \includegraphics[width=0.75\linewidth]{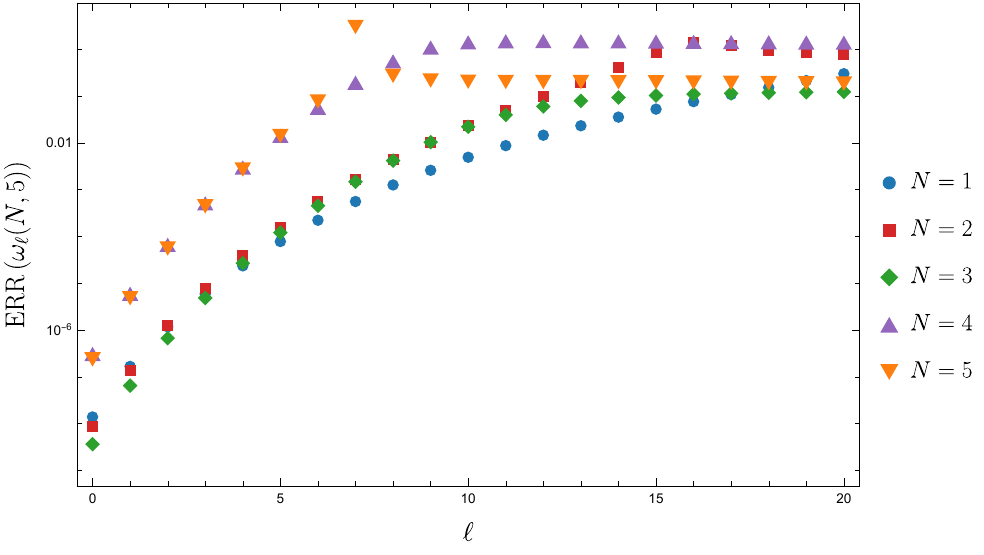}
   
         \caption{Log plots of the error estimates ERR($\omega_\ell(N, q_* =5)$) for $\ell=0, \dots,20$ and $N=1,\dots,  5$ obtained from considering the relative difference between resumming $\omega$ for $98$ and $100$ terms.}
          \label{fig:logerrorplotsderivs}
 \end{figure}

To demonstrate the effectiveness of the large-$q$ expansion \eqref{large q expansion of omega}, we now discuss how to obtain $\omega(N,0)$ solely from such an expansion. 
The resummed large-$q$ expansion allows us to evaluate $\omega(N,q)$ at some finite positive value, say $q_*$. Assuming that the resummation is performed within a fixed sector of analyticity, derivatives with respect to $q$ can be obtained by differentiating the large-$q$ expansion term by term and resumming the resulting series at $q_*$.
In this way, one can generally compute $M$ derivatives of $\omega(N,q)$ evaluated at $q_*$ from the large-$q$ expansion (see appendix \ref{app:Airy} for a worked out example). With these derivatives, one can obtain an approximation to the Taylor series of $\omega(N,q)$ around $q_*$:
\begin{equation} \label{omega series around q*}
\omega(N,q) = \sum_{\ell=0}^{+\infty} \omega^{}_{\ell}(N,q_*)\frac{(q-q_*)^{\ell}}{\ell !}.
\end{equation}
The point $q=0$ may or may not lie within the radius of convergence of this Taylor series. Supposing that we included $M$ terms in the Taylor series \eqref{omega series around q*}, we can approximate the analytic continuation of \eqref{omega series around q*} by its $\lfloor M/2 \rfloor$ diagonal Pad\'e approximant and we can evaluate the resulting rational function at $q=0$. However, there is a limitation to this method: the values of 
 $\omega_{\ell}(N, q_*)$ obtained from resummation become less accurate for higher $\ell$. To estimate the accuracy of $\omega_{\ell}(N, q_*)$, we consider the relative difference of $\omega_{\ell}(N, q_*)$ obtained by resumming 98 and 100 terms in the large-$q$ expansion.  We call this quantity $\text{ERR}(\omega_\ell(N, q_*))$. The results for $\ell =0, \dots, 20$ and $q_*=5$ are shown for $N=1,\dots,5$ in figure \ref{fig:logerrorplotsderivs}.
We then include $M$ terms in the sum \eqref{omega series around q*}, where $M$ is the first natural number such that the error of $\text{ERR}(\omega_{M+1}(N, q_*))$ is greater than some tolerance.
The results of this procedure for $q_* =5$, $N=1, \dots, 5$ with a tolerance of $10^{-2}$ are shown in table \ref{tab:omega at 0}. These results demonstrate a proof of concept rather than a systematic evaluation of $\omega(0)$. Indeed, this blanket approach is naive. As mentioned above, our large-$q$ expansions for $\omega$ are effectively expansions in small $(N/q)^{4/3}$, which means that we should fix $N/q$ and select different values of $q_*$ according to the values of $N$. By comparing the values of $\omega_\ell(N,q_*)$ for a range of $q_*$ more accurate results can be obtained.
\begin{table}[h!]
    \centering
    \begin{tabular}{|c|c|c|}
    \hline
         $N$ & large-$q$ $\omega(0)$ & $\omega_\text{num}$(0) \\ \hline
        1 & $3.00022-2.59152 \rmi$ & $3.11945-2.74668 \rmi$ \\ \hline
        2 & $5.07666-4.72323 \rmi$ & $5.16952-4.76357 \rmi$ \\ \hline
        3 & $7.15466-6.84669 \rmi$ & $7.18793-6.76956 \rmi$ \\ \hline
        4 & $9.11644-8.58581 \rmi$ & $9.19720-8.77248 \rmi$ \\ \hline
        5 & $10.7519-10.7495 \rmi$ & $11.2027-10.7742 \rmi$ \\ \hline
    \end{tabular}
    \caption{Comparison of $\omega(0)$ computed from the large-$q$ expansion via the procedure discussed in the text against the numerically obtained values $\omega_{\text{num}}(0)$. Despite the naive approach taken, the results show qualitative agreement.}
    \label{tab:omega at 0}
\end{table}

\subsection{Large-\texorpdfstring{$q$}{} as a seed}

As shown in the previous subsection, the large-$q$ expansion produces good results even for small values of $q$. However, the analytic structure of $\omega$ is most effectively probed by combining the SW approach with the large-$q$ WKB expansion.
Recall that the SW approach provides $\omega(N,q)$ as a solution to a polynomial equation whose degree was determined by the order of truncation of the small-$t$ expansion. In section \ref{sec:gauge}, this approach was used to calculate $\omega(N,q)$ at fixed values of $q$. This approach can also be used to construct the local expansion of $\omega(N,q)$ around a point $q_*$:\footnote{This expansion is understood away from branch points and for a simple selected root of the truncated spectral equation. Indeed, $\omega(q)$ contains cuts, see \cite{Grozdanov:2019uhi}, and the ansatz \eqref{omega series around q*} is not valid when $q_*$ is a branch point.} as in \eqref{omega series around q*}.
To determine the coefficients $\omega_{\ell}(N,q_*)$, we substitute \eqref{omega series around q*} into the polynomial obtained from the SW approach and expand in powers of $q-q_*$.
The leading coefficient $\omega_{0}(N,q_*)$ satisfies the original polynomial equation at $q=q_*$. Once this root has been chosen, the higher coefficients are determined recursively: the first equation in which $\omega_{\ell}(N,q_*)$ appears is linear in $\omega_{\ell}(N,q_*)$. Thus, no further polynomial equations have to be solved for $n\geq 1$.
The only remaining ambiguity is the choice of the correct root for $\omega_{0}(N,q_*)$ from the polynomial equation. 
The resummed large-$q$ WKB expansion removes this ambiguity. It provides an independent analytic seed for the physical root in the domain where the resummed large-$q$ expansion is accurate:
\begin{equation}\label{resummedlargeqomega}
\omega_{0}(N,q_*) = \mathcal{S}_{0^-}\left[q\left(1 +  \sum_{\ell=1}^\infty \omega_\ell(N,\infty)q^{-\frac{4}{3}\ell}\right)\right](q_*).
\end{equation}
Once this seed is fixed, the SW approach determines the Taylor coefficients $\omega_{\ell}(N,q_*)$ and hence the local analytic continuation of the physical branch in the vicinity of $q_*$.

\begin{figure}[h!]
    \centering
    \begin{subfigure}{0.48\textwidth}
        \centering
        \includegraphics[width=\linewidth]{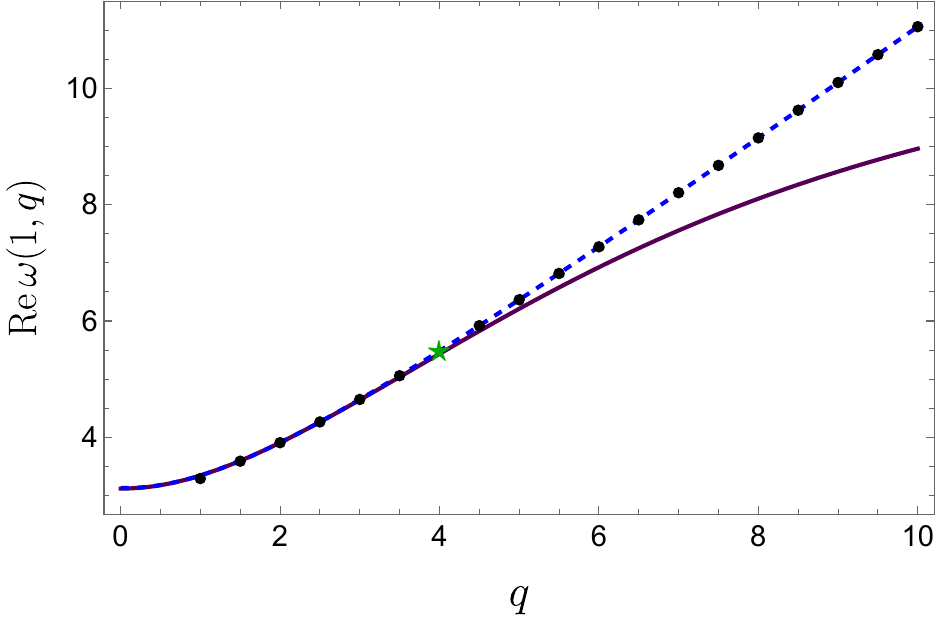}
        \label{fig:re_omega}
    \end{subfigure}
    \hfill
    \begin{subfigure}{0.48\textwidth}
        \centering
        \includegraphics[width=\linewidth]{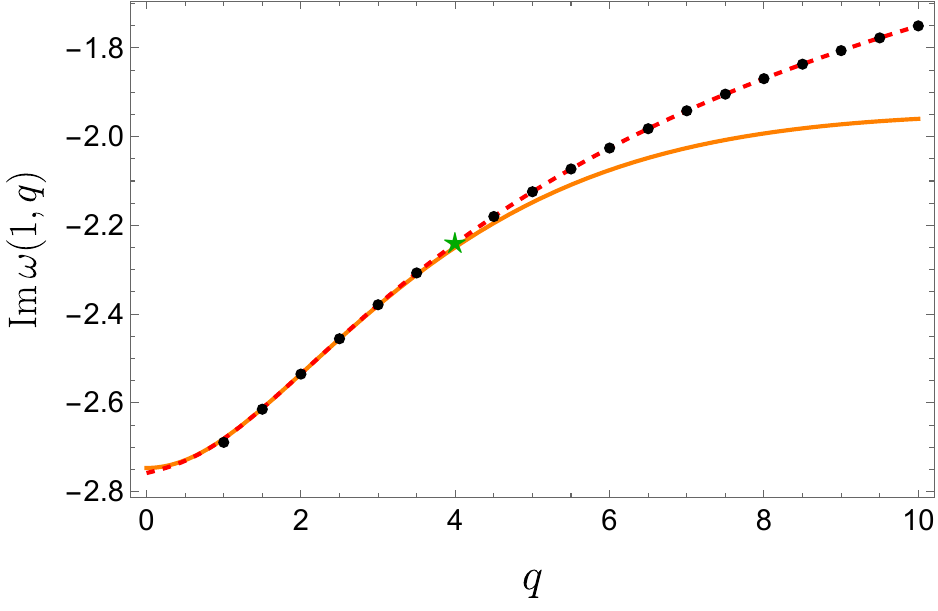}
        \label{fig:im_omega}
    \end{subfigure}
    
    \caption{Comparison of the Pad\'e-resummed QNM $\omega(1,q)$ and results of the resummed large-$q$ expansion \eqref{resummedlargeqomega}. In both panels, black dots denote the results \eqref{resummedlargeqomega} at different values of $q_*$, and the dot corresponding to $q_*=4$ is replaced by a green star.  \textbf{Left panel}: comparison between the real part of \eqref{omega series around q*} with $q_*=0$ (purple) and \eqref{omega series around q*} with $q_*=4$ (dashed blue). \textbf{Right panel}: comparison between the imaginary part of \eqref{omega series around q*} with $q_*=0$ (orange) and \eqref{omega series around q*} with $q_*=4$ (dashed red). }
    \label{fig:omega_pade}
\end{figure}

We show in figure~\ref{fig:omega_pade} the effectiveness of the procedure. In this example, the dispersion relation of the fundamental overtone $\omega(1,q)$ is first expanded around $q_*=4$ using the SW truncation, with the physical seed fixed by the resummed WKB result, and is then analytically continued towards $q=0$ by means of a Pad\'e approximant. We compare this curve with the corresponding expansion constructed around $q_*=0$, again using the SW truncation and a WKB seed. In both cases we compute the Taylor expansion \eqref{omega series around q*} up to order $10$. Finally we included a direct evaluation of $\omega_0(1,q)$ obtained from the resummation of the series \eqref{large q expansion of omega} for various values of $q\in[1/2,10]$.

The fact that the analytic continuation with Padé approximant towards $q=0$ of the expansion around $q_*=4$ provided by the SW approach agrees with the analytic continuation (again with Padé approximant) of the expansion around $q_*=0$ is non-trivial. 
Indeed, the Padé approximant provides only one possible analytic continuation of the local Taylor series, and this Taylor series itself is obtained from a finite-order truncation of the SW continued fraction. Thus, there are two distinct approximations involved: the truncation of the complex curve, which is intrinsic to the SW implementation, and the subsequent Padé approximation used to continue the local series beyond its immediate domain of convergence. The latter step need not preserve the accuracy of the former.

In particular, if the analytic structure of $\omega(1,q)$ in the relevant region of the complex $q$-plane were more complicated, the two Padé continuations could in principle disagree in an intermediate region. Such a mismatch could signal that the rational approximation is not resolving the correct analytic continuation, or that different branches of the spectrum are being mixed, for instance near a branch point or a crossing with another overtone.
In figure \ref{fig:omega_pade} we see that the two plots start showing a slight disagreement towards $q=0,4$ --  this is natural as they are analytic continuations of Taylor series containing (only!) 10 terms, centred at either $q_*=0$ or $q_*=4$. The value of the QMNs determined via the truncated curve will naturally interpolate between the two lines. 

Such a disagreement would not invalidate the SW approach itself. Rather, it would indicate that the Padé continuation is not the optimal way of following the physical branch in that region. In such a situation one can instead use a stepwise Taylor-series method.
Starting from some point, say $q_*=4$, where the WKB seed unambiguously selects the desired root, one first constructs a local Taylor expansion from the SW truncation,
\begin{equation}
\omega(1,q)=\sum_{\ell=0}^{\ell_{\text{truncation}}}\omega_\ell(1,4)\,\frac{(q-4)^\ell}{\ell!}.
\end{equation}
Then, one moves to a nearby point within its radius of convergence, say $q_*^{(1)}=q_*-\epsilon=4-\epsilon$, for some small $\epsilon>0$, and repeats the procedure starting from the seed
\begin{equation}
\omega_0(1,4-\epsilon)=\sum_{\ell=0}^{\ell_{\text{truncation}}}\omega_\ell(1,4)\,\frac{(-\epsilon)^\ell}{\ell!}.
\end{equation}
The procedure is repeated $k$ times, where $k$ is such that the distance between $q=0$ and $q_*^{(k)}$ is less than $\epsilon$.
This provides a more controlled way of tracking the same branch through the complex $q$-plane (in the presented discussion, we are moving along the real line, but the path can be more general). This procedure is analogous to the one we present in appendix \ref{app:Airy}, where we show the procedure for the simpler example of the Airy function.

Finally, figure~\ref{fig:omega_pade} provides a direct test of the resummed large-$q$ expansion against the SW--Pad\'e construction. The dashed curves are obtained from the order-$10$ Taylor expansion around $q_*=4$ and use the resummed value only at $q_*=4$ (the green star) to select the physical root. The resummed values at all other wave numbers are not used in constructing these curves. Their close agreement with the dashed curves for $q\gtrsim1$, in both the real and imaginary parts of $\omega(1,q)$, therefore provides a nontrivial consistency check that the two procedures track the same QNM branch. Although the direct large-$q$ resummation begins to lose accuracy around $q=1/2$, its results remain remarkably close to the SW--Pad\'e curves. The main conclusion is therefore clear: after resummation, the large-$q$ expansion remains effective down to small wave numbers.

\section{Discussion and outlook}

We have developed a unified analytic framework for the quasinormal spectrum of the helicity-$\pm2$ perturbations of the planar Schwarzschild-AdS$_5$ black brane. The starting point was the observation that the scalar-channel boundary value problem, with ingoing behaviour at the horizon and normalisability at the AdS boundary, can be written as a Heun problem with four regular singular points, see \eqref{heunnormal} and the dictionary \eqref{dictioHeun}. This allowed us to reinterpret the standard truncation of the spectral problem in a more structural way. In particular, the Heun connection formula expresses the relevant connection coefficients in terms of the Nekrasov--Shatashvili free energy, and the QNM condition becomes the quantisation of the corresponding composite monodromy parameter, see \eqref{quantcondHeun2}. In this language, the truncation procedure is the truncation of the continued-fraction representation generated by the three-term recurrence relation of the Heun Floquet solution. The physical black-brane problem is recovered at the finite value $t=1/2$, and the Seiberg--Witten formulation makes it possible to analyse systematically when the small-$t$ expansion can be trusted. In particular, the Padé analysis of the instanton series shows that, as either the wave number $q$ or the overtone number $N$ is increased, the relevant singularities in the $t$-plane approach the circle $|t|=1/2$. This explains why the method remains effective at moderate values of $q$ and $N$, but becomes increasingly inefficient in the large wave number or large-overtone regime.

This limitation motivates the second, complementary description developed in the paper. In the large-$q$ regime the same differential equation becomes a singularly perturbed problem, and the natural expansion parameter is $q^{-1}$. We therefore introduced the Liouville--Green ansatz in section \ref{sec:WKB method} and analysed the associated exact-WKB problem. The analytic continuation between the boundary and the horizon is controlled by the Stokes graph of the WKB curve and by the two periods $\alpha$ and $\beta$ defined in \eqref{ABperiods}. Imposing the QNM boundary conditions gives the exact quantisation conditions \eqref{quantcondWKBscalar}.
These conditions are exact in the sense of exact WKB: they are formulated in terms of the full quantum periods and include, through the dependence on $\beta$, the exponentially small contributions, which are invisible in a purely perturbative expansion in $q^{-1}$. This is also the point at which resurgence becomes necessary. Indeed, the perturbative large-$q$ expansions of the periods are not convergent series but asymptotic expansions with factorially growing coefficients. Their resummation is therefore part of the definition of the WKB answer. In section \ref{sec:resurgenceofQNMs} we showed how Borel--Laplace resummation gives meaning to these divergent series, and how the non-perturbative terms controlled by the period $\beta$ cancel the corresponding ambiguities. In this way, the exact-WKB quantisation condition provides not only the perturbative large-$q$ expansion, but also its non-perturbative completion.

Remarkably, the resulting WKB description is effective well beyond the strict large-$q$ regime. In particular, as presented in section \ref{sec:results}, after resummation it gives accurate values of the quasinormal frequencies even in the vicinity of $q=0$. The analytic continuation towards $q=0$ can also be performed by combining the large-$q$ WKB seed with the Seiberg--Witten approach at finite $q$. Crucially, in this context the WKB framework provides a natural starting point for continuing the physical branch towards smaller wave numbers and removes the ambiguity associated with finite-order truncations of the continued fraction. Indeed, at a given truncation order, the truncated spectral condition has as many candidate roots as the degree of the corresponding polynomial, and the truncation procedure alone does not intrinsically select the physical branch. The large-$q$ WKB solution instead provides a well-defined seed, fixed by the Stokes geometry and by the exact quantisation condition. Thus, the two approaches are complementary: the Seiberg--Witten formulation explains and controls the finite-$q$ truncation, while exact WKB identifies, resums and non-perturbatively completes the physical large-$q$ branch.

Let us stress that the resurgent analysis carried out in this work was performed for real values of $q$. If the wave number is analytically continued to complex values, the Stokes geometry in the $q$-plane is expected to change, and different non-perturbative sectors may become relevant as the argument of $q$ is varied. We leave a systematic analysis of this complex-$q$ Stokes phenomenon for future work.

A natural extension of the analysis presented in this work is to apply the same framework to the other channels of gravitational perturbations, in particular to the shear and sound sectors. The shear channel is the closest analogue of the problem considered here, and for this reason it appears to be the most immediate extension for the present methods. In addition to the gapped modes analogous to the helicity $\pm 2$ sector, it contains a hydrodynamic quasinormal mode, and therefore provides a distinguished setting in which quasinormal mode quantisation and the analytic structure of the dispersion relation can be studied simultaneously. 

The dispersion relation of the shear mode has been studied extensively \cite{Policastro:2002se, Kovtun:2005ev, Withers:2018srf, Grozdanov:2019kge, Grozdanov:2019uhi, Heller:2020hnq, Arnaudo:2026axr}. In the long-wavelength, low-frequency regime, it is obtained by imposing on the same wave solution both normalisability at the AdS boundary and ingoing behaviour at the horizon. The resulting spectral condition can be encoded in different but equivalent ways, for instance through the zeros of a truncated spectral curve \cite{Grozdanov:2019uhi}, or through perturbative bulk solutions expressed in terms of multiple polylogarithms \cite{Arnaudo:2026axr}. 

The fact that both boundary conditions can be imposed on the same wave solution is a consequence of the analysis performed in this work, where the relevant Heun connection problem is formulated so as to relate the local basis adapted to the AdS boundary to the one adapted to the horizon within a single patch of the 4-punctured sphere. The sound channel is structurally more involved, but it would be especially interesting to understand how far the same framework can be pushed there. In particular, the sound mode has been studied analytically in four dimensions in terms of perturbative bulk solutions involving multiple polylogarithms \cite{Aminov:2023jve}. It would be very interesting to clarify whether this functional structure admits a direct interpretation from the exact-WKB point of view, namely whether the iterated-integral structure underlying WKB periods, connection formulae and their resurgent completion can, after matching to the hydrodynamic regime, be reorganised in terms of multiple polylogarithms.

Recent developments make the large-frequency regime especially interesting from the AdS/CFT viewpoint. Starting from the early observations of \cite{Fidkowski:2003nf, Festuccia:2005pi}, analytically continued thermal correlators can carry subtle signatures of the black-hole singularity, encoded in geodesics that probe the region behind the horizon and, in suitable limits, in large-frequency asymptotics of the correlator itself. More recently, this picture has been reformulated directly in CFT language: in particular, the stress-tensor sector of the thermal OPE develops singularities precisely at the locations associated with bulk bouncing geodesics \cite{Ceplak:2024bja, Buric:2025fye, Barrat:2025twb, Ceplak:2025dds}. In parallel, complementary high-frequency, spectral and quasinormal-mode analyses have sharpened the same connection: non-perturbative large-frequency corrections to thermal correlators were shown to be governed by null geodesics bouncing off the singularity, and the associated sectors were made explicit through exact-WKB/transseries analyses of thermal spectral functions and Fourier coefficients \cite{Afkhami-Jeddi:2025wra, Dodelson:2025jff, Jia:2025jbi, Arnaudo:2026der, Jia:2026ryl}. 
Our large-frequency WKB analysis of the scalar/helicity-$\pm 2$ master field around planar Schwarzschild-AdS$_5$ fits naturally into this framework: it isolates directly, on the gravity side, the same high-frequency data that recent holographic/OPE studies identify as the boundary imprint of the black-hole interior.

\acknowledgments 

It is a pleasure to thank Giulio Bonelli, Fabrizio Del Monte, Alba Grassi, Chris Howls, Pavel Kovtun, David Sola Gil, Alessandro Tanzini, and Benjamin Withers for useful discussions. IA and MS would also like to thank the Isaac Newton Institute for hosting them during the "Applicable Resurgence Asymptotics" Programme, during the early stages of this work. 
IA was partially supported by an UKRI EPSRC Early Career Fellowship EP/S004076/1. PA is supported by the Royal Society grant URF\textbackslash R\textbackslash 231002, `Dynamics of holographic field theories'. AR is supported by UKRI STFC consolidated grant ST/X000583/1, “New Frontiers in Particle Physics, Cosmology and Gravity”. MS is supported by the National Science Centre, Poland, under Grants No. 2021/41/B/ST2/02909 and 2025/59/B/ST2/02995.


\appendix

\section{Relations between Seiberg-Witten and Heun}\label{app:SW}

In this appendix, we describe the relations between the Heun equation and the quantum Seiberg-Witten curve of the four dimensional $\mathcal{N}=2$ $SU(2)$ gauge theory with $N_f=4$.
We follow the notations in \cite{Zenkevich:2011zx, Aminov:2020yma}, and in particular within this appendix we use the parameter $\hbar$ as the usual parameter describing the quantum curves. This is related to the equivariant parameters $\epsilon_1,\epsilon_2$ so that $\hbar\to 0$ corresponds to $\epsilon_1\to 0$, but it has nothing to do with the parameter $\hbar$ introduced in \eqref{hbar}. We write the quantum curve as
\begin{equation}
\begin{aligned}
&-\left(1+t\right) \hbar^2 \partial^2_x \Psi(x)  + \sqrt{t}\,\rme^{x}\left(\hbar\,\partial_x + M_1+\frac{\hbar}2\right)\left(\hbar\,\partial_x + M_2+\frac{\hbar}2\right)\Psi(x) \\
&+\sqrt{t} \, \rme^{-x} \left(\hbar\,\partial_x + M_3-\frac{\hbar}2\right) \left(\hbar\,\partial_x + M_4-\frac{\hbar}2\right)\Psi(x)= \left(1-t\right)\lambda^2 \Psi(x),
\end{aligned}
\end{equation}
where $t$ is the instanton counting parameter, and the parameters $M_i$ are related to the masses $m_i$ of the hypermultiplets by
\begin{equation}\label{MSub}
M_i=m_i-\frac{t \sum_{j=1}^4m_j}{2 (t+1)},\quad i=1,2,3,4.
\end{equation}
If we redefine the function
\begin{equation}
\begin{aligned}
\Psi(x)=&\,\rme^{\frac{1}{2} x \left(1-\frac{t (M_1+M_2+M_3+M_4)}{(t-1) \hbar }\right)} \left(1-\sqrt{t} \rme^{-x}\right)^{\frac{1}{2} \left(\frac{t (M_1+M_2)+M_3+M_4}{\hbar -t \hbar }-1\right)}\times\\
&\left(1-\sqrt{t} \rme^x\right)^{\frac{M_1+M_2+t (M_3+M_4-\hbar )+\hbar }{2 (t-1) \hbar }}\,\psi(x),
\end{aligned}
\end{equation}
and the variable as
\begin{equation}
y=\sqrt{t}\,\rme^{-x},
\end{equation}
the differential equation satisfied by $\psi(y)$ is given by
\begin{equation}\label{SWquantumODE}
\begin{aligned}
&\hbar^2\,\psi''(y)
-\frac{1}{4 y^2 \hbar ^2 \left(t-(t+1) y+y^2\right)^2}\Bigl[y^3 \left((4 M_3 M_4+\hbar^2) (t+1)-4 \lambda ^2 (t-1)\right) \\
&+t^2 \left((M_1-M_2)^2-\hbar ^2\right) +t y \left(4 M_1 M_2 (t+1)-4 \lambda ^2 (t-1)+(t+1) \hbar ^2\right)\\
&-y^2 \left(4 \lambda ^2+2 M_1 t (2 M_2-M_3-M_4)
-2 M_2 t (M_3+M_4)+4 M_3 M_4 t-4 \lambda ^2 t^2+t^2 \hbar ^2+\hbar ^2\right)\\
&+y^4 \left(M_3^2-2 M_3 M_4+M_4^2-\hbar ^2\right)\Bigr]\,\psi(y)=0,
\end{aligned}
\end{equation}
and it can be written in the Heun's form \eqref{heunnormal} with
\begin{equation}
\begin{aligned}
a_0^2&=\frac{(M_1-M_2)^2}{4 \hbar ^2},\\
a_1^2&=\frac{(t (M_1+M_2)+M_3+M_4)^2}{4 (t-1)^2 \hbar ^2},\\
a_t^2&=\frac{(M_1+M_2+t (M_3+M_4))^2}{4 (t-1)^2 \hbar ^2},\\
a_{\infty}^2&=\frac{(M_3-M_4)^2}{4 \hbar ^2},\\
c&=\frac{1}{4 (t-1)^3 \hbar ^2}\Bigl[M_1^2 (4 t-2)+4 M_1 t^2 (M_2+M_3+M_4)+M_2^2 (4 t-2)+4 M_2 t^2 (M_3+M_4)\\
&\ \ \ +2 t \left(2 M_3 M_4 t^2+t (M_3-M_4)^2+2 M_3 M_4\right)+4 \lambda ^2 (t-1)^3-(t-1)^3 \hbar ^2\Bigr].
\end{aligned}
\end{equation}
By substituting \eqref{MSub} and $\hbar=1$, we get
\begin{equation}\label{fundmasses}
\begin{aligned}
a_0^2&=\frac{(m_1-m_2)^2}{4},\\
a_1^2&=\frac{(m_3+m_4)^2}{4},\\
a_t^2&=\frac{(m_1+m_2)^2}{4},\\
a_{\infty}^2&=\frac{(m_3-m_4)^2}{4},\\
c&=\frac{1}{4\left(t^2-1\right)}\Bigl[-4 \lambda ^2+m_1^2 \left(t^2-2\right)+2 m_1 t (m_2 (t+2)+m_3+m_4)+m_2^2 \left(t^2-2\right)\\
&\ \ \ +2 m_2 t (m_3+m_4)-m_3^2 t^2+2 m_3 m_4 t^2+4 m_3 m_4 t-m_4^2 t^2+4 \lambda ^2 t^2-t^2+1\Bigr],
\end{aligned}
\end{equation}
that provide the standard relations between the hypermultiplet masses and the monodromy parameters, and where $\lambda^2$ is related to $a^2$.

The above relations are obtained by fixing $\hbar=1$, therefore they provide results which are exact in $\hbar$ and perturbative in $t$. The WKB approach, instead, relies on a small-$\hbar$ expansion, and the relation to the quantization of the A-period of the 2-torus covering the 4-punctured sphere can be understood in the classical limit $\hbar\to 0$.

In the limit $\hbar\to 0$, the differential equation \eqref{SWquantumODE} reduces to a 4th degree polynomial in $y$
\begin{equation}
\begin{aligned}
&(M_3-M_4)^2\,y^4+\left[4 M_3 M_4 (t+1)-4 \lambda ^2 (t-1)\right]y^3 \\
&+4 t \left[M_1 M_2 (t+1)-\lambda ^2 (t-1)\right]y
+t^2 (M_1-M_2)^2 \\
&+\left[2 M_1 t (-2 M_2+M_3+M_4)+2 M_2 t (M_3+M_4)-4 M_3 M_4 t+4 \lambda ^2 \left(t^2-1\right)\right]y^2=0,
\end{aligned}
\end{equation}
which is indeed the classical SW curve, having turning points in its zeros. 

For a description of the mathematical relations between the two regimes, we refer to \cite{Gaiotto:2012rg, Kidwai:2017ygi}. Other relevant works for the $N_f=0$ case are \cite{Mironov:2009uv, Kashani-Poor:2015pca}.

\section{Recurrence relations for Heun and its connection coefficients}

This appendix review two related approaches to solve the connection problem for the Heun equation we are concerned about in this manuscript.

\subsection{Three-term recurrence relation}\label{app:3term}

We describe the procedure introduced in appendix A in \cite{Lisovyy:2021bkm} to obtain the instanton series expansion for the accessory parameter $c$. 

The construction relies on the three-term recurrence relation satisfied by the Heun functions. To conveniently write this, we start by redefining the Heun equation by introducing the wave function
\begin{equation}
w(y)=y^{-\frac{1}{2}+a_0}(y-1)^{-\frac{1}{2}+a_1}(y-t)^{-\frac{1}{2}+a_t}\psi(y),
\end{equation}
so that the Heun equation becomes
\begin{equation}
w''(y)+\left(\frac{\gamma}{y}+\frac{\delta}{y-1}+\frac{\epsilon}{y-t}\right)w'(y)+\frac{\alpha\,\beta\,z-q_{\text{Heun}}}{y\,(y-1)\,(y-t)}\,w(y)=0,
\end{equation}
where
\begin{equation}
\begin{aligned}
\alpha&=1-a_0-a_1-a_t+a_{\infty},\\
\beta&=1-a_0-a_1-a_t-a_{\infty},\\
\gamma&=1-2a_0,\\
\delta&=1-2a_1,\\
\epsilon&=1-2a_t,\\
q_{\text{Heun}}&=\frac{1}{2}+t\left(a_0^2+a_1^2+a_t^2-a_{\infty}^2\right)-a_t-a_1\,t+a_0\left[2a_t-1+t\left(2a_1-1\right)\right]+\left(1-t\right)c.
\end{aligned}
\end{equation}
In the space of solutions diagonalizing the composite monodromy around $y=0$ and $y=t$, we consider a solution in the annulus $|t|<|y|<1$ of the form:
\begin{equation}\label{floquet}
w(y)=\sum_{k\in\mathbb{Z}}d_k\,y^{k+\nu},
\end{equation}
where the parameter $\nu$ is related to the previously introduced parameter $a$ by
\begin{equation}
\nu=a+a_0+a_t-\frac{1}{2}.
\end{equation}
The coefficients $d_k$ in \eqref{floquet} satisfy a three-term recurrence relation of the form
\begin{equation}
A_k\,d_{k-1}-B_k\,d_k+t\,C_{k}\,d_{k+1}=0,
\end{equation}
where
\begin{equation}
\begin{aligned}
A_k&=(k-1+\alpha+\nu)(k-1+\beta+\nu),\\
B_k&=q_{\text{Heun}}+\left(k+\nu\right)[(k-1+\gamma+\nu)(1+t)+t\,\delta+\epsilon],\\
C_k&=(k+1+\nu)(k+\gamma+\nu).
\end{aligned}
\end{equation}
In the small $t$ regime, the three-term recurrence relation gives rise to an equation involving continued fractions which determines the parameter $c$ as an expansion in $t$ at fixed $a$:
\begin{equation}
B_0=\frac{t\,C_0\,A_1}{B_1-\frac{t\,C_1\,A_2}{B_2-\dots}}+\frac{t\,A_0\,C_{-1}}{B_{-1}-\frac{t\,A_{-1}\,C_{-2}}{B_{-2}-\dots}}.
\end{equation}

For concreteness, we present the first three coefficients of the expansion
\begin{equation}
c=\sum_{i\ge 0}c_i\,t^i
\end{equation}
when setting $a_t=1$ and $a_\infty=0$ as in \eqref{dictioHeun}:
\begin{equation}
\begin{aligned}
c_0=\frac{3}{4}-a^2+a_0^2,
\end{aligned}
\end{equation}
\begin{equation}
\begin{aligned}
c_1= \frac{\left(3+4 a^2-4 a_0^2\right) \left(4 a^2+4 a_1^2-1\right)}{8-32 a^2},
\end{aligned}
\end{equation}
\begin{equation}
\begin{aligned}
c_2= &-\frac{1}{512 \left(a^2-1\right) \left(4 a^2-1\right)^3}\Bigl[13312 a^{10}-256 a^8 \left(56 a_0^2-72 a_1^2+37\right)\\
&+128 a^6 \left(8 a_0^4+a_0^2 \left(176-96 a_1^2\right)+8 a_1^4-96 a_1^2-87\right)\\
&-32 a^4 \left(24 a_0^4 \left(8 a_1^2+1\right)+24 a_0^2 \left(8 a_1^4-44 a_1^2+15\right)+216 a_1^4+504 a_1^2-307\right)\\
&+4 a^2 \left(16 a_0^4 \left(80 a_1^4+48 a_1^2+3\right)-32 a_0^2 \left(56 a_1^4+114 a_1^2-19\right)+2096 a_1^4+2304 a_1^2-623\right)\\
&+\left(4 a_1^2-1\right) \left(16 a_0^4 \left(28 a_1^2+1\right)+a_0^2 \left(184-992 a_1^2\right)+348 a_1^2-207\right)\Bigr].
\end{aligned}
\end{equation}

\subsection{Alternative approach for the tower of QNMs}\label{app:alternativeHeun}

There is an alternative (but equivalent) way of dealing with the Heun connection problem involving two nearby singular points, which is described in \cite{Lisovyy:2022flm}. The connection problem involves the local solutions centered at $y=0$ and $y=1$ with the assumption that the other singular points lie outside the unit disc centered at the origin. In this case, the connection coefficient is defined by a recurrence relation as defined in Theorem B in \cite{Lisovyy:2022flm} \footnote{Theorem B in \cite{Lisovyy:2022flm} holds in principle for the indicial parameters $a_0,a_1$ not being elements of $\mathbb{Z}/2$, in order to avoid the presence of logarithmic singularities, which would require a modification of the basis of local solutions. However, for the quantization of QNMs the coefficient which is asked to vanish remains the same (and only the coefficient in front of the selected solution at $y=1$ changes). For a discussion on this argument see \cite{Jia:2024zes}.}.

We briefly recall the content of the Theorem. Assuming $|t|>1$ and $a_0,a_1\notin\mathbb{Z}/2$, denoting with $\lambda=1/t$, and writing the connection formula between the Frobenius solutions \eqref{frobeniussol} around $y=0$ and $y=1$:
\begin{equation}
\psi_{\theta}^{(0)}(y)=\sum_{\theta'=\pm}\mathcal{C}_{\theta \theta'}\psi_{\theta'}^{(1)}(y),
\end{equation}
the connection coefficient $\mathcal{C}_{\theta \theta'}$ can be expressed as
\begin{equation}
\begin{aligned}
\mathcal{C}_{\theta \theta'}=&\,\frac{\Gamma\left(1-2\theta a_0\right)\Gamma\left(2\theta' a_1\right)}{\Gamma\left(\frac{1}{2}+\theta' a_1-\theta a_0+\sqrt{\frac{1}{4}+c+a_\infty^2-a_t^2}\right)\Gamma\left(\frac{1}{2}+\theta' a_1-\theta a_0-\sqrt{\frac{1}{4}+c+a_\infty^2-a_t^2}\right)}\\
&\times\exp\sum_{k=1}^{\infty}\log\left(1-\lambda\alpha_{k-1}-\frac{\lambda\beta_k}{1-\lambda\alpha_{k}-\frac{\lambda\beta_{k+1}}{1-\dots}}\right),
\end{aligned}
\end{equation}
where
\begin{equation}
\begin{aligned}
\alpha_k&=-\frac{\left(k+\frac{1}{2}-a_0-a_t\right)^2+\frac{1}{4}+c-a_0^2-a_t^2}{\left(k+\frac{1}{2}-a_0+a_1\right)^2-\frac{1}{4}-c-a_\infty^2+a_t^2},\\
\beta_k&=\frac{k(k-2a_0)\left[\left(k-a_0+a_1-a_t\right)^2-a_\infty^2\right]}{\left[\left(k+\frac{1}{2}-a_0+a_1\right)^2-\frac{1}{4}-c-a_\infty^2+a_t^2\right]\left[\left(k-\frac{1}{2}-a_0+a_1\right)^2-\frac{1}{4}-c-a_\infty^2+a_t^2\right]}.
\end{aligned}
\end{equation}

To use such an approach, we consider the differential equation \eqref{heunnormal}, and we redefine the variable as $Y=2y$, so that
\begin{equation}
y=\{0,t\equiv\frac{1}{2},1,\infty\}\mapsto Y=\{0,1,2,\infty\}.
\end{equation}

\begin{figure}[h!]

\begin{subfigure}{0.45\textwidth}
        \centering
        \includegraphics[width=6.5cm]{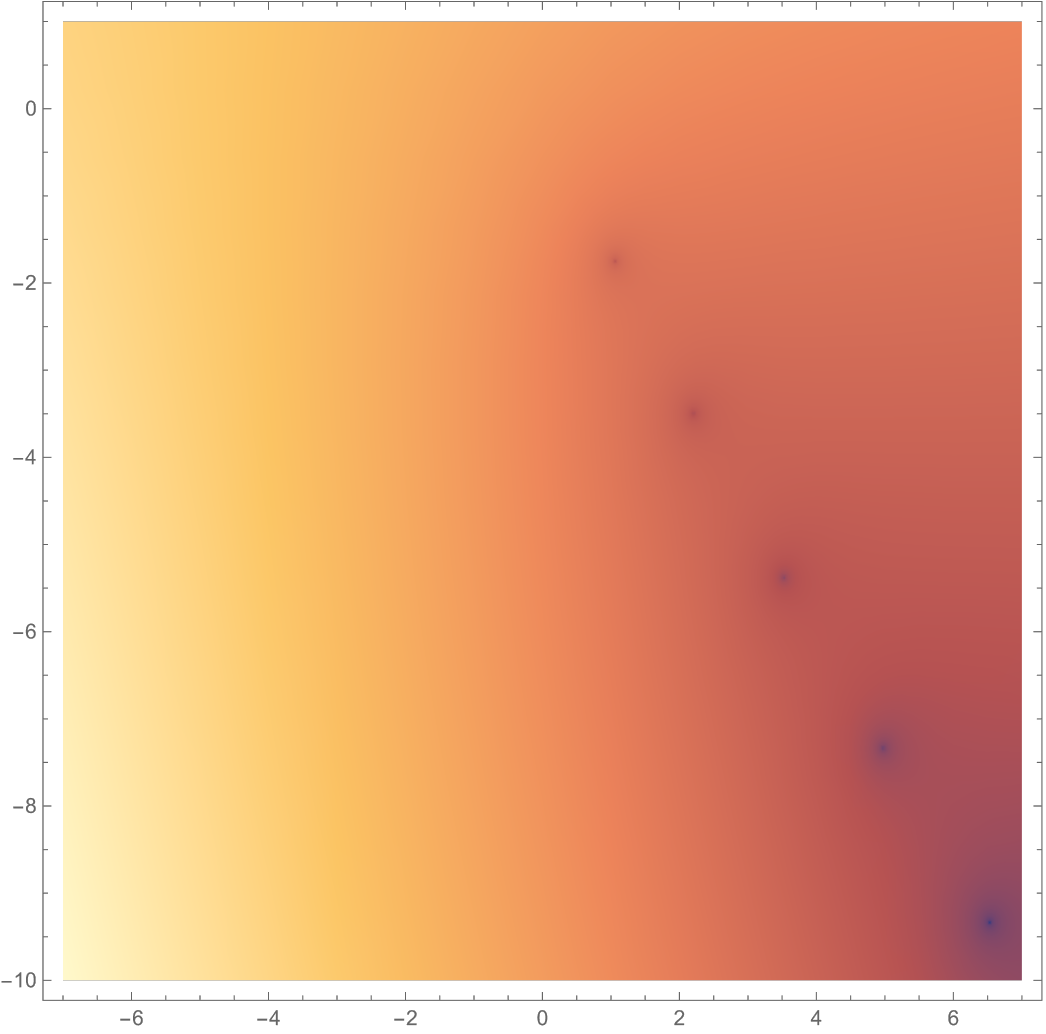}

    \end{subfigure}
    \hfill
        \begin{subfigure}{0.45\textwidth}
        \centering
        \includegraphics[width=6.5cm]{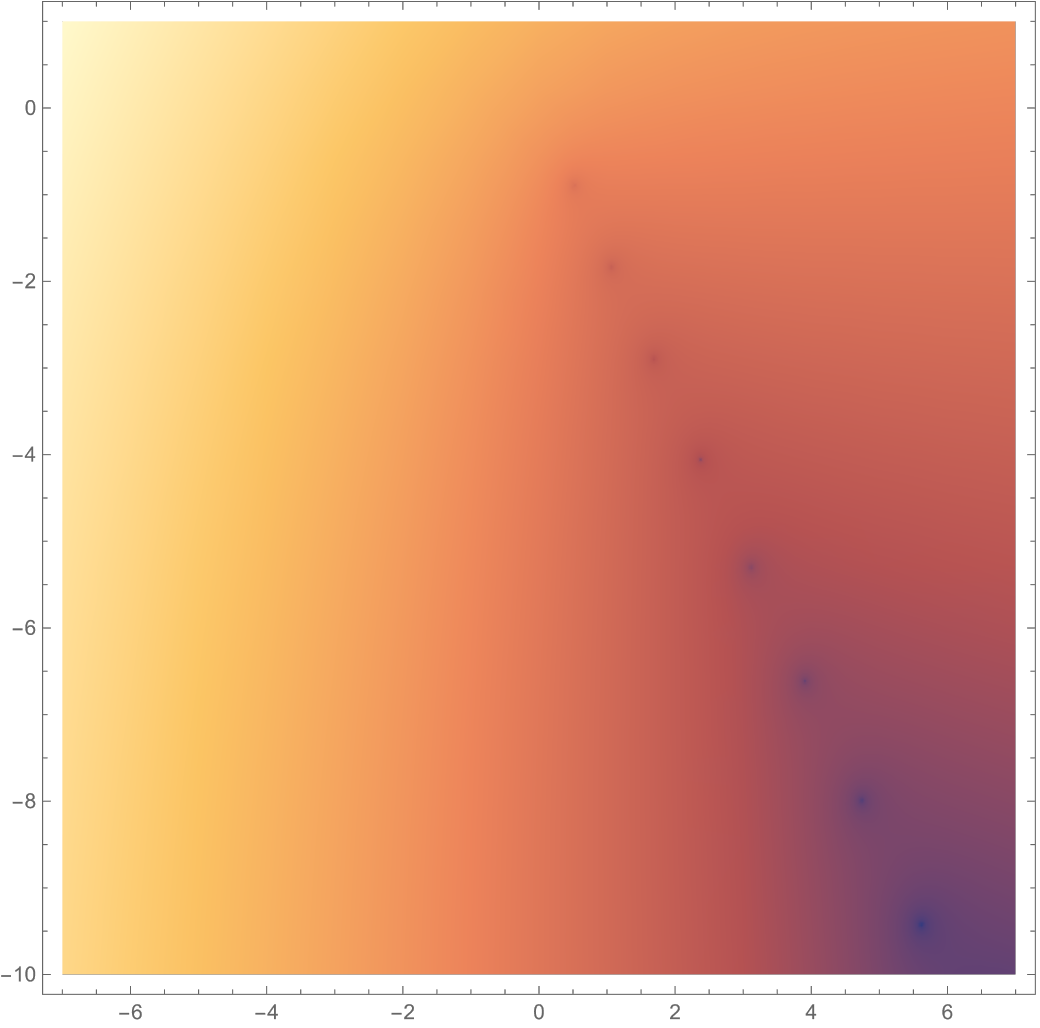}

    \end{subfigure}

\caption{Tower of QNMs for 100\,000 terms in the recurrence relation. Left panel is for $q=10$ and right panel is for $q=80$.}
\label{towerplot}
\end{figure}

By using 100\,000 terms in the recurrence relation in appendix \ref{app:alternativeHeun}, we can plot the $\omega_{\text{table}}$ plane to see the tower of QNMs. 
We present the results for $q=10$ and $q=80$ in figure \ref{towerplot}.

\section{Nekrasov-Shatashvili functions}\label{app:NS}

In this appendix, we collect the notations and conventions used for the NS functions and the Heun connection problem. The relevant theory is $\mathcal{N}=2$ $SU(2)$ gauge theory with $N_f=4$ fundamental hypermultiplets. 

If $Y$ is a Young diagram, we denote with $(Y_1\ge Y_2\ge\dots)$ the heights of its columns and with $(Y'_1\ge Y'_2,\dots)$ the lengths of its rows. For every Young diagram $Y$ and for every box $s=(i,j)$, we denote the arm length and the leg length of $s$ with respect to the diagram $Y$ as
\begin{equation}
A_Y(i, j) = Y_j -  i, \quad L_Y(i, j) =Y'_i - j.
\end{equation}

We introduce the main contributions coming into play for the definition of the instanton partition function of $\mathcal{N}=2$ $SU(2)$  gauge theory with fundamental matter. Let us denote with $\vec{Y}=\left( Y_1, Y_2 \right)$ a pair of Young diagrams and with $| \vec{Y} | = | Y_1 | + | Y_2 |$ the total number of boxes. We denote with $\vec{a}=(a_1,a_2)$ the v.e.v. of the scalar in the vector multiplet and with $\epsilon_1,\epsilon_2$ the parameters characterizing the $\Omega$-background. We define the hypermultiplet and vector contribution as
\begin{equation}
\begin{aligned}
    z_{\text{hyp}} \left( \vec{a}, \vec{Y}, m \right) &= \prod_{k= 1,2} \prod_{(i,j) \in Y_k} \left[ a_k + m + \epsilon_1 \left( i - \frac{1}{2} \right) + \epsilon_2 \left( j - \frac{1}{2} \right) \right] \,, \\
    z_{\text{vec}} \left( \vec{a}, \vec{Y} \right) &= \prod_{i,j=1}^2\prod_{s\in Y_i}\frac{1}{a_i-a_j-\epsilon_1L_{Y_j}(s)+\epsilon_2(A_{Y_i}(s)+1)} \\
    &\hspace{1.3cm}\prod_{t\in Y_j}\frac{1}{-a_j+a_i+\epsilon_1(L_{Y_i}(t)+1)-\epsilon_2A_{Y_j}(s)}\,.
\end{aligned}
\end{equation}
We fix the convention $\epsilon_1=1$ and $\vec{a}=(a,-a)$. 
Let us denote with $m_1,m_2,m_3,m_4$ the masses of the four hypermultiplets related to the gauge parameters $a_0,a_t,a_1,a_{\infty}$ by
\begin{equation}\label{gaugemasses}
\begin{aligned}
m_1&=-a_t-a_0,\\
m_2&=-a_t+a_0,\\
m_3&=a_{\infty}+a_1,\\
m_4&=-a_{\infty}+a_1,
\end{aligned}
\end{equation}
as in \eqref{fundmasses}.
Moreover, the instanton counting parameter $t$ satisfies $t=e^{2\pi i\tau}$, where $\tau$ is related to the gauge coupling by 
\begin{equation}
\tau=\frac{\theta}{2\pi}+\rmi\frac{4\pi}{g_{\rm YM}^2}.
\end{equation}
The instanton part of the NS free energy is then given as a power series in $t$ by
\begin{equation}
F(t)=\lim_{\epsilon_2\to 0}\epsilon_2\log\Biggl[(1-t)^{-2\epsilon_2^{-1}\left(\frac{1}{2}+a_1\right)\left(\frac{1}{2}+a_t\right)}\sum_{\vec{Y}}t^{|\vec{Y}|}z_{\text{vec}} \left( \vec{a}, \vec{Y} \right)\prod_{i=1}^4z_{\text{hyp}} \left( \vec{a}, \vec{Y}, m_i \right)\Biggr].
\end{equation}

\section{Deriving the exact quantisation condition}\label{app:WKBquantcond} 

We will now present the derivations pertaining the exaxt quantisation condition \eqref{quantcondWKBscalar}, including deriving the connection formulas and imposing boundary conditions. We also provide the general form of the WKB period $\alpha$.

\subsection{Connection Formula}

In this section, we derive the connection formula that describes the analytic continuation of our WKB solutions from $z=0$ to $z=1$. Enforcing the boundary conditions on this connection formula will lead to the exact quantisation condition (EQC) \eqref{quantcondWKBscalar}.

To derive this connection formula we must track each Stokes line crossed and the corresponding Stokes jump of our WKB basis of solutions. Recall that we wish to find the connection formula for two Stokes graphs, the plus Stokes graph and the minus Stokes graph -- see figure \ref{fig:Stokes graphs plus and minus}. For each graph, the order in which we cross the Stokes lines and the precise Stokes jump may be different. As such, we will need to consider each Stokes graph and the corresponding analytic continuation separately.

For both Stokes graphs, we start with WKB solutions $\psi_0^\pm$ near $z=0$. These WKB solutions are  normalised at $0$ as in \eqref{normalisationturningpoint}. We then consider analytically continuing $\psi^{\pm}_0(z)$ between different regions delimited by Stokes lines, we label these regions $D_i$ for the plus Stokes graph and $C_i$ for the minus Stokes graph. More precisely, we consider analytically continuing $\psi^{\pm}_0(z)$ from $D_1$ (or $C_1$) to $D_2$ (or $C_2$) and then from there to  $D_3$ (or $C_3$). Once we have tracked our WKB solutions $\psi_0^\pm(z)$ to $D_3$ (or $C_3$), care must be taken to analytically continue to $1$, as we will cross infinitely many Stokes lines. To describe these jumps across Stokes lines and to keep track of the regions in the complex plane, we denote by $\psi^\pm_{a,C}(z)$ the solution normalised at the turning point $a$ in the region $C$.

Before discussing the analytic continuation $\psi_0^\pm(z)$, we discuss a useful way to determine whether a Stokes line is positive or negative \cite{AOKI:2021hmu, Miyachi:2025dyk}. Consider a Stokes line that ends at a pole $p$ of the potential, where $p$ is order $2$. Denote the leading term of the large-$q$ expansion of $S_\text{odd}$ by $S_{\text{odd,-1}}$. Then, one can expand the integral of $S_\text{odd,-1}$ near $p$ to find
\begin{equation}
    \int^z S_{\text{odd},-1}\, dz' \sim \rho (\log|z-p| + \rmi \arg(z-p)),
\end{equation}
where $\rho = \text{Res}_{z=p} S_{\text{odd},-1}$. Along the Stokes line the imaginary part of the above integral will be $0$ and for $z$ sufficiently close to $p$, it is clear that the real part of this integral will negative if $\mathrm{Re}\,\rho >0$ and positive if $\mathrm{Re}\, \rho <0$. The sign of the real part of the integral of $ S_{\text{odd},-1}$ will not change along the Stokes line\footnote{Provided the Stokes line does not pass through a branch cut.}; thus, the sign of $\mathrm{Re}\, \rho$ is sufficient to determine whether a given Stokes line is positive or negative. In our case, we have two double poles located at $z= -1,1$ with residues equal to $- qs/4$ and $- \rmi qs/4$ respectively\footnote{Note the opposite sign convention to section \ref{sec:setup}.}. Recall that we assumed $\mathrm{Re}(s)>0$ and $\mathrm{Im}(s)<0$, which means that any Stokes line ending at $z=\pm1$ on the first sheet must be a positive Stokes line. However, when the WKB solutions are analytically continued through a branch cut, $\psi^{\pm}(z)$ swap roles. This means that any negative/positive Stokes line must become a positive/negative Stokes line once it passes through the branch cut. For our problem, this means that any negative Stokes line must pass through a branch cut as it cannot end at $z=1$ or $z=-1$ on the first sheet.

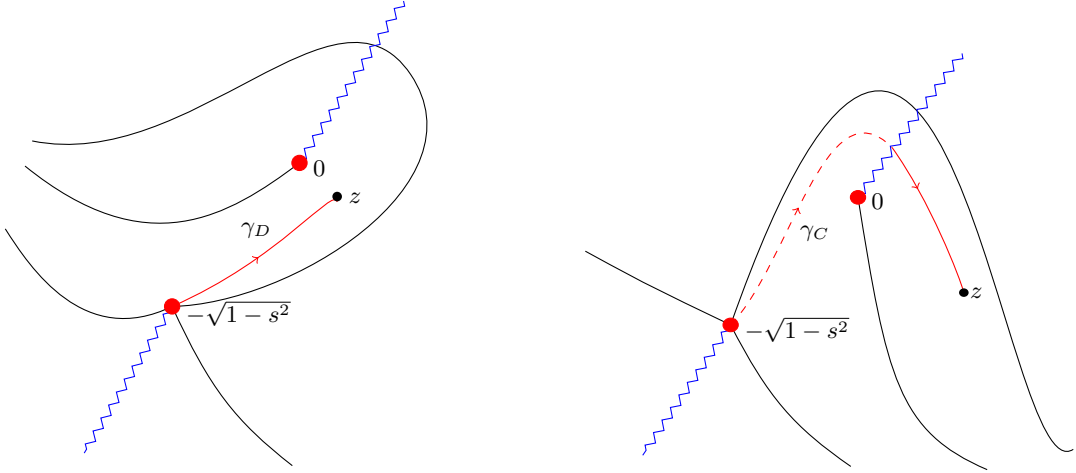
\begin{figure}[h]
    \centering   
    \begin{subfigure}{0.48\textwidth}
\begin{tikzpicture}[x=0.75pt,y=0.75pt,yscale=-0.9,xscale=0.9]
\path[use as bounding box] (120,80) rectangle (420,350);
 \node at (260,261) {\footnotesize$-\sqrt{1-s^2}$};

\draw [color=blue,draw opacity=1,decorate,decoration={zigzag, segment length = 2mm,amplitude=0.5mm} ]   (223,257.25) -- (174,339) ;

\draw    (223,257.25) .. controls (289,256) and (397,189) .. (356,128) .. controls (315,67) and (234,183) .. (145,165) ;

\draw [color=red  ,draw opacity=1,postaction={
    decorate,
    decoration={
      markings,
      mark=at position 0.5 with {\arrow{>}}
    }
  }  ]   (223,257.25) .. controls (283,231) and (313,190) .. (315,199) ;
\fill (315,196) circle (2pt);
\node  at (325,197) {\footnotesize$z$};
\node  at (305,180) {\footnotesize$0$};
\node  at (270,215) {\footnotesize$\gamma_D$};
  ;

\draw    (223,257.25) .. controls (244,303) and (258,321) .. (290,346) ;

\draw    (130,214) .. controls (144,236) and (174,281) .. (223,257.25) ;

\draw  [color=red  ,draw opacity=1 ][fill={rgb, 255:red, 255; green, 0; blue, 0 }  ,fill opacity=1 ] (218.75,257.25) .. controls (218.75,254.9) and (220.65,253) .. (223,253) .. controls (225.35,253) and (227.25,254.9) .. (227.25,257.25) .. controls (227.25,259.6) and (225.35,261.5) .. (223,261.5) .. controls (220.65,261.5) and (218.75,259.6) .. (218.75,257.25) -- cycle ;

\draw    (141,179) .. controls (209,234) and (254,207.25) .. (294,177.25) ;

\draw [color=blue ,draw opacity=1,decorate,decoration={zigzag, segment length = 2mm,amplitude=0.5mm} ]   (352,87) -- (294,177.25) ;
 
\draw  [color=red  ,draw opacity=1 ][fill={rgb, 255:red, 255; green, 0; blue, 0 }  ,fill opacity=1 ] (289.75,177.25) .. controls (289.75,174.9) and (291.65,173) .. (294,173) .. controls (296.35,173) and (298.25,174.9) .. (298.25,177.25) .. controls (298.25,179.6) and (296.35,181.5) .. (294,181.5) .. controls (291.65,181.5) and (289.75,179.6) .. (289.75,177.25) -- cycle ;
\end{tikzpicture}
\end{subfigure}
\begin{subfigure}{0.48\textwidth}
\begin{tikzpicture}[x=0.75pt,y=0.75pt,yscale=-0.8,xscale=0.9]
\path[use as bounding box] (120,80) rectangle (420,350);

\draw [color=blue  ,draw opacity=1,decorate,decoration={zigzag, segment length = 2mm,amplitude=0.5mm} ]   (223,257.25) -- (174,339) ;
 
\draw    (223,257.25) .. controls (343,-125.75) and (374,365.25) .. (414,335.25) ;
 
\draw    (366,348) .. controls (316,324) and (310,290) .. (294,177.25) ;
 
\draw [color=blue  ,draw opacity=1,decorate,decoration={zigzag, segment length = 2mm,amplitude=0.5mm} ]   (352,87) -- (294,177.25) ;
 
\draw  [color=red  ,draw opacity=1 ][fill={rgb, 255:red, 255; green, 0; blue, 0 }  ,fill opacity=1 ] (289.75,177.25) .. controls (289.75,174.9) and (291.65,173) .. (294,173) .. controls (296.35,173) and (298.25,174.9) .. (298.25,177.25) .. controls (298.25,179.6) and (296.35,181.5) .. (294,181.5) .. controls (291.65,181.5) and (289.75,179.6) .. (289.75,177.25) -- cycle ;
 
\draw [color=red  ,draw opacity=1, postaction={
    decorate,
    decoration={
      markings,
      mark=at position 0.3 with {\arrow{>}}
    }
  } ]   (313,146) .. controls (334,179) and (351,228) .. (353,237) ;

\fill (353,237) circle (2pt);
\node  at (360,237) {\footnotesize$z$};
\node  at (305,180) {\footnotesize$0$};
\node  at (270,200) {\footnotesize$\gamma_C$};
 
\draw [color=red  , dashed,draw opacity=1,postaction={
    decorate,
    decoration={
      markings,
      mark=at position 0.5 with {\arrow{>}}
    }
  } ]   (223,257.25) .. controls (256,219) and (276,104) .. (313,146) ;

\draw    (223,257.25) .. controls (244,303) and (258,321) .. (290,346) ;

\draw    (142,211) .. controls (181,235) and (182,235) .. (223,257.25) ;

\draw  [color=red  ,draw opacity=1 ][fill=red  ,fill opacity=1 ] (218.75,257.25) .. controls (218.75,254.9) and (220.65,253) .. (223,253) .. controls (225.35,253) and (227.25,254.9) .. (227.25,257.25) .. controls (227.25,259.6) and (225.35,261.5) .. (223,261.5) .. controls (220.65,261.5) and (218.75,259.6) .. (218.75,257.25) -- cycle ;
 \node at (260,260) {\footnotesize$-\sqrt{1-s^2}$};
\end{tikzpicture}
\end{subfigure}
    \caption{Schematic picture of the paths defining the solution normalised at $-\sqrt{1-s^2}$. We represent the paths starting from $z$ on the second sheet encircling $-\sqrt{1-s^2}$ and returning to $z$ on the first sheet with a single line, with non-dashed representing paths which return to $z$ on the first sheet and dashed representing paths which return to $z$ on the second sheet.}
    \label{fig:betajumpfigures}
    \end{figure}

We now describe the first part of the analytic continuation of $\psi_0^\pm(z)$ for both Stokes graphs. We start with the plus Stokes graph. The first Stokes line crossed emanates from $-\sqrt{1-s^2}$, encircles $z= 0$ through the branch cut, and ends at $z=-1$, see the left graph of figure \ref{fig:Stokes graphs plus and minus}. To cross this Stokes line, we must first change the normalisations of our WKB solutions from $0$ to $-\sqrt{1-s^2}$:
\begin{equation}
   \left( \begin{matrix}
        \psi_{0,D_1}^+\\
        \psi_{0,D_1}^-
    \end{matrix}\right) = \left(\begin{matrix}
       \rme^{-\beta/2} & 0 \\
       0 &\rme^{\beta/2}
    \end{matrix} \right) \left( \begin{matrix}
        \psi_{-\sqrt{1-s^2},D_1}^+\\
        \psi_{-\sqrt{1-s^2},D_1}^-
    \end{matrix}\right).
\end{equation}
Note that the sign difference in the exponentials between this formula and \eqref{change of norm matrix} is due to our definition of $\beta$, \eqref{ABperiods}. We then cross the negative Stokes line in the clockwise direction
\begin{equation}
   \left( \begin{matrix}
        \psi_{-\sqrt{1-s^2},D_1}^+\\
        \psi_{-\sqrt{1-s^2},D_1}^-
    \end{matrix}\right) = \left(\begin{matrix}
       1 & 0 \\
       -\rmi &1
    \end{matrix} \right) \left( \begin{matrix}
        \psi_{-\sqrt{1-s^2},D_2}^+\\
        \psi_{-\sqrt{1-s^2},D_2}^-
    \end{matrix}\right).
\end{equation}
Finally, we change the normalisation of our solutions back to $0$:
\begin{equation}
   \left( \begin{matrix}
        \psi_{-\sqrt{1-s^2},D_2}^+\\
        \psi_{-\sqrt{1-s^2},D_2}^-
    \end{matrix}\right)
   = \left(\begin{matrix}
       \rme^{\beta/2} & 0 \\
       0 &\rme^{-\beta/2}
    \end{matrix} \right) \left( \begin{matrix}
        \psi_{0,D_2}^+\\
        \psi_{0,D_2}^-
    \end{matrix}\right).
\end{equation}
The total Stokes jump is then  
\begin{equation}
   \left( \begin{matrix}
        \psi_{0,D_1}^+\\
        \psi_{0,D_1}^-
    \end{matrix}\right)
   = \left(\begin{matrix}
       1 &  0\\
       -\rmi\rme^{\beta} &1
    \end{matrix} \right) \left( \begin{matrix}
        \psi_{0,D_2}^+\\
        \psi_{0,D_2}^-
    \end{matrix}\right).
\end{equation}
Apart from this first jump, the Stokes jumps for both Stokes graphs will be the same. Thus, we consider the first Stokes jump of the minus Stokes graph before continuing. The first Stokes line for the minus Stokes graph emanates from $-\sqrt{1-s^2}$, but starts on the second sheet, passes through the branch cut emanating from $0$ onto the first sheet, then encircles $z=1$ and ends at $z=-1$, see the right graph of figure \ref{fig:Stokes graphs plus and minus}. This is different from the plus Stokes graph, where the first Stokes line started on the first sheet and is crossed on the first sheet. This difference in topology will result in different a Stokes jump. Specifically, for the minus Stokes graph we need to change the normalisation of $\psi^\pm_0(z)$ from $0$ to the point $-\sqrt{1-s^2}$ on the second sheet. This solution is normalised along the contour $\gamma_C$, see figure \ref{fig:betajumpfigures}, and this changes the normalisation factors:
\begin{equation}
   \left( \begin{matrix}
        \psi_{0,C_1}^+\\
        \psi_{0,C_1}^-
    \end{matrix}\right) = \left(\begin{matrix}
       \rme^{\beta/2} & 0 \\
       0 &\rme^{-\beta/2}
    \end{matrix} \right) \left( \begin{matrix}
        \psi_{-\sqrt{1-s^2},C_1}^+\\
        \psi_{-\sqrt{1-s^2},C_1}^-
    \end{matrix}\right),
\end{equation}
where the solutions $\psi_{-\sqrt{1-s^2},C_1}^\pm$ are defined along $\gamma_C$ whereas the solutions $\psi_{-\sqrt{1-s^2},D_1}^\pm$ were defined along $\gamma_D$\footnote{The depictions of the contours $\gamma_C$ and $\gamma_D$ shown in figure \ref{fig:betajumpfigures} are schematic, in reality $\gamma_C$ and $\gamma_D$ should start on one sheet, loop around $-\sqrt{1-s^2}$ and return to $z$ on the other sheet.}.
 
We can then cross the Stokes line. Although this starts as a minus Stokes line, we want to cross it after it passes through the cut. As mentioned above, when WKB solutions cross the cut they pass onto a different sheet and $\psi^{\pm}$ swap roles. Therefore, the negative Stokes line becomes a positive Stokes line once it crosses onto the first sheet. Crossing this Stokes line in the anticlockwise direction and then changing normalisation back to $z=0$, the final result is
\begin{equation}
   \left( \begin{matrix}
        \psi_{0,C_1}^+\\
        \psi_{0,C_1}^-
    \end{matrix}\right) = \left(\begin{matrix}
       1 &  \rmi \rme^{\beta} \\
      0& 1
    \end{matrix} \right) \left( \begin{matrix}
        \psi_{0,C_2}^+\\
        \psi_{0,C_2}^-
    \end{matrix}\right).
\end{equation}

\begin{figure}
  
\centering

\tikzset{every picture/.style={line width=0.75pt}}        

\begin{tikzpicture}[x=0.75pt,y=0.75pt,yscale=-1,xscale=1]

\draw [color = blue  ,draw opacity=1,decorate,decoration={zigzag, segment length = 2mm,amplitude=0.5mm}  ]   (283,135.75) -- (208,250) ;

\draw   (367.85,193.99) .. controls (367.69,191.4) and (368.46,189.01) .. (370.15,186.81) .. controls (372.05,184.34) and (374.69,182.64) .. (378.08,181.74) .. controls (381.85,180.73) and (385.83,180.88) .. (390,182.2) .. controls (394.58,183.64) and (398.61,186.24) .. (402.07,189.99) .. controls (405.84,194.08) and (408.32,198.86) .. (409.5,204.32) .. controls (410.78,210.23) and (410.35,215.99) .. (408.2,221.63) .. controls (405.9,227.67) and (401.98,232.63) .. (396.45,236.5) .. controls (390.56,240.63) and (383.71,242.9) .. (375.93,243.31) .. controls (367.67,243.75) and (359.58,242.04) .. (351.65,238.17) .. controls (343.28,234.1) and (336.3,228.22) .. (330.75,220.54) .. controls (324.89,212.46) and (321.44,203.57) .. (320.39,193.84) .. controls (319.29,183.65) and (320.97,174.02) .. (325.46,164.95) .. controls (330.14,155.46) and (337.2,147.95) .. (346.65,142.38) .. controls (356.49,136.58) and (367.53,133.73) .. (379.76,133.83) .. controls (392.48,133.93) and (404.7,137.2) .. (416.4,143.64) .. controls (428.56,150.32) and (438.47,159.49) .. (446.14,171.12) .. controls (454.08,183.17) and (458.51,196.19) .. (459.42,210.19) .. controls (460.36,224.66) and (457.41,238.15) .. (450.59,250.66) .. controls (443.54,263.58) and (433.32,273.67) .. (419.96,280.92) .. controls (406.17,288.4) and (390.94,291.83) .. (374.25,291.22) .. controls (357.08,290.58) and (340.74,285.75) .. (325.24,276.73) .. controls (309.3,267.44) and (296.46,255.01) .. (286.68,239.41) .. controls (276.64,223.39) and (271.25,206.24) .. (270.48,187.97) .. controls (269.68,169.23) and (273.89,151.88) .. (283.07,135.91) .. controls (283.07,135.91) and (283.07,135.91) .. (283.07,135.91) ;

\draw  [color=red  ,draw opacity=1,postaction={
    decorate,
    decoration={
      markings,
      mark=at position 0.5 with {\arrow{<}}
  } } ] (291.59,184.44) .. controls (294.72,167.63) and (301.94,152.7) .. (313.2,139.65) .. controls (324.64,126.4) and (338.71,116.89) .. (355.41,111.1) .. controls (372.35,105.22) and (389.71,103.98) .. (407.49,107.35) .. controls (425.51,110.78) and (441.51,118.5) .. (455.51,130.51) .. controls (469.7,142.68) and (479.9,157.62) .. (486.1,175.32) .. controls (492.4,193.26) and (493.74,211.6) .. (490.14,230.35) .. controls (486.49,249.35) and (478.27,266.2) .. (465.45,280.89) .. controls (452.48,295.78) and (436.55,306.44) .. (417.68,312.89) .. controls (398.57,319.42) and (379.02,320.74) .. (359.04,316.85) .. controls (338.82,312.91) and (320.87,304.17) .. (305.22,290.62) .. controls (289.37,276.91) and (278.01,260.12) .. (271.14,240.27) .. controls (264.18,220.18) and (262.75,199.66) .. (266.86,178.72) .. controls (269.56,164.97) and (274.4,152.24) .. (281.39,140.53) ;

\fill (291.59,184.44) circle (2pt);

\fill (367.85,193.99) circle (2pt);
\node  at (305,180) {\footnotesize$z$};
\node  at (260,120) {\footnotesize$\sqrt{1-s^2}$};
\node  at (380,193.99){\footnotesize$1$};
\draw    (189,173) .. controls (225,180) and (243,165.75) .. (283,135.75) ;

\draw    (283,135.75) .. controls (297.23,115.33) and (338.21,83.27) .. (386,79.6) .. controls (433.79,75.93) and (480,86.6) .. (505,106.6) ;

\draw  [color=red  ,draw opacity=1 ][fill={rgb, 255:red, 255; green, 0; blue, 0 }  ,fill opacity=1 ] (278.75,135.75) .. controls (278.75,133.4) and (280.65,131.5) .. (283,131.5) .. controls (285.35,131.5) and (287.25,133.4) .. (287.25,135.75) .. controls (287.25,138.1) and (285.35,140) .. (283,140) .. controls (280.65,140) and (278.75,138.1) .. (278.75,135.75) -- cycle ;
\node  at (305,220) {\footnotesize$D_4/C_4$};
\node  at (345,180) {\footnotesize$D_5/C_5$};
\end{tikzpicture}

    \caption{Schematic picture depicting the contour that defines WKB solutions normalised at $\sqrt{1-s^2}$. The normalisation is such that the WKB solutions can cross from $D_4/C_4$ to $D_5/C_5$. The Stokes lines have been truncated for clarity. The contour is in red and represents a contour starting from $z$ on the second sheet, encircling $\sqrt{1-s^2}$ and returning to $z$ on the first sheet. The integral of $S_\text{odd}$ along this contour can be deform to a loop around $z=1$ and a straight line from $\sqrt{1-s^2}$ to $z$.}
    \label{fig:encricling0oncecontour}
\end{figure}
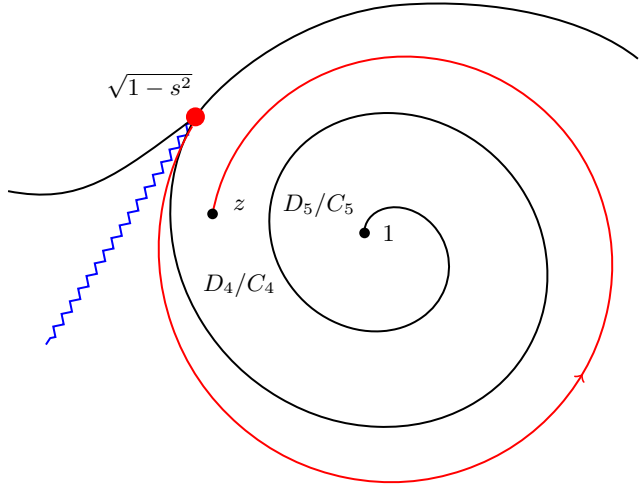

The next Stokes line the WKB solutions cross is the same for both Stokes graphs, this is the Stokes line separating $D_2$ (or $C_2$) and $D_3$ (or $C_3$). This line starts from $\sqrt{1-s^2}$, encircles the pole at $z=1$ once, and ends at $z=-1$. To cross this Stokes line, we first change the normalisations from $z=0$ to $z=\sqrt{1-s^2}$. The path defining the change in normalisation starts from $z=0$ and ends at $z=\sqrt{1-s^2}$ but passes $z=1$ to right. This contour can be deformed into two contours encircling $z=1$, one on the first sheet in the anticlockwise direction and one on the second sheet in the clockwise direction, as well as $\gamma_\alpha$ the contour defining the $\alpha$ period, see figure \ref{fig:alpha and beta cycle}. Any integral of $S_\text{odd}$ along a path is invariant under the inversions of the sheets followed by inverting the direction of the path. Therefore, the appropriate matrix that changes the normalisation is
\begin{equation}
   \left( \begin{matrix}
        \psi_{0,D_2}^+\\
        \psi_{0,D_2}^-
    \end{matrix}\right) = \left(\begin{matrix}
       e^{\alpha/2+q\pi s/2} & 0 \\
      0& e^{-\alpha/2-q\pi s/2}
    \end{matrix} \right) \left( \begin{matrix}
        \psi_{\sqrt{1-s^2},D_2}^+\\
        \psi_{\sqrt{1-s^2},D_2}^-
    \end{matrix}\right),
\end{equation}
where the factors $e^{\pm\pi s/2}$ are contributions from the residue at $z=1$ on both sheets. We can now cross the minus Stokes line in the clockwise direction and change our normalisation back to $0$ with the inverse of the above matrix: 
\begin{equation}
   \left( \begin{matrix}
        \psi_{0,D_2}^+\\
        \psi_{0,D_2}^-
    \end{matrix}\right) = \left(\begin{matrix}
       1 & 0 \\
      -\rmi \rme^{-\alpha-q\pi s}&1 
    \end{matrix} \right) \left( \begin{matrix}
        \psi_{0,D_3}^+\\
        \psi_{0,D_3}^-
    \end{matrix}\right).
\end{equation}
The same connection formula holds for the minus Stokes graph with $C$ and $D$ swapped. Finally, we analytically continue from $D_3$ (or $C_3$) to $D_4$ (or $C_4$) to $D_5$ (or $C_5$) etc until we reach $z=1$ by passing through the same spiralling Stokes line infinitely many times. For both graphs, we can cross the first segment of this Stokes line by changing normalisations to $\sqrt{1-s^2}$:
    \begin{equation}
   \left( \begin{matrix}
        \psi_{0,D_3}^+\\
        \psi_{0,D_3}^-
    \end{matrix}\right) = \left(\begin{matrix}
       \rme^{\alpha/2} & 0 \\
      0& \rme^{-\alpha/2}
    \end{matrix} \right) \left( \begin{matrix}
        \psi_{\sqrt{1-s^2},D_3}^+\\
        \psi_{\sqrt{1-s^2},D_3}^-
    \end{matrix}\right).
\end{equation}
This is a positive Stokes line and we cross it in the anti-clockwise direction, thus the total Stokes jump is
\begin{equation}
   \left( \begin{matrix}
        \psi_{0,D_3}^+\\
        \psi_{0,D_3}^-
    \end{matrix}\right) = \left(\begin{matrix}
       1 &  ie^{\alpha} \\
     0&1 
    \end{matrix} \right) \left( \begin{matrix}
        \psi_{0,D_4}^+\\
        \psi_{0,D_4}^-
    \end{matrix}\right).
\end{equation}
The next line segment that the WKB solutions $\psi^\pm(z)$ cross is a segment of the same Stokes line that we crossed before but now we cross this segment after the Stokes line has encircled $z=1$ once. For our solutions to cross this Stokes line, the normalisations must respect the encirclement of $z=1$. This means the normalisation factors will have contributions from the residue at $z=1$ (see figure \ref{fig:encricling0oncecontour}):
\begin{equation}
   \left( \begin{matrix}
        \psi_{0,D_4}^+\\
        \psi_{0,D_4}^-
    \end{matrix}\right) = \left(\begin{matrix}
       e^{\alpha/2-q \pi s/2} &  0 \\
     0& e^{-\alpha/2+q \pi s/2} 
    \end{matrix} \right) \left( \begin{matrix}
        \psi_{\sqrt{1-s^2},D_4}^+\\
        \psi_{\sqrt{1-s^2},D_3}^-
    \end{matrix}\right).
\end{equation}
Again, the Stokes line is positive and we cross it in the anti-clockwise direction. After changing the normalisation of the WKB solutions back to $0$ using the inverse of the above matrix, the net Stokes jump is 
\begin{equation}
   \left( \begin{matrix}
        \psi_{0,D_4}^+\\
        \psi_{0,D_4}^-
    \end{matrix}\right) = \left(\begin{matrix}
      1 &   \rmi\rme^{\alpha-q \pi s} \\
     0& 1
    \end{matrix} \right) \left( \begin{matrix}
        \psi_{0,D_5}^+\\
        \psi_{0,D_5}^-
    \end{matrix}\right).
\end{equation}

Note that we always change our normalisation back to a direct path to $z=0$ (i.e, with no encirclements of $z=1$). This is because we are considering the analytic continuation directly from $z=0$ to $z=1$, we could consider analytic continuations which wrap around $z=1$ any number of times, but we should expect these analytic continuations to yield different results.

The next line segment that the WKB solutions cross is a segment of same Stokes line again, but it has encircled $z=1$ twice. The segment after that is crossed after the Stokes line has encircled $z=1$ three times. This repeats inductively, and the $n^{th}$ crossing occurs after the Stokes line has encircled $n$ times $z=1$. This means that the path defining the normalisation at $z = \sqrt{1-s^2}$ must encircle $n$ times $z=1$ if the solutions are crossing the $n^{th}$ line segment. If we denote the region to the left of the $n^{th}$ line segment by $D_{n+3}$ and the region to the right of the $n^{th}$ line segment by $D_{n+4}$, the connection formula to cross only the $n^{th}$ line segment is given by
\begin{equation}
   \left( \begin{matrix}
        \psi_{0,D_{3+n}}^+\\
        \psi_{0,D_{3+n}}^-
    \end{matrix}\right) = \left(\begin{matrix}
      1 &   \rmi \rme^{\alpha-nq \pi s} \\
     0& 1
    \end{matrix} \right) \left( \begin{matrix}
        \psi_{0,D_{4+n}}^+\\
        \psi_{0,D_{4+n}}^-
    \end{matrix}\right),
\end{equation}
where $\psi_{0,D_{3+n}}^\pm$ are the WKB solutions normalised at $0$ valid in the sector of the spiral reached after crossing $n-$line segments. The total connection formula for crossing this line and the $n-1$ preceding lines is 
\begin{align}
      \left( \begin{matrix}
        \psi_{0,D_3}^+\\
        \psi_{0,D_3}^-
    \end{matrix}\right) &= \left(\begin{matrix}
      1 &   \rmi \rme^{\alpha} \\
     0& 1
    \end{matrix} \right) \left(\begin{matrix}
      1 &   \rmi \rme^{\alpha-\pi qs} \\
     0& 1
    \end{matrix} \right) \dots\left(\begin{matrix}
      1 &   \rmi \rme^{\alpha-n\pi qs} \\
     0& 1
    \end{matrix} \right) \left( \begin{matrix}
        \psi_{0,D_1}^+\\
        \psi_{0,D_1}^-
    \end{matrix}\right)  \\
    &= \left(\begin{matrix}
      1 &   \rmi \rme^\alpha\sum_{m=0}^n \rme^{-m\pi q s}\\
     0& 1
    \end{matrix} \right)\left( \begin{matrix}
        \psi_{0,D_{4+n}}^+\\
        \psi_{0,D_{4+n}}^-
    \end{matrix}\right).
\end{align}
Recall that the Stokes graphs were derived under the assumption that $\mathrm{Re} (s) >0$. Thus,  we can take the limit $n\to \infty$ to obtain:
\begin{equation}
    \left( \begin{matrix}
        \psi_{0,D_3}^+\\
        \psi_{0,D_3}^-
    \end{matrix}\right) = \left(\begin{matrix}
      1 &   \rmi\rme^\alpha (1-\rme^{-\pi qs})^{-1}\\
     0& 1
    \end{matrix} \right)\left( \begin{matrix}
        \psi_{0,D_\infty}^+\\
        \psi_{0,D_\infty}^-
    \end{matrix}\right).
\end{equation}
Combining this with the connection formula for $D_1$ (or $C_1$) to $D_2$ (or $C_2$) and $D_2$ (or $C_2$) to  $D_3$ (or $C_3$), the total connection formula is 
\begin{equation}
\begin{aligned} \label{plusgraphconnenctionformula}
    \left( \begin{matrix}
        \psi_{0,D_1}^+\\
        \psi_{0,D_1}^-
    \end{matrix}\right) 
   &= \left(\begin{matrix}
       1 & 0 \\
       -\rmi\rme^{\beta} &1
    \end{matrix} \right) 
    \left(\begin{matrix}
       1 & 0 \\
      -\rmi\rme^{-\alpha-q\pi s}&1 
    \end{matrix} \right)
    \left(\begin{matrix}
      1 &   \rmi \rme^\alpha (1-e^{-\pi qs})^{-1}\\
     0& 1
    \end{matrix} \right)\left( \begin{matrix}
        \psi_{0,D_\infty}^+\\
        \psi_{0,D_\infty}^-
    \end{matrix}\right)\\
 &=\left(\begin{matrix}
      1 &   \rmi\rme^\alpha (1-\rme^{-\pi qs})^{-1}\\
     -\rmi(\rme^{\beta}+\rme^{-\alpha -\pi  qs})& (1-\rme^{-\pi qs})^{-1}(\rme^{\alpha+\beta}+\rme^{-\pi qs})+1
    \end{matrix} \right) \left( \begin{matrix}
        \psi_{0,D_\infty}^+\\
        \psi_{0,D_\infty}^-
    \end{matrix}\right) 
\end{aligned}
\end{equation}
for the plus Stokes graph, and 
\begin{equation}\label{minusgraphconnenctionformula}
 \left( \begin{matrix}
        \psi_{0,C_1}^+\\
        \psi_{0,C_1}^-
    \end{matrix}\right) =   \left(\begin{matrix}
      1+\rme^{\beta-\alpha-q\pi s} &    \rmi(1-\rme^{-\pi qs})^{-1}(\rme^\alpha+\rme^{\beta-\pi qs})\rmi \rme^{\beta}\\
     -\rmi\rme^{-\alpha -\pi  qs}& \rme^{-\pi qs}(1-\rme^{-\pi qs})^{-1}+1
    \end{matrix} \right)\left( \begin{matrix}
        \psi_{0,C_\infty}^+\\
        \psi_{0,C_\infty}^-
    \end{matrix}\right) 
\end{equation}
for the minus Stokes graph.

\subsection{Boundary Conditions}\label{app:boundary conditions}

Now that we have the connection formula from $z=0$ to $z=1$, we impose the boundary conditions \eqref{boundcond2}. At $z=1$, this is straightforward as it simply requires expanding the WKB solution normalised at 0 locally around $z=1$
\begin{equation}
    \psi^\pm_0 (z) \sim (z-1)^{\mp \rmi qs/4+1/2}+ \dots
\end{equation}
We can naively take the $z\to 1$ expansion because the large-$q$ and $z \to 1$ limits do not compete\footnote{This can be seen from the structure of $S_{\text{odd},\ge 0}$, since they do not have a pole at $z=1$.}, and so it is enough to consider the leading term only. Therefore, the boundary condition selects the local behaviour of only $\psi^+_0(z)$ near $z=1$.

The behaviour near $z=0$ is not so simple because we expect a competition between the limit large-$q$ limit and the small-$z$ limit. Explicitly, near $z=0$ and for $n \geq 1$ the integral of $S_{\text{odd}, n}$ behaves as
\begin{equation}
    \int^z S_{\text{odd}, n}(z') \, dz' \sim q^{-n} z^{-n/2}.
\end{equation}
This means that as the solutions approach $z=0$, they cross a threshold where $z^{-n/2} \gg q^{n}$, and the large-$q$ WKB expansion cannot be trusted anymore.

We assume that the correct local behaviour at $z=0$ requires the following linear combination of solutions 
\begin{equation}
    \psi_0^+ + K\psi^{-}_0.
\end{equation}
where $K$ is some $z$ independent constant, which we determine below. Then, the boundary conditions \eqref{boundcond2}, as well as \eqref{plusgraphconnenctionformula} and \eqref{minusgraphconnenctionformula}, yield the following EQCs:
\begin{equation}
    \rmi(1-\rme^{-\pi q s})^{-1}(\rme^{\alpha}-\rme^{\beta -\pi qs})-\rmi\rme^{\beta}+K \left( \rme^{-\pi qs}(1-\rme^{-\pi qs})^{-1}+1\right)=0
\end{equation}
for the plus Stokes graph, and
\begin{equation}
    \rmi \rme^{\alpha}(1-\rme^{-\pi qs})^{-1}+K\left((1-\rme^{-\pi qs})^{-1}(-\rme^{\alpha+\beta}+\rme^{-\pi qs})+1\right)=0
\end{equation}
for the minus Stokes graph.
Multiplying these conditions by the appropriate factors, we can simplify them to 
\begin{equation}\label{app: QC's with C}
    \begin{aligned}
    \rme^{\alpha}-\rme^{\beta}-\rmi K=0,\\
    \rme^{-\alpha}-\rme^{\beta}+\rmi K^{-1}=0.       
    \end{aligned}
\end{equation}

We now find the correct constant $K$ using the \emph{uniform WKB method} \cite{e22a9a57-b6b7-33c4-ac51-655992a56ecf, 10.1093/oso/9780198505730.001.0001}. The idea is to uniformly describe the behaviour of our solution near some point $a$ where the WKB expansion breaks down. In our case $a=0$. To do this, we rewrite the spectral problem (i.e. both the differential equation and the resulting connection formula) in a simpler form, preserving the local behaviour of the solution. This is achieved by hiding the complexity of the problem in the definition of a new variable $\xi(z)$.
More concretely, starting from the problem 
\begin{equation}
    \psi''(z) -V(z)\psi(z)=0,
\end{equation}
we analyse the structure around $z=0$ and we define a new simplified potential $W(\xi(z))$ which preserves the local structure, but simplifies the global one (for example, by reducing the number of singularities), and we rewrite the problem in the form
\begin{equation}
    \varphi''(\xi) -W(\xi)\varphi(\xi)=0,
\end{equation}
where $\xi(z)$ is the new variable, such that the relation between $W(\xi(z))$ and $V(z)$ is 
\begin{equation} \label{change of var UWKB}
    \frac{1}{2}\{\xi(z), z \}-\left(\xi'(z)\right)^2 W(\xi(z)) +V(z) =0,
\end{equation}
where $\{\xi(z), z\}$ is the Schwarzian derivative 
\begin{equation}
    \{\xi(z), z \} =\frac{\xi'''(z)}{\xi'(z)} - \frac{3}{2}\left(\frac{\xi''(z)}{\xi'(z)} \right)^2,
\end{equation}
and the wave functions are related by the redefinition
\begin{equation} \label{wavefunc redef}
    \psi(z) = \frac{\sigma}{\sqrt{\xi'(z)}}\varphi(\xi(z)),
\end{equation}
where $\sigma$ is some $z$-independent quantity.

Applying this to our case, we choose the reference potential $W(\xi)$ to be of Bessel type, and we preserve the index of the singular point $z^*=0$ which is mapped to $\xi=0$. More precisely, we take the simplified differential equation to be
\begin{equation} \label{uniform wkb bessel equation}
    \varphi''(\xi) + \left( \frac{q^2}{4\xi} - \frac{\mu^2 -1}{4\xi^2}\right)\varphi(\xi)=0
\end{equation}
with index $\mu=2$.

Our new coordinate $\xi=\xi(z)$, admits a perturbative expansion in the large WKB parameter $q$:
\begin{equation}
    \xi(z) =\xi_0(z) + q^{-1}\xi_1(z) + \dots
\end{equation}
where each $\xi_i(z)$ obeys a first order differential equation. For example, $\xi_0(z)$ obeys 
\begin{equation}
    \left(\xi'_0(z)\right)^2 W_0(\xi_0(z)) -V_0(z) =0,
\end{equation}
with $W_0(\xi) =1/(4\xi)$. Each $\xi_i(z)$ will contain an integration constant which we set by requiring $\xi_i(0)=0$ for all $i\geq0$. This condition ensures that $\xi(0)=0$.

The solution satisfying the boundary condition at $z=0$ is given in terms of the Bessel function of the first kind by $q\sqrt{\xi}\,J_2(q\sqrt{\xi})$. Then, using \eqref{wavefunc redef}, the linear combination of the original WKB solutions with the correct behaviour as $z \to 0$ must be proportional to the large-$q$ expansion of
\begin{equation}\label{besselsolution}
\,\frac{\,\xi^{1/2}(z)}{(\xi')^{1/2}(z)} J_2(q\xi^{1/2}(z)),
\end{equation}
near $0$.
To find the required linear combination, we first rewrite \eqref{besselsolution} in terms of Bessel functions of the third kind
\begin{equation} \label{Bessel function combination}
      \frac{\xi^{1/2}}{2(\xi')^{1/2}} \left(H_1^{(2)}(q\xi^{1/2})+H_2^{(2)}(q\xi^{1/2})\right).
\end{equation}
Then, let $\varphi_0^\pm$ be the large-$q$ WKB solutions of \eqref{uniform wkb bessel equation} normalised at $0$. One can show that these solutions are have the following relation to  large-$q$ expansions of the Bessel function of third kind:\footnote{For real $z$ and $q$.} 
\begin{equation}
\begin{aligned} \label{wavefunc redef and bessel}
     \varphi_0^+(\xi) &= -\sqrt{\pi} \xi^{1/2} H_1^{(2)}(q\xi^{1/2}), \\
   \varphi_0^-(\xi) &= -i\sqrt{\pi} \xi^{1/2} H_1^{(2)}(q\xi^{1/2}).  
\end{aligned}
\end{equation}
With the normalisation around $\zeta=0$ for $\varphi_0^\pm(\zeta(z))$ and around $z=0$ for $\psi_0^\pm(z)$, one can show that
\begin{equation}\label{wavefunc rel}
    \psi^\pm_0(z) = \frac{1}{\sqrt{\xi'(z)}}\varphi_0^{\pm}(\xi(z)),
\end{equation} 
i.e. $\sigma=1$ in \eqref{wavefunc redef} for these choices of basis. Finally, combining \eqref{wavefunc redef and bessel} and \eqref{wavefunc rel}, the condition that our WKB solutions $\psi_0^\pm (z)$ are proportional to \eqref{Bessel function combination} near $z=0$, requires that we select the combination
\begin{equation}
      \psi^+_0 +i \psi^-_0
\end{equation}
near $z=0$.

Thus, we conclude that $K=\rmi$. Substituting this value into \eqref{app: QC's with C}, we find the EQCs for the QNM frequencies:
\begin{equation}
    \begin{aligned}
        \rme^{-\alpha} +\rme^{\beta}+1=0,\\
        \rme^{\alpha}+\rme^\beta+1=0.
    \end{aligned}
\end{equation}
Rearranging and taking the logarithm, we obtain 
\begin{equation}
    \begin{aligned}
       \alpha =-2\pi \rmi \left(N+\frac{1}{2}\right) -\log(1+\rme^{\beta}),\\
        \alpha =2\pi \rmi \left(N+\frac{1}{2}\right) +\log(1+\rme^{\beta}),
    \end{aligned}
\end{equation}
with $N \in\mathbb{Z}$. Recall that these quantisation conditions were derived with assumptions placed on $s$. The above quantisation conditions are consistent with these assumptions only if we restrict $N<0$ in the first condition and $N>0$ in the second. As a result, we can combine the two quantisation into 
\begin{equation}
     \alpha =2\pi \rmi \left(N+\frac{1}{2}\right) \mp\log(1+\rme^{\beta})
\end{equation}
which is precisely the quantisation condition we presented in the main text.

\subsection{Computing the periods
}
\label{app:genericS}

In this section, we describe the general form of the period 
\begin{equation}
    \alpha = \sum_{m=0}^\infty q^{1-2m}\alpha_m
\end{equation}
where
\begin{equation}
    \alpha_m=\oint_{\gamma_\alpha} S_{\mathrm{odd},\,2m-1}(z)\,\mathrm{d}z .
\end{equation}

The first integral to compute is
\begin{equation}
q\,\oint_{\gamma_\alpha} S_{\mathrm{odd},-1}(z)\,\mathrm{d}z .
\end{equation}
Defining $A^2= 1-s^2$ and $u=z/A$, this can be rewritten as
\begin{equation}\label{leadingperiod2}
q\,\frac{A^{3/2}}{2}\oint_0^{1} u^{-1/2}\,(1-u^2)^{1/2}\,(1-A^2\,u^2)^{-1}\mathrm{d}u.
\end{equation}
This integral is of the generic form
\begin{equation} 
P(k, m, n, A^2)=\oint_\gamma u^k (1-u^2)^m (1-A^2 u^2)^n \, du,
\end{equation}
where $k ,  m \in \mathbb{Z} +1/2$, $n \in \mathbb{Z}_{\le 0}$, and where $\gamma$ is the anticlockwise contour around the branch cut from $u=0$ to $u=1$.

Using the Pochhammer contour representation of the hypergeometric function, the result of the integration can be written as 
\begin{equation}
P(k,m,n,A^2) = \frac{\rme^{\rmi\pi\left(\frac{1}{2} + \frac{k}{2} + m\right)} \,4 \pi^2\,\,{}_2F_1\left(-n, \frac{k}{2} +\frac{1}{2}; \frac{3}{2} +\frac{k}{2} +m; A^2\right)}{2(1+\rme^{\rmi\pi k})\Gamma\left(\frac{1}{2}-\frac{k}{2}\right)\Gamma(-m)\Gamma\left(\frac{3}{2} +\frac{k}{2} +m\right)}\,,
\end{equation}
which for the case in \eqref{leadingperiod2} gives
\begin{equation}q\alpha_0 =
q\,A^{3/2}\,\frac{\Gamma\left(\frac{1}{4}\right)\Gamma\left(\frac{3}{2}\right)}{2\,\Gamma \left(\frac{7}{4}\right)}\,{}_2F_1\left(1,\frac{1}{4};\frac{7}{4};A^2\right).
\end{equation}

We then prove that $\alpha_i$, $i \geq 1$, can be written as sums of integrals of the form 
\begin{equation} \label{form of sodd periods2}
P(k, m)=\oint_\gamma u^k (1-u^2)^m \, du,
\end{equation}
where $k , \, m \in \mathbb{Z} +1/2$,  and where $\gamma$ is the anticlockwise contour around the branch cut from $u=0$ to $u=1$. 

Such an integral can be evaluated as Beta functions,
\begin{equation}
P(k, m) = \frac{2 \rme^{\pi \rmi (m + k/2 +3/2)} \sin(\pi(m+1))\sin(\pi (k/2 + 1/2))}{(\rme^{\pi \rmi k} + 1)} B(m+1, k/2 +1/2).
\end{equation}
To prove this, we start by writing explicitly the first few orders of $S^+$:
\begin{equation}\label{firstS+jterms}
\begin{aligned}
S^+_{-1}(z)&=\sqrt{\frac{s^2-1+z^2}{4 z \left(z^2-1\right)^2}},\\
S^+_{0}(z)&=\frac{s^2 \left(5 z^2-1\right)+3 z^4-4 z^2+1}{4 z \left(z^2-1\right) \left(s^2+z^2-1\right)},\\
S^+_{1}(z)&=\frac{\left(17-10 s^2\right) z^4+15 \left(s^2-1\right)^2+3 \left(3 s^4+8 s^2-11\right) z^2+z^6}{16 z^{3/2} \left(s^2+z^2-1\right)^{5/2}},\\
S^+_{2}(z)&=\frac{1}{16 z^2 \left(s^2+z^2-1\right)^4}\Bigl[3 s^6 \left(9 z^4+2 z^2+5\right)+s^4 \left(-77 z^6+45 z^4+45 z^2-45\right)\\
&\ \ +s^2 \left(17 z^8+28 z^6+18 z^4-108 z^2+45\right)+\left(z^2-1\right)^2 \left(z^6-21 z^4+27 z^2-15\right)\Bigr].
\end{aligned}
\end{equation}
The first few terms of $S^+_i$ motivate the assumption that for $i \geq 1$, $S^+_i$ is of the form
\begin{equation} \label{general S_i}
    S^+_i(z)=\frac{P_{2i+1}(z^2)}{z^{1+i/2 }(z^2-A^2)^{1+3i/2}}
\end{equation}
where $P_n(z)$ is a polynomial of degree $n$.
We prove this by induction. The base of the induction is provided by the explicit expressions for $S^+_1$ and $S^+_2$ in \eqref{firstS+jterms}, which are both of the form \eqref{general S_i}. We then assume that \eqref{general S_i} holds for all $1\leq j\leq i$, with $i\geq 2$, and prove it for $S^+_{i+1}$.
For $i \geq 2$, $V_i(z)=0$ and the recurrence relation for $S_{i+1}$ is
\begin{equation}
    S^+_{i+1} =-\frac{1}{2 S^+_{-1}}\left((S_i^+)'+\sum_{j=0}^iS^+_j S^+_{i-j} \right).
\end{equation}
Substituting the induction hypothesis \eqref{general S_i} and the value of $S_0(z)$ calculated into the above equation, we obtain
\begin{equation}
     S^+_{i+1}=\frac{P_{2i+3}(z^2)}{z^{(3+i)/2}(z^2-A^2)^{(5+3i)/2}},
\end{equation}
where, writing $x=z^2$, the numerator is given by
\begin{equation}\label{polynomial}
\begin{aligned}
P_{2i+3}(x)=&-(x-1)\Bigg[2x(x-A^2)P'_{2i+1}(x)-\left(\left(3+\frac{7i}{2}\right)x-\left(1+\frac{i}{2}\right)A^2\right)P_{2i+1}(x) \\
&+\sum_{j=1}^{i-1}P_{2j+1}(x)P_{2(i-j)+1}(x)
\Bigg]-\frac12\left(3x^2+(1-5A^2)x+A^2\right)P_{2i+1}(x),
\end{aligned}
\end{equation}
where the prime denotes differentiation with respect to $x$. Then if for $1\leq j \leq i$, $P_{2j+1}(z)$ is a polynomial of degree $2j+1$ the RHS of \eqref{polynomial} is a polynomial in $z^2$ of degree $2i+3$. Hence, \eqref{general S_i} is true, and we can use the recurrence relation \eqref{polynomial} to compute many terms of $S^+(z)$ and thus $S_{\text{odd}}(z)$.
The important point for the period computation is that, for $i\geq 1$, no negative powers of $(z^2-1)$ occur in \eqref{general S_i}. Therefore, after setting $z=A u$, the higher-order contributions to the period do not contain factors of the form $(1-A^2u^2)^n$ with $n<0$. The only branch-point structure relevant for these terms is generated by $u=0$ and $u=1$, and the corresponding integrals reduce to the Beta-type integrals \eqref{form of sodd periods2}.
Equation \eqref{general S_i} implies that, for $i\geq 1$, the coefficients entering $S_{\mathrm{odd}}$ are linear combinations of terms of the form
\begin{equation}
 \frac{z^{2r}}{z^{1+i/2}(z^2-A^2)^{1+3i/2}},
\end{equation}
with coefficients polynomial in $A^2$. Setting $z=A u$, each such term becomes an overall power of $A$ multiplying an integral of the form
\begin{equation}
\oint_\gamma u^k(1-u^2)^m\,\mathrm{d}u,
\end{equation}
with $k,m\in \mathbb{Z}+1/2$. These integrals are precisely those evaluated in \eqref{form of sodd periods2}, and hence reduce to Beta functions. Applying this to the coefficient
\begin{equation}
\alpha_m=\oint_{\gamma_\alpha}S_{\mathrm{odd},2m-1}(z)\,\mathrm{d}z
\end{equation}
gives a finite expansion of the form
\begin{equation}
    \alpha_m = \sum_{k=0}^{2m-1}c_{m,2m-k}A^{2k-3m}.
\end{equation}
The constants $c_{m,2m-k}$ are coefficients obtained from the Beta-function integrals and from the polynomial coefficients generated by the recurrence relation. Combining these terms, the full period can therefore be organised as in \eqref{largeqexpansionofalpha}.

\section{Details on large \texorpdfstring{$q$}{} and \texorpdfstring{$N$}{} expansions} \label{app:large q and N expansion}

In this appendix we present some technical aspects of the large$-q$ series for the QNM frequencies which were used in the main text.

\subsection{Composition of series and Bell polynomials} \label{app:CompOfseries}

Throughout the derivation of the large-$q$ series of $s$, we often need the $n^{th}$ term of a composition of two series. This can be derived by specialising Fa\'a di Bruno's formula. 
Let 
\begin{equation}
    \begin{aligned}
        g(x) &= \sum_{n=0}^\infty g_n x^n,\\
         f(x) &= \sum_{n=1}^\infty f_n x^n.
    \end{aligned}
\end{equation}
Then, the series $g(f(x))$ is given by 
\begin{equation} \label{FaadiBruno}
    g(f(x)) = g_0 + \sum_{n=1}^\infty\left(\sum_{k=1}^n g_k k!B_{n,k}(f_1, 2!f_2, \dots, (n-k+1)!f_{n-k+1})\right) \frac{x^n}{n!} ,
\end{equation}
where $B_{n,k}$ are called the \emph{partial exponential Bell polynomials} \cite{bell}. These polynomials may seem difficult to work with; however, they obey recurrence relations, which make them fast to compute if the arguments of polynomials are numerical values. The recurrence relation is given by 
\begin{equation}\label{PartialBellRecRel}
\begin{aligned}
    B_{n+1,k+1}(f_1, \dots,(n-k+1)! f_{n-k+1} ) = \sum_{i=0}^{n-k}\binom{n}{i}&\,(i+1)!\,f_{i+1} \\
    & B_{n-i,k}(f_1,  \dots, (n-k-i+1)! f_{n-k-i+1}).
\end{aligned}
\end{equation}
where
\begin{equation}
    \begin{aligned}
        B_{0,0}&=1;\\
        B_{n,0}&=0, \, n\geq1;\\
        B_{0,k}&=0, \, k\geq 1.
    \end{aligned}
\end{equation}

The composition \eqref{FaadiBruno} simplifies if $g(x) =\exp(x)$. In this case, one has
\begin{equation}
    \exp\left(f(x)\right) = \sum_{n=0}^\infty B_n(f_1,2! f_2, \dots, n! f_n)\frac{x^n}{n!}
\end{equation}
where $B_n$ are the \emph{complete exponential Bell polynomials}. These are given by 
\begin{equation}
    B_n(f_1, \dots, n! f_n)= \sum_{k=0}^nB_{n,k}(f_1, 2! f_2, \dots,(n-k+1)!f_{n-k+1})
\end{equation}
and satisfy the recurrence relation
\begin{equation}
    B_{n+1}(f_1, \dots, n! f_n)= \sum_{i=0}^n\binom{n}{i} (i+1)!f_iB_{n-i}(f_1, \dots, (n-i)! f_{n-i}),
\end{equation}
where
\begin{equation}
    B_0=1.
\end{equation}

We can use this technology to compute $D_l (E^{4k/3})$.
Using Fa\'a di Bruno's formula \eqref{FaadiBruno} with 
\begin{equation}
\begin{aligned}
    g(x) &= (1+x)^{4k/3},\\
    f(x)&=\sum_{n=1}^\infty E_nx^n,
\end{aligned}
\end{equation}
the composition $g(f(x))$ is given by
\begin{equation}\label{EPowerExp}
   \left(1+\sum_{n=1}^\infty E_nx^n\right)^{4k/3} = 1 + \sum_{l=1}^\infty \left(\sum_{p=1}^l \binom{4k/3}{p}p! B_{l,p}(E) \right)\frac{x^l}{l!}
\end{equation}
where $B_{l,p}(E)$ is given by equation \eqref{BellpolyE}. From equation \eqref{EPowerExp}, we obtain
\begin{equation}
    D_l\left(E^{4k/3}\right) = \frac{1}{l!}\sum_{p=0}^l \binom{4k/3}{p}p!B_{l,p}(E),
\end{equation}
and we can compute this recursively using the recurrence relation \eqref{PartialBellRecRel}.

\subsection{Deriving the transseries for \texorpdfstring{$\mathcal{E}$}{}}

To calculate the non-perturbative corrections to the large-$N$ expansion \eqref{eq:Large_N_expansion_of_mathcal_E} of $\mathcal{E}$, we upgrade this expansion to a transseries
\begin{equation} \label{E small delta expansion}
    \mathcal{E}^\pm = \sum_{n=0}^\infty\delta^n \mathcal{E}^{(n)}_\pm.
\end{equation}   
$\mathcal{E}_\pm^{(0)}$ is the perturbative series \eqref{eq:Large_N_expansion_of_mathcal_E} for both $+$/$-$ and $\delta$ is a quantity that is non-perturbative for large-$N$ to be specified. 

To find $\mathcal{E}_\pm^{(n)}$ for $n\geq1$, we first note that $\mathcal{E}_\pm^{(0)}(N+1/2)$ is the series inverse to $\alpha_0$, see \eqref{alphanexpanded}, meaning that
\begin{equation} \label{mathcal E inverse}
    \mathcal{E}_\pm^{(0)}\left(N+\frac{1}{2}\right)=\alpha_0^{-1}\left(N+\frac{1}{2}\right).
\end{equation}
Including the non-perturbative corrections \eqref{T transseries} to the perturbative condition $\alpha_0$ \eqref{hbar expansion of alpha}, yields the following equation obeyed by $\mathcal{E}^\pm$
\begin{equation} \label{alpha0 np condition}
    \alpha_0(\mathcal{E}^\pm)= 2 \pi \mathrm{i}\left(N+\frac{1}{2}\right)\mp \log\left(1+\rme^{\beta_0(\mathcal{E^\pm})}\right).
\end{equation}
Then equation \eqref{alpha0 np condition} together with \eqref{mathcal E inverse} implies that the transseries $\mathcal{E}^\pm$ obeys the implicit equation
\begin{equation} \label{mathcal E inverse 2}
\mathcal{E}^{\pm}\left(N+\frac{1}{2}\right)=\mathcal{E}^{(0)}_\pm\left(N+\frac{1}{2}\mp \frac{1}{2\pi \rmi}\log\left(1+\rme^{\beta_0(\mathcal{E}^\pm)}\right)\right).
\end{equation}
Then $\mathcal{E}^{(n)}_\pm$ can be found by expanding both sides for small $\delta$ and equating the $n^{th}$ coefficients of these expansions. To expand the RHS of \eqref{mathcal E inverse 2} we need to calculate the small $\delta$ expansion of $\log\left(1+\rme^{\beta_0(\mathcal{E^\pm})}\right)$. The small $\delta$ expansion of $\beta_0(\mathcal{E^\pm)}$ is of the form 
\begin{equation}
    \beta_0(\mathcal{E}^\pm) = \sum_{n=0}^\infty\beta_0^{(n)}\delta^{n},
\end{equation}
where 
\begin{equation}
    \beta_0^{(n)}= \frac{1}{n!} \sum_{k=0}^n \left.\frac{d^k\beta_0(x)}{dx^k} \right |_{x=\mathcal{E}_\pm^{(0)}(N+1/2)}B_{n,k}\left(1!\mathcal{E}_\pm^{(1)},\dots, (n-k+1)! \mathcal{E}_\pm^{(n-k+1)}\right).
\end{equation}
Using this, the small $\delta$ expansion of $\rme^{\beta_0(\mathcal{E})}$ can be written as
\begin{equation}
    \rme^{\beta_0(\mathcal{E}^\pm)}= \delta\sum_{n=0}^\infty \delta^n
B_n\left(\beta_0\right),
\end{equation}
where
\begin{equation}
    B_n\left(\beta_0\right) = B_n\left(1! \beta_0^{(1)},\dots, n! \beta_0^{(n)}\right).
\end{equation}
and we have defined $\delta$ as the series
\begin{equation}
   \delta \equiv \exp\left(\beta_0^{(0)}\left(\mathcal{E}_\pm^{(0)}\right) \right),
\end{equation}
which is indeed non-perturbative for large $N$ as
\begin{equation}
\beta_0^{(0)}\left(\mathcal{E}_\pm^{(0)}\right) = -2 \pi \left(N+\frac{1}{2}\right) + \mathcal{O}\left(\left(N+\frac{1}{2}\right)^{-1}\right).
\end{equation}
Finally, we can write the small delta expansion of the logarithm 
\begin{equation} \label{log small delta expansion}
\log\left(1+\rme^{\beta_0}\right)= \sum_{n=1}^\infty L_n \delta^n,
\end{equation}
where the coefficients $L_n$ are given by
\begin{equation}
    L_n= \frac{1}{n!}\sum_{k=1}^n(-1)^{k+1}(k-1)!B_{n,k}\left(1, 2B_1(\beta_0),\dots, (n-k+1)B_{n-k}(\beta_0)\right).
\end{equation}
Expanding both sides of \eqref{mathcal E inverse 2} for small $\delta$ then gives the following recursive formula for $\mathcal{E}^{(n)}_\pm$
\begin{equation} \label{mathcal E transseries rec}
\mathcal{E}_\pm^{(n)}\left(N+\frac{1}{2}\right)=\frac{1}{n!} \sum_{k=1}^n \left. (\mp1)^k\frac{d^k \mathcal{E}_\pm^{(0)}(x)}{d x^k}\right \vert_{x=N+1/2} B_{n,k}(1! L_1, \dots, (n-k+1)! L_{n-k+1}).
\end{equation}

To calculate the resummed transseries sector $\mathcal{S}_\pm [\mathcal{E}^{(n)}_\pm ](N+1/2)$ we first calculate $\mathcal{S}_\pm [\mathcal{E}^{(0)}_\pm](N+1/2)$ and its first $n$ derivatives. Next, we calculate $\mathcal{S}_\pm [\beta_0](x)$ and its first $n$ derivatives all evaluated at $x = \mathcal{S}_\pm [\mathcal{E}^{(0)}](N+1/2)$. The resummation of $\delta$ is then found by exponentiating $\mathcal{S}_\pm [\beta_0](x)$. Finally, the recurrence relation \eqref{mathcal E transseries rec} can be used to compute  $\mathcal{S}_\pm[\mathcal{E}^{(n)}_\pm ](N+1/2)$.

\section{Interpolation via analytic continuation: the Airy function example}\label{app:Airy}

In this appendix we will use the Airy function to exemplify how one can analytically continue the results from an asymptotic region $\kappa\gg1$
all the way to $\kappa=0$ making use of the Taylor-series method of analytic continuation, see \cite{NIST:DLMF} \S $3.7(ii)$.

The linear ODE describing the Airy function is 
\begin{equation}
Z^{''}\left(\kappa\right)-\kappa\,Z(\kappa)=0.\label{eq:Airy-ODE}
\end{equation}
If we take an ansatz of the form $Z\left(\kappa\right)=\mathrm{e}^{-\frac{1}{2}A\,\kappa^{\alpha}}\Phi\left(\kappa\right)$
where 
\begin{equation}
\Phi\left(\kappa\right)\simeq\kappa^{\beta}\sum_{\ell=0}^{+\infty}a_{\ell}\,\kappa^{-\alpha\,\ell},
\end{equation}
we easily find from (\ref{eq:Airy-ODE}) that there are two values of $A$ allowed
\begin{equation}
A_{\pm}=\pm A\:,\quad A=\frac{4}{3}.
\end{equation}
We will also find $\alpha=3/2$ and $\beta=-1/4$. 
This, together with the fact that we are studying a linear ODE, tells us that we will have two independent solutions.
 The full solution will be a two-parameter transseries given by
\begin{equation}
Z\left(\kappa,\sigma_{1},\sigma_{2}\right)=\sigma_{1}\,Z_{\mathrm{Ai}}(\kappa)+\sigma_{2}\,Z_{\mathrm{Bi}}(\kappa)\,,\label{eq:Airy-transseries}
\end{equation}
 where
\begin{align}
Z_{\mathrm{Ai}}(\kappa) & =\frac{1}{2\sqrt{\pi}\kappa^{1/4}}\,\mathrm{e}^{-\frac{1}{2}A\kappa^{3/2}}\,\Phi_{-\frac{1}{2}}\left(\kappa\right);\nonumber \\
Z_{\mathrm{Bi}}\left(\kappa\right) & =\frac{1}{2\sqrt{\pi}\kappa^{1/4}}\,\mathrm{e}^{\frac{1}{2}A\kappa^{3/2}}\,\Phi_{\frac{1}{2}}\left(\kappa\right),\label{eq:Airy-expansion-sectors}
\end{align}
and 
\begin{equation}
\Phi_{\pm\frac{1}{2}}\left(\kappa\right)\simeq\sum_{n=0}^{+\infty}a_{n}^{(\pm1/2)}\,\kappa^{-\frac{3}{2}n}\,,\label{eq:Airy-asymptotic-sectors}
\end{equation}
with coefficients 
\begin{equation}
a_{n}^{(\pm1/2)}\equiv(\pm1)^{n}\,a_{n}\,,\quad a_{n}=\frac{1}{2\pi}A^{-n}\frac{\Gamma\left(n+\frac{5}{6}\right)\Gamma\left(n+\frac{1}{6}\right)}{n!}.\label{eq:Airy-coefficients}
\end{equation}
These coefficients are factorially divergent
with a subleading exponential growth. An analysis of the resurgent properties of the Airy solution can be found, e.g., \cite{10.1007/978-88-7642-613-1_1}.

We will focus on the solution of the differential equation given by a resummation of the $Z_{\mathrm{Ai}}(\kappa)$ series. That solution
can be approximated at a neighbourhood of a finite value $\kappa_{0}$ by a (convergent) Taylor series
\begin{equation}
\mathrm{Ai}(\kappa)=\sum_{\ell=0}^{+\infty}\frac{b_{\ell}^{(0)}}{\ell!}(\kappa-\kappa_{0})^{\ell}.\label{eq:Airy-Taylor-series-kappa0}
\end{equation}
The coefficients $b_{k}^{(0)}$ obey a recursion relation which can be determined by using the above expansion in the original ODE (\ref{eq:Airy-ODE}). The recursion relation is:
\begin{equation}
b_{\ell+2}^{(0)}=\ell\,b_{\ell-1}^{(0)}+\kappa_{0}\,b_{\ell}^{(0)}\:,\:\ell\ge0\,,\label{eq:recursion-k0}
\end{equation}
with $b_{0}^{(0)}$ and $b_{1}^{(1)}$ being initial conditions.

Here we are interested in showing that 
\begin{enumerate}
\item we can determine the coefficients of this Taylor series through resummations
of the $Z_{\mathrm{Ai}}(\kappa)$ series, for a value of $\kappa_{0}\sim\mathcal{O}(1)$;
\item once we have the coefficients of this Taylor series at $\kappa_{0}$ we can perform analytic continuation at the way to $\kappa=0$ in two distinct ways:
\begin{enumerate}
\item determining a Padé approximant and evaluating at $\kappa=0$ 
\item using the Taylor-series method of analytic continuation to ``walk'' all the way to $\kappa=0$.
\end{enumerate}
\end{enumerate}
%


The coefficients of the Airy function at $\kappa=\kappa_{0}$ are given by its derivatives evaluated at that point:
\begin{equation}
b_{k}^{(0)}=\left.\frac{d^{k}\mathrm{Ai}(\kappa)}{d\kappa^{k}}\right|_{\kappa_{0}}\,.
\end{equation}
We know that the value of the function at $\kappa_{0}$ can be obtained from the asymptotic expansion through Borel resummation
\begin{equation}
\mathrm{2\sqrt{\pi}\kappa^{1/4}\,\mathrm{e}^{\frac{1}{2}A\kappa^{3/2}}Ai}(\kappa)=\mathcal{S}_{\theta}\Phi_{-\frac{1}{2}}\left(\kappa\right)=\int_{0}^{\mathrm{e}^{\mathrm{i}\theta}\infty}d\xi\,\mathcal{B}\left[\Phi_{-\frac{1}{2}}\right]\left(\xi\right)\mathrm{e}^{-\kappa^{3/2}\,\xi}.
\end{equation}
The derivatives of the function at this point can be determined iteratively in much the same way, by noting that
\begin{align}
\frac{d^{\ell}}{d\kappa^{\ell}}\left(\mathrm{2\sqrt{\pi}\kappa_{}^{1/4}\,\mathrm{e}^{\frac{1}{2}A\kappa^{3/2}}Ai}(\kappa)\right) & =\frac{d^{\ell}}{d\kappa_{}^{\ell}}\int_{0}^{\mathrm{e}^{\mathrm{i}\theta}\infty}d\xi\,\mathcal{B}\left[\Phi_{-\frac{1}{2}}\right]\left(\xi\right)\mathrm{e}^{-\kappa^{3/2}\,\xi}\nonumber \\
 & =-\frac{3}{2}\frac{d^{\ell-1}}{d\kappa^{\ell-1}}\kappa^{1/2}\int_{0}^{\mathrm{e}^{\mathrm{i}\theta}\infty}d\xi\,\xi\,\mathcal{B}\left[\Phi_{-\frac{1}{2}}\right]\left(\xi\right)\mathrm{e}^{-\kappa^{3/2}\,\xi}\,,
\end{align}
and iterating the procedure until we have the expression for the $\ell^{\mathrm{th}}$
derivative of $\mathrm{Ai}$ at $\kappa$.
If up to $N$ terms of the Borel transform are available, we use a diagonal Pad{\'e} approximant to analytically continue it along the path of integration.

\subsection{Analytic continuation to \texorpdfstring{$\kappa=0$}{} via Padé approximants}

For simplicity we shall work with the normalised version of the Airy function
\begin{equation}
\mathrm{Ai_{R}}(\kappa):=\mathrm{2\sqrt{\pi}\kappa^{1/4}\,\mathrm{e}^{\frac{1}{2}A\kappa^{3/2}}Ai}(\kappa)\,.
\end{equation}
This normalised function will inherit the same behaviour as the original Airy function. In a slight abuse of notation, we will assume that
the Taylor series in (\ref{eq:Airy-Taylor-series-kappa0}) is the Taylor series for the normalised function.

\begin{figure}
    \centering
    \includegraphics[width=0.6\linewidth]{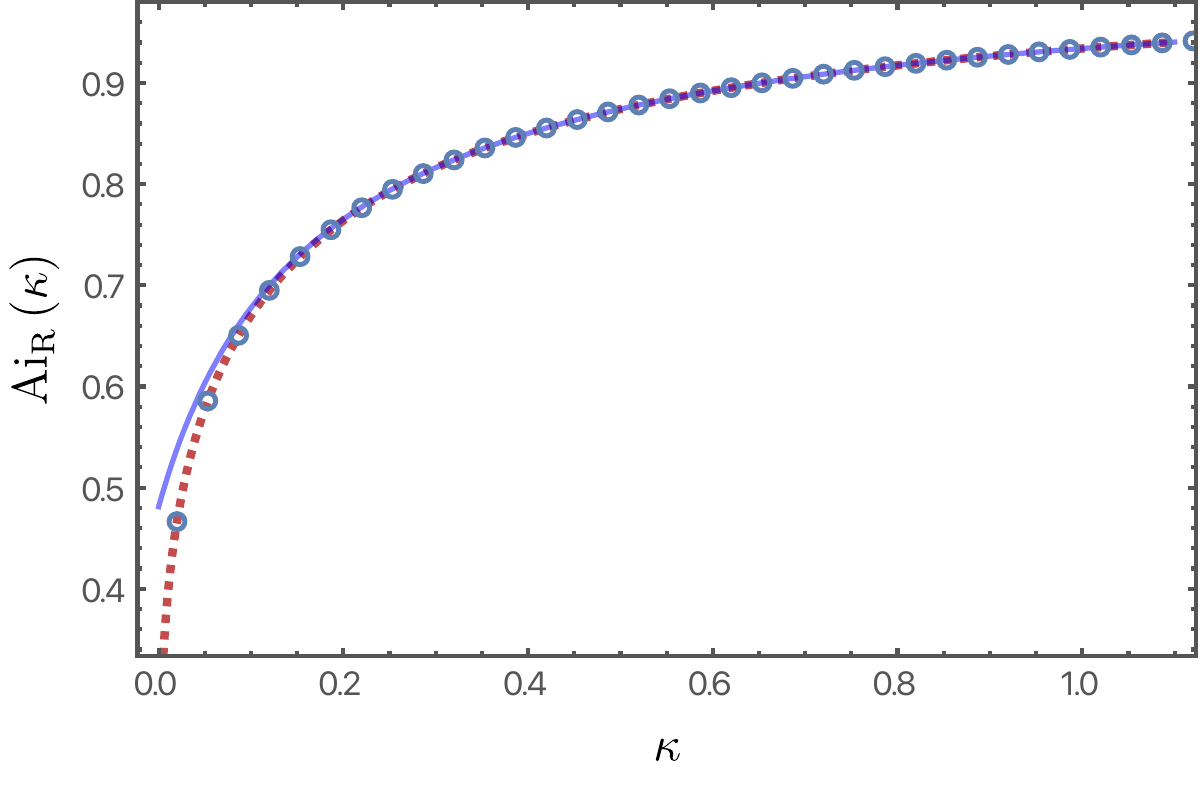}
    \caption{Comparison of different analytic continuations (in blue) and exact numerical value of the Airy function (dashed red line) in the range $\kappa\in[0,1]$, starting from a Taylor series of 10 terms at $\kappa=4^{2/3}$. The blue line shows the Padé approximant method and the blue circles show the Taylor-series method results.}
    \label{fig:Airy-an-cont}
\end{figure}

Taking $\kappa_{0}=4^{2/3}$ and 60 terms in the expansion (\ref{eq:Airy-asymptotic-sectors})
and approximating the Borel transform by a diagonal Padé approximant, the Borel resummations will give us the coefficients $b_{k}^{(0)}$
of (\ref{eq:Airy-Taylor-series-kappa0}). We then calculate up to 10 terms in the Taylor expansion and determe the Padé approximation of that series, to compare to the exact function $\mathrm{Ai_{R}}(\kappa)$. We present this comparison close to the origin, in the range $\kappa\in[0,1]$, in figure \ref{fig:Airy-an-cont}. We
find that the Padé approximant correctly approximates the solution away from the origin (and closer to $\kappa_{0}$) but we do not have enough terms in the Taylor expansion to determine the value of the function at $\kappa=0$.

\subsection{Analytic continuation to \texorpdfstring{$\kappa=0$}{} via the Taylor-series method}

The Taylor-series method of analytic continuation uses the Taylor expansion at a point $\kappa_{0}$ and the fact that we have an ODE to keep on re-calculating the higher derivatives in the Taylor expansion at a nearby point $\kappa_{1}=\kappa_{0}+\varepsilon$,
where $|\varepsilon|$ is a small (complex) iteration. The method works by using the known Taylor expansion at $\kappa_{0}$ to approximate
the initial conditions needed to solve the ODE at $\kappa_{1}$. The Airy ODE is of second order and thus we need both $\mathrm{Ai_{R}}(\kappa_{1})$ and $\mathrm{Ai_{R}}'(\kappa_{1})$. Once we have these values, we can use the ODE to generate all higher derivatives using the recursion relations (\ref{eq:recursion-k0}) at $\kappa=\kappa_{1}$. This procedure can be repeated keeping error propagation to a minimum, as long as we keep the iteration step $\varepsilon$ small. This step can be taken in any direction in the complex plane, which is particularly important if attempting analytic continuation around a singularity. The results
of this procedure are also shown in figure \ref{fig:Airy-an-cont}, where we started at $\kappa_0=4^{2/3}$ and used a small step $\varepsilon=-1/30$. We can see that the analytic continuation can be done accurately along the real axis all the way to zero. 

Note that an identical procedure can be performed if one has an implicit function, such as the truncated curve determined in the main text. In that case, we are interested in solving for the zeros $\omega(q)$
of a high order polynomial in both $\omega$ and $q$. The initial condition is given by the choice of a zero $\omega_{0}(q_0)$ at some value $q=q_0$, and then the Taylor series of $\omega(q)$ around $q_{0}$ can be determined recursively from the curve via a series of linear equations for the coefficients of the Taylor expansion.

\bibliographystyle{ytphys}
\bibliography{biblio}

\end{document}